\documentclass[aps,pra,showpacs,10pt,twocolumn,twoside,,nofootinbib]{revtex4-2}
\usepackage{titlesec}
\usepackage[colorlinks=true, citecolor=blue, urlcolor=blue ]{hyperref}

\usepackage{epsfig,newlfont,amssymb,amsfonts,amsmath,bm,subfigure,palatino,mathtools,amsthm,braket,soul,enumitem,color,times,comment,float,array,physics,mathrsfs,bbm}

\usepackage{booktabs}      
\usepackage{tabularx}
\usepackage{array}         
\usepackage{multirow}      
\usepackage{ragged2e}      
\newcolumntype{L}{>{\RaggedRight\arraybackslash}X}
\newcolumntype{C}{>{\Centering\arraybackslash}X}

\usepackage{array}
\usepackage{multirow}
\usepackage{ragged2e}

\usepackage[table]{xcolor}
\usepackage[normalem]{ulem}
\titleclass{\subsubsubsection}{straight}[\subsection]
\usepackage[utf8]{inputenc}

\newcounter{subsubsubsection}[subsubsection]
\renewcommand\thesubsubsubsection{\thesubsubsection.\arabic{subsubsubsection}}

\titleformat{\subsubsubsection}
  {\normalfont\normalsize\bfseries}{\thesubsubsubsection}{1em}{}
\titlespacing*{\subsubsubsection}
{0pt}{3.25ex plus 1ex minus .2ex}{1.5ex plus .2ex}

\makeatletter
\renewcommand\paragraph{\@startsection{paragraph}{5}{\z@}%
  {3.25ex \@plus1ex \@minus.2ex}%
  {-1em}%
  {\normalfont\normalsize\bfseries}}
\renewcommand\subparagraph{\@startsection{subparagraph}{6}{\parindent}%
  {3.25ex \@plus1ex \@minus .2ex}%
  {-1em}%
  {\normalfont\normalsize\bfseries}}
\def\toclevel@subsubsubsection{4}
\def\toclevel@paragraph{5}
\def\toclevel@subparagraph{6}
\def\l@subsubsubsection{\@dottedtocline{4}{7em}{4em}}
\def\l@paragraph{\@dottedtocline{5}{10em}{5em}}
\def\l@subparagraph{\@dottedtocline{6}{14em}{6em}}
\makeatother

\addtocontents{toc}{\protect\thispagestyle{empty}}
\makeatletter
\renewcommand\@dotsep{140}   
\makeatother

\usepackage[T1]{fontenc}

\usepackage{booktabs}
\usepackage{array}

\usepackage{hyperref}
\usepackage{braket}

\begin{document}

\title{Journey in quantum metrology and sensing from foundations to applications:  a review}

\author{Priya Ghosh$^1$, Tanoy Kanti Konar$^{1,2}$, Debraj Rakshit$^1$, Aditi Sen(De)$^1$, Ujjwal Sen$^1$}
\affiliation{$^1$Harish-Chandra Research Institute, 
Chhatnag Road, Jhunsi, Allahabad 211 019, India\\
Homi Bhabha National Institute, Training School Complex, Anushakti Nagar, Mumbai 400 094, India}
\affiliation{$^{2}$Instytut Fizyki Teoretycznej, Wydzia\l{} Fizyki, Astronomii i Informatyki Stosowanej, Uniwersytet Jagiello\'nski, \L{}ojasiewicza 11, PL-30-348 Krak\'ow, Poland}

\begin{abstract}

We present a review on quantum metrology and sensing, from its foundations to current applications. Highlights of the review include consideration of both frequentist and Bayesian approaches to  parameter estimation; single as well as multiparameter estimation; estimation for different encoding processes comprising unitary as well as noisy channels, quantum thermometry, and channels involving indefinite causal order; different estimation strategies incorporating also recent advances like quantum error correction-aided methods and reservoir engineering; usefulness of quantum Fisher information to detect resources; applications of quantum metrology in diverse arenas covering quantum many-body sensors, sensing protocols in atomic ensembles, atom-photon systems, and continuous-variable systems, quantum imaging, quantum illumination, atomic clocks and atom interferometry, etc; and experimental realizations of quantum sensors in different physical platforms. 

\end{abstract}

\maketitle

\section{\textbf{Introduction}}
\label{sec:intro}
Since the early developments  of quantum mechanics in the beginning of the twentieth century and their subsequent experimental verifications, the principles of quantum theory have led to the emergence of a new paradigm in physics. The ability to harness quantum mechanical effects has opened up a wide range of possibilities, ranging from fundamental explorations of nature to practical applications that were previously difficult or unachievable.  In particular, the exploitation of quantum resources has led to remarkable advances in several areas, including quantum computation, quantum communication, quantum simulation, and quantum metrology.

A fundamental aspect of science is the fact-based assessment of physical phenomena, which heavily relies on experimental innovations. In every experiment, be it in biology, astrophysics, particle physics, or other domains, sensors are essential for measuring parameters. The primary role of a sensor is to estimate these parameters as precisely as possible. Often, experimental breakthroughs are closely linked to the sensitivity of these sensors. However, due to inherent constraints,
achieving perfect parameter estimation with sensors is fundamentally impossible, with finite resources. In order to acquire the most accurate estimations of physical parameters, including magnetic, electric, and gravitational fields, rotation angles,  noise levels, frequency, temperature, and so on, sensors must be carefully selected and designed to minimize  estimation errors to the greatest extent feasible. To address this, C. W. Helstrom and A. S. Holevo around 1970~\cite{Helstrom67,Helstrom68,Helstrom69,Holevo82}, basing on earlier work in the 1940s by H. Cram{\'e}r and C. R. Rao~\cite{Cramer46,Rao1992}, provided  theoretical tools and methodologies to find sensors that maximize precision in estimation of a parameter or a set of parameters simultaneously present in a process.

At the end of the twentieth century, the discovery that systems with a quantum resource might potentially perform parameter estimation more precisely than systems that lack such a resource
gave rise to a new branch of study, viz. quantum metrology and sensing. From interferometry to magnetometry to thermometry, this quantum advantage has been confirmed in a variety of fields. It has also been experimentally validated in a number of physical platforms, including optical setups, nuclear magnetic resonance (NMR), nitrogen vacancy (NV) centers, and superconducting circuits, to mention a few. Further, real-world applications of quantum sensing protocols include the development of quantum sensing devices for atomic clocks,  crucial, e.g., for high-bit-rate quantum cryptography, high-precision gravity measurements, quantum radar, and gravitational wave detection.

There are several excellent review articles~\cite{Paris2009Jan,Petz2011Jan,Giovannetti11,Toth2014Oct,Degen17,Pezze18,pirandola_rmp,Pirandola2018Dec,Mehboudi19a,Liu20,Sidhu2020,Albarelli20,Demkowicz-Dobrzanski2020Aug,nisq_sesing_review,barbieri-1,
particle_physics_sensing,Jun-Zollar-review,Huang2024Sep,montenegro2024review,cv_metrology,Liu2024Dec,Mukhopadhyay2025Feb,Fadel2025,Mihailescu25b} that address various important fields of quantum metrology and sensing, covering theoretical advances to experimental progresses. These reviews typically focus on specific areas of this field. Some delve into fundamental theoretical concepts of quantum parameter estimation \cite{Petz2011Jan,pirandola_rmp,Toth2014Oct}, including multiparameter estimation~\cite{Liu20,Demkowicz-Dobrzanski2020Aug}, parameter estimation in quantum many-body (QMB) systems~\cite{montenegro2024review,Mihailescu25b}, global parameter estimation~\cite{Mukhopadhyay2025Feb}, and continuous-variable system parameter estimation~\cite{cv_metrology,Fadel2025}. Others concentrate on more specialized topics, such as noisy quantum metrology, estimation strategies, 
geometrical perspectives~\cite{Sidhu2020}, and applications in fields like particle physics~\cite{particle_physics_sensing}, NISQ devices~\cite{nisq_sesing_review}, and many more.
Since its early days, quantum sensing has gradually evolved from foundational studies to deployable technologies and has evolved as one of the fastest-growing directions in quantum science and technology. In parallel, the idealized unitary estimation scenario has increasingly shifted toward more practical, and thus complex, settings that account for environmental noise, many-body effects, and platform-specific protocols.

 The significance of quantum sensors to modern technology demands a review article that traces the journey of theoretical and experimental advances in this field across a range of platforms.
In our review, we 
intend to provide a comprehensive and up-to-date exploration of quantum metrology and sensing. Our work encompasses major aspects of quantum metrology and sensing, including theoretical foundations, cutting-edge techniques, practical applications, experimental progress, emerging technologies, and more, offering a broad perspective on this subject, particularly emphasizing recent significant progress and directions.

The review proceeds along the path outlined below. It begins with an introduction to the fundamentals of classical parameter estimation theory before moving into a rigorous mathematical framework for the quantum scenario, covering both single- and multiparameter cases. Within both classical and quantum settings, we discuss the two key approaches for determining the lower bound on the mean square error of an estimator, viz. the frequentist and Bayesian methods. We explore modern techniques designed to enhance sensitivity in parameter estimation, along with quantum resources that provide a quantum advantage in this context.

Next, we illustrate the geometrical interpretation of the quantum parameter estimation problem for both single and multiple parameters. We then systematically describe the quantum Cram{\'e}r-Rao bound across various encoding processes, including unitary evolution, quantum channels, and indefinite causal order of maps. In many encoding scenarios, directly identifying the fundamental limit on estimation error is highly challenging. To address this, numerous estimation strategies, such as adaptive techniques, Floquet engineering, sequential measurements, and quantum error correction-based methods, have been developed, all of which we discuss in detail.
Since noise is an inherent challenge in any estimation problem, we further explore strategies for mitigating its effects in quantum parameter estimation. 
Our review covers key advances in noisy quantum metrology, including error mitigation techniques and optimal encoding strategies that maintain high precision even in the presence of decoherence.

Beyond these advancements, we highlight the broad range of applications of quantum metrology, from detecting phase transitions and witnessing entanglement to estimating physical quantities in diverse quantum systems. We also provide a detailed exploration of its applications in various fields, including quantum imaging, quantum illumination, atomic clocks, particle physics, many-body systems, non-Hermitian systems, and many more.

Additionally, we offer a comprehensive overview of experimental realizations, covering implementations in photonic setups, ultacold atoms and trapped ions, solid-state platforms, including NMR,  NV centers and quantum dots, and superconducting circuits. 

Finally, we conclude by summarizing the key insights from our review, identifying open challenges, and discussing future near-term research directions in quantum metrology and sensing. Through this review, we aim to provide a comprehensive and up-to-date perspective on quantum parameter estimation, bridging foundational theory, modern technologies, experimental realizations, and emerging applications.

{
  \hypersetup{linkcolor=blue}
  \tableofcontents
}


\section{\textbf{Acronyms and Symbols}}

\begin{enumerate}
    \item $\iota$ $\coloneqq \sqrt{-1}$
    \item $\mathrm{AAAS}$: Active auxiliary-assisted scheme
  \item $\mathrm{AMSE}$: Average mean square error
  \item $\mathrm{BCRB}$: Bayesian Cram\'er-Rao bound
  \item $\mathrm{BEC}$: Bose-Einstein Condensate
    \item $\mathrm{BTC}$: Boundary time crystal
    \item $\mathrm{BH}$: Bose-Hubbard
    \item $\mathrm{cavity-QED}$: Cavity quantum electrodynamics
       \item $\mathrm{CD}$: Counter-diabatic
       \item $\mathrm{CFT}$: Conformal field theory
    \item $\mathrm{CI}$: Classical illumination 
    \item $\mathrm{CoE}$: Curvature of entanglement
    \item $\mathrm{CPDD}$: Continuous phased dynamical decoupling
    \item $\mathrm{CPT}$:  Coherent population trapping
    \item $\mathrm{CPTP}$: Completely positive trace preserving
    \item $\mathrm{CSS}$: Causal superposition scheme
    \item $\mathrm{CTQW}$: Continuous-time quantum walk
    \item $\mathrm{CV}$: Continuous-variable
    \item $\mathrm{DD}$: Dynamical decoupling
      \item $\mathrm{DQPT}$: Dynamical quantum phase transition
     \item $\mathrm{DTC}$: Discrete time crystal
    \item $\mathrm{DTQW}$: Discrete-time quantum walk
    \item $\mathrm{EP}$:  Exceptional-point
    \item $\mathrm{ESR}$: Electron spin resonance
    \item $\mathrm{FH}$: Fermi-Hubbard
    \item $\mathrm{FI}$: Fisher information $(\mathscr{F})$
    \item $\mathrm{FIM}$: Fisher information matrix $(\mathbbm{F})$
    \item $\mathrm{FI MeNoS}$: Fisher information measurement noise susceptibility
     \item $\mathrm{GHZ}$: Greenberger-Horne-Zeilinger
     \item $\mathrm{GKSL}$: Gorini-Kossakowski-Sudarshan-Lindblad
    \item $\mathrm{GRAPE}$: Gradient ascent pulse engineering 
    \item $\mathrm{HCRB}$: Holevo Cram\'er-Rao bound 
    \item $\mathrm{HL}$: Heisenberg limit
     \item $\mathrm{HNKS}$: Hamiltonian-not-in-Kraus-span
    \item $\mathrm{ICO}$: Indefinite causal order
    \item $\mathrm{LR}$: Long-range
    \item $\mathrm{ME}$:  Multiparty entanglement
    \item $\mathrm{MIPT}$: Measurement-induced phase transition
    \item $\mathrm{ML}$: Machine learning
    \item $\mathrm{MZI}$: Mach-Zehnder interferometer 
    \item $\mathrm{MPO}$: Matrix product operators
    \item $\mathrm{MPS}$:  Matrix product states
    \item $\mathrm{MSE}$: Mean square error
    \item $\mathrm{MSEM}$: Mean square error matrix $(\mathbb{T})$
    \item $\mathrm{NCRB}$: Nagoka Cram\'er-Rao bound 
    \item $\mathrm{NHCRB}$: Nagoka Hayashi Cram\'er-Rao bound 
    \item $\mathrm{NMR}$: Nuclear magnetic resonance
     \item $\mathrm{NN}$: Nearest-neighbor
     \item $\mathrm{NV}$: Nitrogen vacancy 
    \item $\mathrm{NV center}$: Nitrogen-vacancy center
     \item $\mathrm{OAT}$: One-axis twisting
    \item $\mathrm{PAS}$: Passive auxiliary scheme
    \item $\mathrm{POVM}$: Positive operator valued measure
    \item $\mathrm{PS}$: Parallel scheme
    \item $\mathrm{PSF}$: Point-spread function
    \item $\mathrm{QCB}$: Quantum Chernoff bound
    \item $\mathrm{QCRB}$: Quantum Cram\'er-Rao bound
    \item $\mathrm{QEC}$: Quantum error correction
    \item $\mathrm{QED}$: Quantum electrodynamics
    \item $\mathrm{QFI}$: Quantum Fisher Information {$(\mathcal{F})$}
     \item $\mathrm{QFIM}$: QFI matrix $(\mathbb{F})$
     \item $\mathrm{QGT}$: Quantum geometric tensor
    \item $\mathrm{QI}$: Quantum illumination 
    \item $\mathrm{QMB}$: Quantum many-body
    \item $\mathrm{QPT}$: Quantum phase transition
    \item $\mathrm{RL}$:  Reinforced learning
    \item $\mathrm{RLD-CRB}$: Right logarithmic derivative Cram\'er-Rao bound 
    \item $\mathrm{SDP}$:  Semi-definite programming
    \item $\mathrm{SLD}$: Symmetric logarithmic derivative 
    \item $\mathrm{SPDC}$: Spontaneous parametric down-conversion
    \item $\mathrm{SLD-CRB}$: Symmetric logarithmic derivative--Cram\'er-Rao bound 
    \item $\mathrm{SNR}$: Signal-to-noise ratio
     \item $\mathrm{SQL}$: Standard quantum limit
    \item $\mathrm{SQUID}$: Superconducting quantum interference device
    \item $\mathrm{SS}$: Sequential scheme
    \item $\mathrm{TAT}$: Two-axis twisting
    \item $\mathrm{TN}$:  Tensor network
 
\end{enumerate}

\newpage

\section{\textbf{Theory of parameter estimation}}
\label{sec:parameter-estimation}

Parameter estimation is a fundamental component of the theory of statistical inference, concerned with {determining the values of unknown parameter(s) pertaining to a model or probability distribution based on experimental or observational data}. At its core, the task involves constructing an \emph{estimator} - a systematic procedure that assigns values of parameter(s) to measured outcomes while satisfying desirable properties such as unbiasedness and efficiency.
Estimation theory can be classified 
based on both the methodology employed and the underlying mechanism of encoding and decoding the parameter(s), leading to the distinction between {\it classical} and {\it quantum} parameter estimation. In the following sections, we describe these estimation frameworks in detail.



\subsection{Classical parameter estimation}
\label{subsec:classical-estimation}

In estimation theory, the available statistical data encodes the parameter, or a set of parameters, that to be determined.
The central objective is to estimate these parameters with the highest attainable precision. This framework is referred to as \textit{classical parameter estimation theory}~\cite{Cramer46,Rao1992,Kay93,trees2001,Lehmann06}. Within this theory, two primary approaches are commonly employed: the \textit{frequentist} and the \textit{Bayesian} methods. In the frequentist framework, the parameter(s) of interest is(are) considered fixed but approximately known. In contrast, the Bayesian approach incorporates prior knowledge, by modeling the parameter(s) as random variable(s) described by a probability distribution.

\subsubsection{Frequentist approach}
\label{subsubsec:classical-frequentist}

We begin this section by reviewing the simpler case of single-parameter estimation within the frequentist framework. We then extend the discussion to the more general setting of multiparameter estimation within the same approach.


\textit{Single parameter estimation in the frequentist approach}~\cite{Cramer46,Rao1992,Kay93,trees2001,Lehmann06}: In single-parameter estimation theory under the frequentist framework, the parameter to be estimated with the best possible precision is denoted by $\theta \in \Theta \subset \mathbb{R}$ with $\theta^*$ being its true but unknown value. The given statistical data, which depends on the parameter $\theta^*$, is represented by a random variable $X$, with its probability distribution represented by $p(X|\theta^*)$. To estimate the parameter of interest with best possible precision for a given statistical data, a function is introduced, known as the \textit{estimator}. Let us denote the estimator by $\tilde{\theta}(X)$ 
which
is a function of the 
random variable $X$. 
Since X is random, the estimator $\tilde{\theta}(X)$ is itself a random variable.
The \textit{error in estimation} is defined as the 
deviation of 
the estimator from the true parameter value, $\tilde{\theta}(X) - \theta^*$. A common measure of estimation accuracy in the parameter estimation problem is the \textit{mean squared error (MSE)} 
defined as
\begin{equation*}
   \text{MSE} (\tilde{\theta}, \theta^*) \coloneqq \int_{X \in \Omega}  dX \left(\tilde{\theta}(X)-\theta^*\right)^2 p(X|\theta^*),
\end{equation*}
where $\Omega$ represents the sample space of the random variable $X$.

In the frequentist approach, the goal is to \textit{minimize the MSE} over all possible estimators, thereby identifying the estimator that achieves the best possible precision in estimating the parameter. To derive fundamental limits on this precision, the frequentist method imposes two conditions on the estimator:
\begin{enumerate}
    \item \textbf{Unbiasedness:} The expected value of the estimator should be equal to the true parameter value, i.e.,
    \begin{equation*}
        \int_{X \in \Omega}   dX \hspace{1.5 mm} p(X|\theta^*) \hspace{1.5 mm} \tilde{\theta}(X) = \theta^*.
    \end{equation*}
    
    \item \textbf{Locality:} The estimator satisfies the condition
    \begin{equation*}
        \int_{X \in \Omega} dX \hspace{1.5 mm} \tilde{\theta}(X) \hspace{1.5 mm} \frac{\partial p(X|\theta)}{\partial \theta} \bigg|_{\theta = \theta^*} = 1.
    \end{equation*}
\end{enumerate}
Together, these conditions define a \textit{locally unbiased estimator}, meaning the estimator is unbiased in a neighborhood around the true value of the parameter. Under these assumptions, 
the MSE turns out to be the variance of the estimator, 
denoted by $\Delta^2\tilde{\theta}$.
A fundamental lower bound on  $\Delta^2[\tilde{\theta}, \theta^*]$ for the parameter of interest optimizing over all locally unbiased estimators, known as the~\textit{Cram\'er-Rao bound}, is given by
\begin{equation}
    \Delta^2[\tilde{\theta}, \theta^*] \geq \frac{1}{\mathscr{F} (\theta^*)},
\end{equation}
where the~\textit{Fisher information (FI)}, $\mathscr{F} (\theta^*)$, is defined as
\begin{equation}
\label{eqn-classical-FI}
\mathscr{F} (\theta^*) \coloneqq  \int_{X \in \Omega}   dX \left( \frac{\partial \log p(X|\theta)}{\partial \theta} \big|_{\theta = \theta^*} \right)^2 p(X|\theta^*).
\end{equation}
This bound sets an achievable limit on the precision of any unbiased parameter estimation for a given statistical data. 
The derivation of the Cram\'er-Rao bound relies on the following key ingredients:
\begin{itemize}
    \item The local unbiasedness conditions of the estimator,
    \item the normalization condition on $p(X|\theta^*)$,
    \item the \textit{regularity condition}, 
    allowing the
    interchange of differentiation and summation, and
    \item the \textit{Cauchy-Schwarz inequality}.
\end{itemize}

When $M$ independent and identically distributed 
(i.i.d.) data sets
are provided, the lower bound on the variance of the local unbiased estimator is given by $ \text{MSE} \geq \frac{1}{M \mathscr{F}}$;
it implies that as more i.i.d. datasets are allowed, the parameter can be estimated more accurately with the scaling beling $1/M$.
The condition for an estimator saturating the lower bound is
\begin{equation}
\label{eqn-estimator}
    \frac{d}{d\theta} \left(\log p(X|\theta) \big|_{\theta = \theta^*}\right) = \lambda (\theta) (\tilde{\theta} (X) - \theta^*).
\end{equation}
Here $\lambda(\theta)$ is a proportionality function that, in general, depends on $\theta$.
However, there exists statistical models for which an optimal estimator saturating the Cram\'er-Rao bound
does not exist for finite sample sizes, although such optimal estimators always exist asymptotically.

It has been shown that if one wishes to estimate a function of a parameter, 
$g(\theta^*)$, the Cram\'er-Rao bound generalizes to
\begin{equation}
\Delta^2[\tilde{\theta}, \theta^*]  \geq \frac{\left( \frac{\partial g(\theta)}{\partial \theta} \big|_{\theta = \theta^*} \right)^2}{M \mathscr{F}(\theta^*)},
\end{equation}
where $\mathscr{F} (\theta^*)$ is the FI associated with the Cram\'er-Rao bound for estimation of $\theta^*$.

\textit{Multiparameter estimation in the frequentist approach}~\cite{Cramer46,Rao1992,Kay93,trees2001,Lehmann06}: We now turn our attention to simultaneous estimation of multiple parameters
within the frequentist framework. Here, the objective is to infer a set of parameters, $\vec{\theta} \coloneqq (\theta_1, \theta_2, \ldots, \theta_d) \in  \mathbb{R}^d$ , simultaneously with the highest 
attainable
precision, where the true values 
are denoted by $\vec{\theta}^*$. As in the single-parameter case, the statistical data described by a probability distribution \( p(X| \vec{\theta}^*) \) 
encodes the dependence of the data on the underlying parameters.

For each parameter $\theta_i^*$, we can define a corresponding estimator $\tilde{\theta}_i(X)$, where $i \in \{1, 2, \ldots, d\}$, denoting the corresponding vector as $\vec{\tilde{\theta}}$. In the multiparameter scenario, the estimation error is characterized by a matrix. The mean square error matrix (MSEM) for locally unbiased estimators is defined as
\begin{align*}
    &\mathrm{T}_{ij}(\vec{\tilde{\theta}}, \vec{\theta}^*) \coloneqq \nonumber \\ &\int_{X \in \Omega} dX \hspace{1.5 mm} p(X| \vec{\theta}^*) \left(\tilde{\theta}_i(X) - \theta_i^* \right)\left(\tilde{\theta}_j(X) - \theta_j^* \right),
\end{align*}
which quantifies both individual uncertainties and their mutual correlations.

The central goal in multiparameter estimation
is to ``minimize" MSEM over all admissible estimators for a given statistical data where each estimator satisfies the local unbiasedness and normalization conditions. 
{Under these conditions,
the achievable precision is limited by the~\textit{multiparameter Cram\'er-Rao bound}, imposing a matrix inequality on the MSEM,}
\begin{equation}
    \mathrm{T}(\vec{\tilde{\theta}}, \vec{\theta}^*) \succeq \frac{\mathbbm{F}^{-1}( \vec{\theta}^*)}{M},
\end{equation}
Here, $\mathbbm{F}_{ij} ( \vec{\theta}^*)$ denotes the Fisher information matrix (FIM) whose matrix elements are given by
\begin{widetext}
    \begin{align}
    \label{eqn-FIM}
    \mathbbm{F}_{ij} ( \vec{\theta}^*) \coloneqq \int_{X \in \Omega} dX \left( \frac{\partial \log p(X| \vec{\theta})}{\partial \theta_i} \big|_{\vec{\theta} = \vec{\theta}^*} \right)\left( \frac{\partial \log p(X| \vec{\theta})}{\partial \theta_j} \big|_{\vec{\theta} = \vec{\theta}^*} \right) p(X| \vec{\theta}^*),
\end{align}
\end{widetext}
and the matrix inequality ``$\succeq$" indicates that the difference between the two matrices is positive semidefinite.

The derivation of the multiparameter Cram\'er-Rao bound is based on the following ingredients:
\begin{itemize}
\item Local unbiasedness of each estimator $\tilde{\theta}_i(X)$,
\item the normalization condition of the probability distribution $p(X| \vec{\theta}^*)$,
\item the~\textit{regularity condition}, which permits interchange of differentiation and summation,
\item the application of the \textit{Cauchy–Schwarz inequality}, and
\item the assumption that the FIM, $\mathbb{F}$, is invertible.
\end{itemize}

\textit{Properties of the {Fisher information matrix}:}  
The FIM possess the following properties:  
\begin{enumerate}
    \item \textit{Additivity:} The FI is additive for independent random variables, {leading} to linear scaling, i.e., $\mathscr{F}_N$  =  $N \mathscr{F}$.
    \item \textit{Real-valued and Symmetric:} The FIM is a real and symmetric matrix, i.e.,  $\mathbbm{F}_{ij} = \mathbbm{F}_{ji}$.  
    \item \textit{Positive Semi-definiteness:} The FIM is a positive semi-definite matrix, ensuring the  valid lower bound given by the multiparameter Cramér–Rao bound.
\end{enumerate}

\subsubsection{Bayesian approach}
\label{subsubsec:classical-Bayesian}

{In the frequentist framework discussed above in Sec.~\ref{subsubsec:classical-frequentist}, the parameter to be estimated is treated as a fixed but unknown quantity, known only approximately. 
Here, we discuss single as well as multiple parameter estimation within the ``Bayesian'' framework, an alternative approach for parameter estimation.
In contrast to the frequentist approach, the Bayesian one~\cite{Cramer46,Rao1992,Kay93,trees2001,Lehmann06,peter2012}  treats the parameter of interest as a random variable and incorporates prior knowledge about the parameter (to be estimated) in the form of a probability distribution.}

\emph{Single-parameter estimation in Bayesian framework}~\cite{Cramer46,Rao1992,Kay93,trees2001,Lehmann06,peter2012}: Let $\theta \in \Theta \subset \mathbb{R}$ denotes the parameter to be estimated, which is now treated as a random variable,
having a known prior distribution $p(\theta)$.
The statistical data that carries information about this parameter is represented by another random variable $X$, whose conditional probability distribution is $p(X|\theta)$.

An estimator, $\tilde{\theta}(X)$, defined as a function of the observed data $X$, can be used to estimate the value of the parameter.
Given that the prior distribution is known, the central quantity of interest in Bayesian estimation is the \textit{average mean square error of the estimator }$(\text{AMSE})$, which is obtained by averaging over both, the parameter and the data, AMSE is given by
\begin{equation*}
    \text{AMSE} (\tilde{\theta},\theta) \coloneqq \int_\theta \hspace{1.5 mm} d\theta p(\theta) \int_X dX \left( \tilde{\theta}(X) - \theta \right)^2 p(X|\theta).
\end{equation*}
Under suitable regularity conditions, the fundamental lower bound on the AMSE is provided by the \textit{Bayesian Cram\'er-Rao bound} (BCRB), which reads as
\begin{equation}
    \text{AMSE}  (\tilde{\theta},\theta) \geq \frac{1}{\tilde{\mathscr{F}} + \mathscr{I}}.
\end{equation}
Here $\tilde{\mathscr{F}}$ denotes the Fisher information averaged over the prior distribution $p(\theta)$ and is defined as
$\tilde{\mathscr{F}} \coloneqq \int_\theta d\theta \hspace{1.5 mm} p(\theta) \mathscr{F}(\theta)$,
where $\mathscr{F}(\theta)$ is the Fisher information evaluated at a specific value of $\theta$ defined for single-parameter estimation within the frequentist framework in Eq.~\eqref{eqn-classical-FI}.
And $\mathscr{I}$ captures the Fisher information associated with the prior distribution itself, and is expressed as
\begin{equation}
\label{eqn-bayesian-FI}
\mathscr{I} \coloneqq \int_\theta d\theta \left( \frac{\partial \log p(\theta)}{\partial \theta} \right)^2 {p(\theta)}.
\end{equation}

\textit{Multiparameter estimation in the Bayesian framework}:~\cite{Cramer46,Rao1992,Kay93,trees2001,Lehmann06,peter2012} In order to extend the estimation of multiple parameters to the Bayesian regime,
let us consider a set of unknown parameters to be estimated by a parameter vector, denoted by \( \vec{\theta} \coloneqq (\theta_1, \theta_2, \dots, \theta_d ) \in \mathbb{R}^d \), with corresponding estimators \( \vec{\tilde{\theta}} \coloneqq ( \tilde{\theta}_1, \tilde{\theta}_2, \dots, \tilde{\theta}_d ) \). The estimation is based on the observed data distribution \( p(X|\vec{\theta}) \) and a known distribution over the parameters, \( p(\vec{\theta}) \).

Within the Bayesian setting, the estimation error is typically quantified by the \textit{average mean square error matrix} (AMSEM), whose elements are defined as
\begin{align*}
    &\text{Cov}_{ij} (\vec{\tilde{\theta}}, \vec{\theta}) \\ &\coloneqq \int d\vec{\theta}\, p(\vec{\theta}) \int dX\, p(X;\vec{\theta})\, \left( \tilde{\theta}_i(X) - \theta_i \right)\left( \tilde{\theta}_j(X) - \theta_j \right).
\end{align*}
Again, multiparameter BCRB provides a lower bound on the {AMSEM} over all possible estimators. It generalizes the single-parameter BCRB to the multiparameter case as
\begin{equation}
    \text{Cov} (\vec{\tilde{\theta}}, \vec{\theta}) \succeq \left( \tilde{\mathbbm{F}} + \mathbbm{I} \right)^{-1},
\end{equation}
{where the elements of the matrices \( \tilde{\mathbbm{F}} \) and \( \mathbbm{I} \) are defined as
\begin{align*}
    \tilde{\mathbbm{F}}_{ij} &\coloneqq \int d\vec{\theta}\, p(\vec{\theta})\, \mathbbm{F}_{ij}(\vec{\theta}), \\
    \mathbbm{I}_{ij} &\coloneqq \int d\vec{\theta}\, p(\vec{\theta})\, \left( \frac{\partial \log p(\vec{\theta})}{\partial \theta_i} \right) \left( \frac{\partial \log p(\vec{\theta})}{\partial \theta_j} \right),
\end{align*}
where \( \mathbbm{F}_{ij}(\vec{\theta}) \) denotes the Fisher information matrix corresponding to the likelihood \( p(X|\vec{\theta}) \), given in Eq.~\eqref{eqn-FIM}.}

Both frequentist and Bayesian  estimation frameworks have their benefits and drawbacks. The frequentist technique is conceptually objective and frequently computationally efficient, especially for huge datasets, because it is entirely data-driven and does not rely on prior assumptions on distribution of the parameters. However, under data-scarce contexts, it may produce unstable or less reliable estimates with limited interpretability of uncertainty, particularly given the fact that it is unable to incorporate prior information. The Bayesian technique, on the other hand, improves performance in small-sample circumstances and increases modeling flexibility by combining prior knowledge with observed data to generate a complete probabilistic description via the posterior distribution. However, its results can be susceptible to the prior selection and usually demand more advanced computational methods.

\subsection{Quantum parameter estimation and quantum Fisher information}
\label{subsec:quantum-estimation}
Following the discussion on classical parameter estimation theory in Sec.~\ref{subsec:classical-estimation}, we now turn to quantum parameter estimation~\cite{Helstrom67,Helstrom68,Helstrom69,Caves1981,Holevo82,Braunstein94}, a framework that exploits the principles of quantum mechanics to achieve estimation efficiencies beyond those attainable in classical settings. This field has a rich history dating back to the 1970s, with pioneering contributions by C.~W. Helstrom and A.~S. Holevo, who established fundamental bounds on estimation errors using quantum mechanical principles~\cite{Helstrom67,Helstrom68,Helstrom69,Holevo82}. 
Before discussing various aspects of quantum parameter estimation in detail, we briefly review the estimation theory from two perspectives: single-parameter estimation and multiparameter estimation.

\subsubsection{Single parameter estimation theory}
\label{subsubsec:quantum-single-parameter}
Let us consider an approximately known but fixed parameter $\theta$ that one wants to estimate here and it is encoded into a given input quantum state $\sigma_\theta$ on $\mathcal{H}_d$\footnote{$\mathcal{H}_d$ is the complex Hilbert space of dimension $d$.}. To gain information about the parameter of interest, one needs to perform a measurement on the encoded state, denoted by $\{\Pi_i\}$, satisfying \(\sum_i \Pi_i = \mathbb{I}_d\) and \(\Pi_i \geq 0, \quad \forall i\).  
According to Born's rule, the probability of obtaining outcome $i$ for this measurement $\{\Pi_i\}$ on the encoded state is given by \( p_{i|\theta} \coloneqq \Tr[\Pi_i \sigma_\theta] \). 
Based on this probability distribution $\{p_i\}$, one can construct an estimator.
Like in the classical case, precision of the estimate is quantified by the variance $\Delta[\theta]:=\langle\theta^2\rangle-\langle\theta\rangle^2$, bounded by the 
quantum Cram\'er-Rao bound (QCRB), given by
\begin{equation}
    \Delta[\theta] \geq \frac{1}{M \mathcal{F}(\theta)},
    \label{eq:qcrb}
\end{equation}
where $M$ denotes the number of independent repetitions of the experiment, and  \(\mathcal{F}(\theta) \coloneqq \Tr[\sigma_\theta L_\theta^2]\) is called the quantum Fisher information (QFI), with $L_\theta$ being the symmetric logarithmic derivative (SLD) defined as
\begin{equation}
    \pdv{\sigma_\theta }{\theta} = \frac{1}{2}(L_\theta \sigma_\theta + \sigma_\theta L_\theta).
\end{equation}
Note that the QFI is obtained by maximizing the classical Fisher information, \(\mathscr{F}(\theta)\), over all POVMs: \( \mathcal{F}(\theta) = \max_{\{\Pi_i\}} \mathscr{F}(\theta)\). The classical Fisher information represents the lower bound of estimation error optimized over all local unbiased estimators for a particular measurement.
While there may exist multiple POVMs that saturate this bound, identifying such optimal measurements is generally a challenging task. A key result in quantum estimation theory is that there always exists an optimal measurement, given by the eigenbasis of SLD operator, that saturates
the quantum Cram\'er–Rao bound. 
It is crucial for single-parameter estimation that QCRB can always be achieved provided both the optimal basis of measurement and optimal estimator are given. One can derive the maximum QFI of an arbitrary state by evoking the SLD basis, which is given as 
\begin{equation}  
L_\theta=2\sum_{m,n}\frac{\langle\lambda_m|\partial_\theta\sigma_\theta|\lambda_n\rangle}{\lambda_m+\lambda_n}|\lambda_m\rangle\langle\lambda_n|,
\end{equation}
where we utilize the spectral decomposition of the encoded state, \(\sigma_\theta=\sum_m\lambda_m|\lambda_m\rangle\langle\lambda_m|\), and the QFI in the SLD basis is given as
\begin{equation}  
\label{eqn-QFI}
\mathcal{F}(\sigma_\theta)=2\sum_{m,n}\frac{|\langle\lambda_m|\partial_\theta\sigma_\theta|\lambda_n\rangle|^2}{\lambda_m+\lambda_n}, \quad \lambda_m+\lambda_n\ne 0.
\end{equation}
This form of QFI can be decomposed into -- (i) the classical part, which appears due to the distribution of the eigenvalues of \(\sigma_\theta\), and (ii) the quantum part, arsing from the coherence between eigenstates. Therefore,  Eq.~\eqref{eqn-QFI} can be rewritten as
\begin{equation}    
\mathcal{F}(\sigma_\theta)=\sum_k\frac{(\partial_\theta\lambda_k)^2}{\lambda_k}+2\sum_{m\ne n}\frac{(\lambda_m-\lambda_n)^2}{\lambda_m+\lambda_n}|\langle\partial_\theta \lambda_m|\lambda_n\rangle|^2.
\end{equation}
For pure states \(\ket{\Phi(\theta)}\), the expression of QFI simplifies as 
\begin{equation}
   \mathcal{F}(\ket{\Phi(\theta)})=4(\langle \partial_\theta\Phi(\theta)|\partial_\theta\Phi(\theta)\rangle-|\langle\partial_\theta\Phi(\theta)|\Phi(\theta)\rangle|^2),
\end{equation}
with \(L_\theta=2\partial_\theta\ket{\Phi(\theta)}\bra{\Phi(\theta)}\),
thereby making it
tractable.
This highlights that the QFI quantifies distinguishability between two infinitesimally close quantum states. In particular, the more sensitively a state varies with respect to the parameter of interest, the easier it becomes to distinguish neighboring states through appropriate measurements, thereby enabling higher accuracy in parameter estimation.

\textit{Standard quantum limit (SQL) and Heisenberg limit (HL).} A common classical technique for noise reduction involves averaging the results from \( N \) independent and identically prepared probes. In the quantum mechanical context, 
this corresponds to preparing a pure quantum system, with \( N \) parties, in a product state  \( |\phi\rangle = \bigotimes_{i=1}^N |\phi_i\rangle. \) Suppose the parameter \( \theta \) is imprinted into the state via a unitary evolution governed by the Hamiltonian
\( H(\theta) = \theta \sum_{i=1}^N h_i, \)
leading to the evolved state
\( |\Phi(\theta)\rangle = \exp[-i H(\theta)] |\psi\rangle. \)
According to the QCRB, the minimum possible variance $\Delta[\hat{\theta}]$
in the estimation of \( \theta \), given \( M \) repetition of experiments, is
\begin{equation}
    \Delta[\hat{\theta}] \ge \frac{1}{M \mathcal{F}(\theta)}.
\end{equation}
Additivity of the QFI immediately implies the \(1/N\) scaling of the variance.  Since the above inequality is derived by optimizing over all possible measurements on the full system, it follows that even collective (entangled) measurements cannot improve upon the \( {N} \) scaling. Hence, it underlines the fact that if QFI of the system scales linearly with the number of parties present in the quantum state \(N\) and there is no advantage over classical system since the QFI is additive. 
However, quantum resources such as entanglement in the probe enable enhanced scaling. In particular, for a general probe state evolving under a Hamiltonian, the QFI may scale as $\mathcal{F}(\theta)$ \(\sim N^\mu\) with \(\mu\) being the scaling component~\cite{Giovannetti06,Giovannetti11}.
The standard quantum limit (SQL)~\cite{Pezze09}, which can be attained by conventional means, 
corresponds to
\(\mu=1\); while when \(\mu>1\), the system exhibits a quantum advantage. Notably, when \(\mu\sim 2\), the scaling reaches the Heisenberg limit (HL). Interestingly, certain quantum systems can surpass the HL, achieving \(\mu\sim 2\), often informally referred to as the super-Heisenberg limit, thereby displaying an even higher quantum advantage~\cite{Boixo07,Boixo08,Mishra21,Yousefjani23,Mihailescu24,Sahoo24b,Adani24,Mondal25b,Yousefjani25,Mihailescu25c}\footnote{Such enhancements may involve additional resources, which may modify the scaling.}.

\subsubsection{Multiparameter estimation}
\label{subsubsec:quantum-multiple-parameter}

A central goal of parameter estimation theory is the simultaneous estimation of multiple parameters. More precisely, instead of a single parameter \(\theta\), the vector of multiparameters, $\vec{\theta}^* \coloneqq (\theta_1^*, \theta_2^*, \ldots, \theta_N^*)$, is to be inferred simultaneously with optimal precision. 
Let \(\vec{\theta}^*\) be encoded in a state \(\sigma_{\vec{\theta}^*}\).
{The matrix elements of the covariance matrix, $\hat{\mathcal{C}}(\vec{\theta}^*,\Pi_i)$, whose matrix elements for a set of POVMs $\{\Pi_i\}$  can be defined as \(\hat{\mathcal{C}}_{ij}=\langle \theta_i^*\theta_j^*\rangle-\langle \theta_i^*\rangle\langle \theta_j^*\rangle\).}
Like the single parameter case, the lower bound, referred to as multiparameter CRB, on the estimation error covariance matrix, $\hat{\mathcal{C}}(\vec{\theta}^*,\Pi_i)$ has also been found. Therefore, under a set of POVMs $\{\Pi_i\}$, 

\begin{eqnarray*}
    \hat{\mathcal{C}}(\vec{\theta}^*,\Pi_i)\ge \frac{\mathbb{F}^{-1}}{M},
    \label{eq:multiparameter_fq}
\end{eqnarray*}
where \(\mathbb{F}\) is an invertible QFI matrix (QFIM) with \(M\) repetitions of the experiment.
In case of multiparameter estimation, the tighter bound can be achieved via two SLD bases. Denoting \({L}_{\theta_i}\), and \({L}_{\theta_j}\) as the SLDs corresponding to the parameters of interest $\theta_i$, and $\theta_j$ respectively, the elements of the QFIM in terms of \({L}_{\theta_i}\) and \({L}_{\theta_j}\) are given by
\begin{equation}
    {\mathbb{F}}_{ij}=\frac{1}{2}\Tr[\sigma_{\theta^*} \{{L}_{\theta_i},{L}_{\theta_j}\}],
    \label{eq:qei_multi}
\end{equation}
where \(\{.,.\}\) denotes the anticommutation of its arguments. In order to obtain a scalar figure of merit for estimating error, we consider a scaler product of the QFIM with a positive definite matrix \(\mathcal{W}\) which leads to the scaler equation, given as
\begin{equation}
\Tr[\mathcal{W}\Delta[\vec{\theta}_*]]\ge \Tr[\mathcal{W}\mathbb{F}^{-1}[\vec{\theta}_*]].
\end{equation}
The matrix \(\mathcal{W}\) is generally referred to as cost or weight matrix. By using spectral decomposition, the QFIM for a given encoded state can be computed analytically, assuming that the weight matrix is full-rank, given as
\begin{equation*}
    \mathbb{F}_{{ij}}=c\sum_{m,n=0}^{d-1}\frac{\text{Re}(\langle\lambda_m|\partial_i\sigma_{\theta^*}|\lambda_n\rangle\langle\lambda_n|\partial_j\sigma_{\theta^*}|\lambda_m\rangle)}{\lambda_m+\lambda_n},
\end{equation*}
where \(\lambda_{i}\)'s are eigenvalues of the state \(\sigma_{\vec{\theta^*}}\) with corresponding eigenstates \(\ket{\lambda_i}\) and $d$ is the dimension of the complex Hilbert space.  However, when the matrix is not of full rank, the divergent terms have to be removed by
invoking the condition \(\lambda_m+\lambda_n>0\).
Therefore, it can be rewritten, in this case, as
\begin{widetext}
    \begin{equation}
      \mathbb{F}_{{ij}} = \sum_{m=0}^{d-1} \frac{(\partial_i \lambda_m) (\partial_j \lambda_m)} {\lambda_m}  + \sum_{\substack{
   m,n=0 \\
   m\neq n \\
   \lambda_m+\lambda_n\neq 0
   }}^{d-1} \frac{2 (\lambda_m -\lambda_n)^2}{\lambda_m + \lambda_n}\text{Re}(\braket{\lambda_m | \partial_i \lambda_n} \braket{\partial_j \lambda_n|\lambda_m}).
\end{equation}
\end{widetext}
Further note that, QFIM of a non-full rank density matrix can be determined by using the support of the density matrix~\cite{Liu14}. Such a formulation can help us to obtain the similar form of QFIM as known for pure states and full rank density matrices.

As discussed above, in contrast to single-parameter estimation, where the objective is to infer a single unknown quantity, multiparameter estimation aims to simultaneously determine several parameters. While in the former case, there always exists an optimal measurement strategy that can, in principle, saturate the CRB under suitable conditions, the multiparameter scenario is inherently more intricate. Specifically, it may not be possible to simultaneously attain the optimal precision for all parameters, and the multiparameter CRB cannot, in general, be saturated. More recently, considering these fundamental limitations, a necessary and sufficient criterion for the simultaneous estimation of two parameters encoded by arbitrary quantum processes is established using qubit probes, thereby identifying the precise conditions under which the multiparameter QCRB remains valid and asymptotically achievable~\cite{Mondal25c}.

\textit{Reparameterization of estimating parameters:} If one wishes to estimate $\vec{\phi^{*}}=f(\vec{\theta^*})$ instead of $\vec{\theta^{*}}$, the QFI matrices for the two different parametrizations are related to each other via a transformation using a Jacobian, $\mathcal{J}$, as
\begin{align*}  \mathbb{F} (\vec{\theta^*})=\mathcal{J}^T \mathbb{F}(\vec{\phi^{*}})\mathcal{J},
\end{align*}
where the elements of the Jacobian $\mathcal{J}$ are given by $\displaystyle{\mathcal{J}_{ij}\coloneqq\partial \phi^*_i/\partial \theta^*_j}$~\cite{Liu20}. In the following, we discuss the properties of the QFI matrices.
\\
\\
\textbf{Properties of quantum Fisher information matrix:} 
The QFI and QFIM exhibit several fundamental properties, which we summarize below.
\begin{enumerate}
    \item \textit{Real and positive semidefinite.} The QFI is nonnegative. And the QFIM is  a
    positive-definite real symmetric matrix.
    
    \item \textit{Invariance under unitaries.} Both QFI and QFIM remain the same by evolving through a unitary which does not depend upon the parameter to be estimated. Specifically, for QFIM, \(\mathbb{F}(\sigma_{\vec{\theta}})=\mathbb{F}(U \sigma_{\vec{\theta}} U^\dagger)\) where \(U\) is independent of \(\sigma_{\vec{\theta}}\).
    
    \item \textit{Additivity.} For product states, both QFI and QFIM are additive, i.e.,
    if \(\sigma_{\vec{\theta}}=\bigotimes_i \sigma_{\vec{\theta_i}}\), \(\mathbb{F}(\sigma_{\vec{\theta}})=\sum_i \mathbb{F}(\sigma_{\vec{\theta_i}})\).
    
    \item \textit{Convexity.} Both QFI and QFIM are convex under probabilistic mixing of quantum states, specifically, for QFIM, we have
    \(\mathbb{F}(p\sigma_{\vec{\theta_1}} +(1-p)\sigma_{\vec{\theta_2}})\le p \mathbb{F}(\sigma_{\vec{\theta_1}})+(1-p) \mathbb{F}(\sigma_{\vec{\theta_2}})\) for \(p\in [0,1]\).
    
    \item \textit{CPTP monotonicity.} The QFI and QFIM can be shown to be monotonic under completely positive and trace-preserving map (CPTP) \(\Lambda\).
    Mathematically,  \(\mathcal{F}(\Lambda(\sigma_{\vec{\theta}}))\le \mathcal{F}(\sigma_{\vec{\theta}})\)
    and
    \(\mathbb{F}(\Lambda(\sigma_{\vec{\theta}}))\le \mathbb{F}(\sigma_{\vec{\theta}})\).
\end{enumerate}

\subsubsubsection{Various bounds on quantum multiparameter estimation}
\label{subsubsubsec:different-bound-multiple-parameter}

In the single-parameter quantum estimation, 
the SLD-CRB provides a tighter and in principle, achievable lower bound on the error.  This naturally led to the believe that its multiparameter generalization would also yield a similar optimality.
Although, the SLD-CRB  can be extended to the multiparameter setting as discussed in the previous section,  it is not generally achievable, due to the noncommutativity of SLD operators. This noncommutativity prevents the simultaneous attainability of the bound for all parameters. As a result, extensive research have been devoted towards finding tighter and achievable lower bounds on  MSEM in multiparameter quantum estimation. Below, we summarize some of the alternative bounds developed within the frequentist framework.

  \textit{Right logarithmic derivative Cram\'er-Rao bound (RLD-CRB)}~\cite{Yuen73,Fujiwara94,He25}.
    The RLD-CRB is analogous to the SLD-CRB, with the key difference being that it uses right logarithmic derivative (RLD) operators instead of symmetric ones. The RLD operator for each parameter from a set of parameters ($\vec{\theta}$) that one wishes to estimate is defined as 
    \begin{equation}
        \pdv{\sigma_{\vec{\theta}}} {\theta_i} = \sigma_{\vec{\theta}} {L}_{\theta_i},
    \end{equation}
    where $\sigma_{\vec{\theta}}$ denotes the encoded state for a set of parameters $\vec{\theta}$ to be estimated and ${L}_{\theta_i}$ denotes the RLD operator corresponding to the parameter $\theta_i \in \vec{\theta}$.     	
    The elements of the corresponding QFIM are given by $R_{ij} = \Tr[\sigma_{\vec{\theta}} {L}_{\theta_i} {L}_{\theta_j}^\dagger] $, and the matrix is complex and Hermitian.
    The RLD-CRB generally provides a tighter bound than the SLD-CRB but may not always correspond to a physically realizable measurement, since RLD operators are, in general, non-Hermitian. Note that the RLD and its QFIM are known for general Gaussian states~\cite{Gao14} 
    which is one of the suitable methods to estimate complex numbers, where the real and imaginary parts are encoded simultaneously, and they naturally form a pair of noncommuting observables. A prominent example of the usefulness of the RLD method is the estimation of the amplitude of the coherent state, say \(\alpha = \alpha_r + i\alpha_I\) where \(\alpha_r\) and \(\alpha_I\) represent two conjugate quadratures of bosonic modes. Similar to the SLD, the scalar cost function on RLD-CRB reads 
    \begin{equation}
    C_{\mathrm{R}} = \Tr\!\left(\mathcal{W}\, \mathrm{Re}(R^{-1}) \right) + \left\| \sqrt{\mathcal{W}}\, \mathrm{Im}(R^{-1})\,
    \sqrt{\mathcal{W}}   \right\|_1 ,
\end{equation}
where $||\cdot ||_1$ denotes the operator trace norm.    

\textit{Holevo Cram\'er-Rao bound} (HCRB). The HCRB~\cite{Holevo82,HCRB-2,Sidhu21} is one of the metrological lower bounds on the weighted mean square error 
which is asymptotically achievable. 
However, such attainability generally requires collective measurements on many copies of the probe state.
The HCRB ($C_\text{HCRB}$), for a positive definite $\mathcal{W}$, is given by 
\begin{align*}
    C_\text{HCRB} \coloneqq \min_{X_i} \{\Tr(\mathcal{W}\text{Re} V) + ||\mathcal{W}\text{Im}V||_1\}
\end{align*}
where $V_{ij} \coloneqq \Tr(X_iX_j \sigma_{\vec{\theta}})$, and the minimization is performed over Hermitian matrices $X_i$, satisfying $\frac{1}{2}\Tr (\sigma_{\vec{\theta}} \{X_i, L_{\theta_j}\}) = \delta_{ij}$. 
Its first term reproduces the multiparameter SLD-CRB 
while
the second term of the HCRB is positive and hence the HCRB is a tighter bound 
compared to the multiparameter SLD-CRB. 
However, there are important exceptions - such as when the encoded quantum state is pure~\cite{Matsumoto02} or when estimating displacement parameters in Gaussian states~\cite{Holevo82} - where the HCRB can be achieved using only separable measurements. 
Furthermore, HCRB is not easy to compute due to the optimization involved in the definition,
although its evaluations
can be formulated as a semi-definite programme. 
Further, the HCRB has both upper and lower bounds in terms of SLD and RLD as
\begin{equation*}
\max \{C_{\text{SLD}},C_{\text{RLD}} \} \leq C_{\text{HCRB}}\leq 2 C_{\text{SLD}}.
\end{equation*}     

The HCRB is not only asymptotically attainable in principle, but also generally tighter than the standard multiparameter QCRB, thereby regarded as the fundamental limit on precision. 
For a positive definite weight matrices $\mathcal{W}$, the SLD-based QCRB becomes attainable if and only if a specific compatibility condition is satisfied, referred to as the~\textit{weak-commutativity criterion} or~\textit{asymptotic compatibility condition}~\cite{Crowley14,Ragy16}.
It states that
that the expectation values of all SLD commutators vanish in the encoded state $\rho_{\vec{\theta^*}}$ in the asymptotic limit,  i.e., for every estimating parameter pair, 
the elements of the skew-symmetric Uhlmann matrix vanishes~\cite{Carollo19,Razavian2020}, i.e.,
\begin{equation}
\label{weak}
D_{ij} \coloneqq \text{Im}\left[\Tr\left(\rho_{\vec{\theta^*}}L_{\theta^*_i}L_{\theta^*_j}\right)\right]=0;\quad\forall\,\theta^*_i\neq \theta^*_j,
\end{equation}
vanishes, making $C_{\text{HCRB}} = C_{\text{SLD}}$.
It 
quantifies asymptotic incompatibility of optimal measurements corresponding to different parameters in quantum multiparameter estimation. In the special case where the encoded state is pure, the Uhlmann matrix reduces to the Berry curvature (discussed in detail in Sec.~\ref{subsec:geometrical interpretation}), allowing the condition to be interpreted geometrically in parameter space. Further developments along these lines can be found in Ref.~\cite{Vidrighin14}.

\textit{Nagoka Cram\'er-Rao bound} (NCRB). The NCRB discussed below are constructed keeping in mind that a collective measurement is required to obtain HCRB  are typically difficult
to implement. Hence, from the experimental feasibility, a lower bound on the mean square error of estimators in two-parameter quantum estimation, NCRB, is derived by optimizing over separable measurements only~\cite{NCRB-1}.
\begin{align}
&C_{\mathrm{NCRB}}
\coloneqq \nonumber \\ 
&\min_X
\left[
\Tr \left(
\mathcal{W}\,\mathrm{Re}V
\right)
+
\sqrt{\det(\mathcal{W})}\,
\Tr (\left|
\sigma_{\vec{\theta}}[X_1,X_2]
\right|)
\right],
\end{align}
where $\Tr(\abs{A})$ denotes the sum of the absolute values of the eigenvalues of $A$, and $X \coloneqq (X_1, X_2)^T$ is the vector of operators, satisfying $\frac{1}{2}\Tr (\sigma_{\vec{\theta}} \{X_i, L_{\theta_j}\}) = \delta_{ij}$,  over which the minimization is performed.
For qubit probes, the NCRB is both achievable and admits a compact analytic expression~\cite{NCRB-1, NCRB-2},
\begin{equation}
C_{\text{NCRB}, d=2} \coloneqq \Tr(\mathcal{W} \mathcal{F}^{-1}) + 2\sqrt{\det[\mathcal{W} \mathcal{F}^{-1}]}.
\end{equation}
For higher-dimensional probes, the NCRB can be casted as a semidefinite programme. Further note that for qubits, NCRB coincides with Gill and Massar bound which is based on classical Fisher information~\cite{Gill2000}, although they differ in higher dimension.

\textit{Nagoka-Hayashi Cram\'er-Rao bound} (NHCRB). The NHCRB is the generalization of the NCRB for multiparameter quantum estimation involving more than two parameters. However, it is generally not attainable~\cite{NHCRB-1,NCRB-2}. Like the NCRB, the NHCRB is derived by optimizing over separable measurements only. Moreover, both NCRB and NHCRB provide tighter bounds than the HCRB.

\textit{Most Informative Cram\'er–Rao Bound.}
By using similar idea of Gill and Massar bound~\cite{Gill2000,Matsumoto02}, 
another attainable precision bound has been found for arbitrary two-parameter estimation with pure encoded states in any Hilbert space dimension~\cite{Yung25}. The resulting bound is computationally much simpler than existing general bounds, making it practically convenient to evaluate.  
In addition, the optimal measurement that saturates this bound is fully characterized. 

Hence, the hierarchy of bounds depends on parameter compatibility, probe-state structure, and measurement restrictions. 
In recent years, there has been an ongoing effort to provide bounds for multiparameter estimation scenarios that are both computationally and experimentally feasible.

Till date, comparatively few studies have addressed quantum multiparameter estimation within the Bayesian framework~\cite{Rubio20}. In contrast to the extensively developed frequentist approach, Bayesian multiparameter studies remain relatively limited, mainly due to the increased mathematical complexity arising from prior distributions. Nevertheless, the Bayesian approach is particularly relevant for realistic finite-data scenarios, where prior information is available, and asymptotic unbiasedness assumptions may not hold. Existing works have begun to explore Bayesian precision bounds, optimal probe states, and adaptive strategies for simultaneous parameter estimation; see Refs.~\cite{NHCRB-bayesian}.

\subsection{Optimization over probe states}
\label{subsec:probe-optimization}

In quantum parameter estimation, any quantum metrological bound provides the best achievable lower bound on the MSE of an estimator for a given input probe by optimizing over the estimators and the quantum measurements. The CRB and its quantum variants are common examples, where the attainable precision  clearly depends on the  input probe. As a result, the structure of the initial quantum state plays a crucial role in determining the performance of a metrological protocol, in addition to the encoding dynamics and measurement technique. Thus, a natural question arises: \emph{``Which probe state maximizes the possible precision for a given estimation task?''} The ultimate accuracy limit for that task and the corresponding optimal probe are obtained by further optimizing the metrological bound over all permitted input states. Determining such optimal states constitutes a central problem in quantum metrology and has been extensively studied in both discrete- and continuous-variable settings as well as in the context of single and multiparameter estimation which we  briefly review here.

\subsubsection{Optimal probes for single parameter estimation}
\label{subsubsec:probe-optimization-single-parameter}

Within the frequentist framework, for a single parameter estimation encoded by a unitary encoding process (discussed in detail in Sec.~\ref{subsec:unitary-encoding}), 
the optimal probe is known to be an equally weighted superposition of the eigenvectors of the encoding Hamiltonian associated with its largest and smallest eigenvalues~\cite{Giovannetti04,Giovannetti06} when one copy of the probe is available.
Moreover, when multiple copies of a probe are available, for local phase and frequency estimation within a frequentist framework, the Greenberger-Horne-Zeilinger (GHZ) state acts as an optimal probe to achieve Heisenberg scaling~\cite{Giovannetti06,Giovannetti11}.
In contrast, for phase estimation under a Bayesian framework with flat priors (i.e., no prior knowledge), the optimal probe state is given by~\cite{wiseman-resource-1}
\begin{equation*}
\ket{\psi} = \sum_{n=0}^N c_n \ket{n}.
\end{equation*}
Here $\{\ket{n}\}$ denotes the eigenstates of the Hamiltonian $H$ corresponding to $(n+1)$-eigenvalues and the coefficients $c_n$ are defined as
\begin{equation}
c_n \coloneqq \sqrt{\frac{2}{N+2}} \sin \bigg[{\frac{(n+1)\pi}{N+2}}\bigg],
\end{equation}
and the error is quantified using the Holevo variance. Notably, it has also been shown that this same probe state achieves Heisenberg scaling in Bayesian phase and frequency estimation even for Gaussian priors of varying widths.

Extending the discussion to continuous-variable (CV) systems,  a number of studies have similarly investigated the problem of finding optimal probes in these platforms (see also Sec. \ref{subsec:CV-metrology}).
For example, the optimal pure probe states for identical bosonic particles in a two-mode interferometric setup for single-parameter estimation within the frequentist framework are identified~\cite{gaussian-optimal-1}. In this scenario, the total particle number fluctuates around a fixed average with a fixed variance. Depending on the properties of the interferometric circuit, the optimal state generalizes either the \(\mathrm{N00N}\) state (discussed in detail in Sec.~\ref{subsubsec:NOON-like-state-preparation}) or the two-mode Schr{\"o}dinger cat state. Furthermore, among pure Gaussian states with a given average particle number, a product state consisting of a squeezed vacuum in one mode and vacuum in the other is identified as the optimal input in that work.
Another result in this direction concerns parameter estimation in a generic multimode passive linear optical circuit (i.e., photon-number-preserving interferometers). The optimal input probe among general Gaussian states within the frequentist framework is found~\cite{Matsubara19}, where the optimal probe is shown to be a single-mode squeezed vacuum in an appropriate basis.
In addition, the problem of finding the optimal input state among all states with a fixed photon number for estimating the relative phase shift between the two arms of an optical interferometer has also been addressed~\cite{Dorner09}, where each arm is subject to photon loss. Using numerical optimization within the frequentist approach, it is shown that although the Heisenberg limit cannot be achieved in the presence of loss, the standard quantum limit  can still be surpassed. Interestingly, the optimal probe state in the lossy scenario is not the \(\mathrm{N00N}\) state, even though in the ideal (lossless) two-mode interferometric setup, the \(\mathrm{N00N}\) state is known to be optimal among all fixed-photon-number probes~\cite{Boto2000}.
On the other hand, within the Bayesian framework, the problem of identifying the optimal pure single-mode Gaussian probe state for estimating an unknown phase rotation of a continuous-variable system using homodyne detection with a Gaussian prior distribution has been addressed~\cite{Rodríguez25}.  
Furthermore, when prior information about the parameter is limited, the optimal probe states are close to coherent states~\cite{Rodríguez25}. However, as more precise prior information about the parameter becomes available, it becomes advantageous to increase the squeezing of the probe state. Interestingly, this transition is not smooth; instead, a sudden jump from squeezed displaced states to squeezed vacuum states is observed. These results indicate that the optimal Gaussian probe state for phase estimation with homodyne detection strongly depends on the prior knowledge of the phase parameter.

Since noise is ubiquitous in quantum systems, any realistic metrological task must take the presence of noise into account. Consequently, several works have focused on identifying optimal probe states for noisy quantum metrology~\cite{Rafal05,Monras06,Olivares09,Gaiba09,Pinel13,Pinel13,Friis15,Frowis16,Mondal25}. For example,
in the problem of frequency estimation of local Hamiltonians, i.e., any Pauli operator $\sum_{i=1}^L \sigma_j^i$, where $j \in \{x,y,z\}$ and $L$ denotes the probe size, in the presence of local uncorrelated dephasing noise, one-axis twisted spin-squeezed states (SSSs) with a specific squeezing parameter turn out to
be asymptotically optimal~\cite{noisy-unitary-estimation-1} within the frequentist framework.
Furthermore, in the finite-$N$ regime, the optimal state has been numerically identified for up to $N=70$ qubits~\cite{Frowis16}, where $N$ denotes the number of available copies of the input probe. In this setting, although one-axis twisted SSS are near-optimal for a certain squeezing parameter, spin-squeezing itself is not a necessary property: optimal states can be non-spin-squeezed. A key feature identified for optimality here is the global structure of the optimal input probe state’s coefficients in the eigenbasis of the Hamiltonian. Additionally, optimal states have been found for frequency estimation under other local noise models, such as local depolarizing noise and spatially and temporally correlated dephasing noise~\cite{Frowis16}.

Beyond conventional unitary and noise parameter encoding, a different paradigm of quantum metrology has also recently been explored~\cite{Mondal25c}. In this approach, the parameter is encoded through the statistics of generalized quantum measurements rather than solely through Hamiltonian dynamics or environmental interactions. Interestingly, it is shown that deliberately discarding part of the measurement record (\emph{forgetting}) can counterintuitively improve estimation precision, revealing that information erasure may enhance metrological performance in certain regimes~\cite{Mondal25c}. There has been substantial further work in this direction. See e.g. Refs.~\cite{Rafal05,Monras06,Olivares09,Gaiba09,Pinel13,Pinel13,Friis15,Chaki25}.

\subsubsection{Best probes for multiparameter estimation}
\label{subsubsec:probe-optimization-multiple-parameter}
Since not all metrological bounds are simultaneously achievable and valid in multiparameter estimation tasks, finding the optimal probes in such scenarios becomes a nontrivial problem. In the context of multiparameter estimation, several works in the literature have been focused on identifying optimal probe states for multiparameter metrological tasks.

For example, in the simultaneous estimation of the three components of a magnetic field under a parallel scheme, with or without an auxiliary, a fundamental precision limit has been established. The estimation error is characterized by the arbitrary weighted sum of the variances associated with each magnetic field component. The optimal input probe state and measurement achieving this bound are identified for an arbitrary spin-\(S\)particle, assuming access to a sufficiently large number of probe copies~\cite{Hou21}. See Refs.~\cite{Bhattacharyya24b,pal25,Saha26b} for additional works in this context.

\subsection{Techniques to handle theory of quantum metrology in complex systems}
\label{subsec:technique}

\begin{figure*}[t]
    \centering
    \includegraphics[scale=0.28]{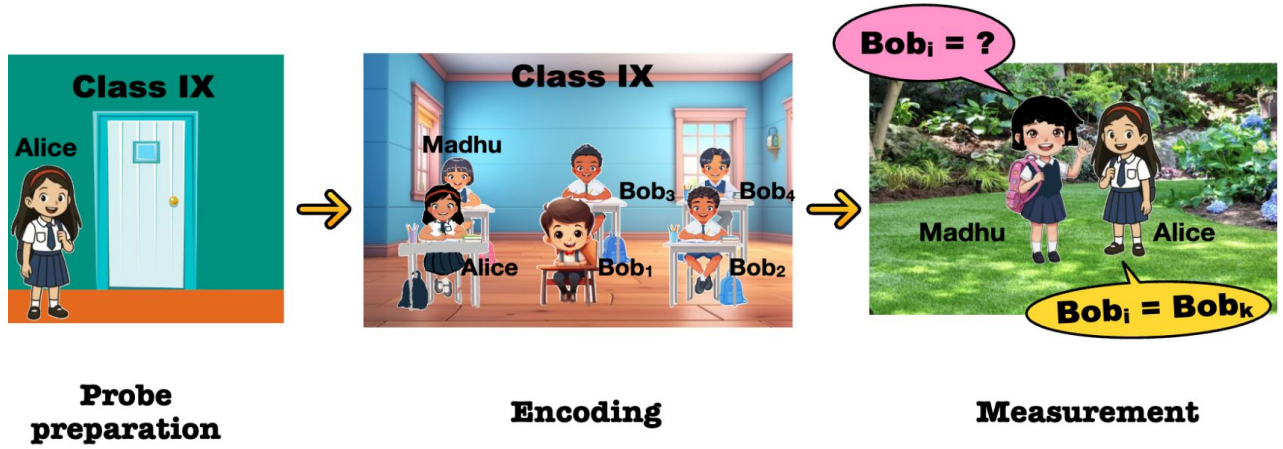}
    \caption{A cartoon illustration of the general idea behind a sensing protocol. Imagine Alice entering her first day of school with a completely unbiased and uncorrelated state of mind, analogous to an initially prepared probe state. As she interacts with her classmates, information from the environment becomes encoded into her (mental) state. In particular, among several boys denoted as Bob\(_i\), Alice gradually develops a preference toward one specific Bob, so that her internal state effectively carries information about a particular Bob\(_i\). Unaware of this process, her friend Madhu notices subtle changes in Alice’s behavior and attempts to infer which Bob has been encoded in Alice’s mind. If Madhu restricts herself to testing only a single predetermined possibility, the protocol resembles classical sensing with a fixed measurement strategy. In contrast, if she optimizes over all possible Bob\(_i\) and extracts the maximum distinguishable information from Alice’s encoded state of mind, the procedure becomes analogous to quantum sensing, where optimized measurements maximize the obtainable information about an unknown parameter. }
    \label{fig:schematic}
\end{figure*}

The ultimate goal of the parameter estimation is to infer an unknown parameter as precisely as possible, which includes several steps: (i) preparation of an input probe state; (ii) parameter encoding through a suitable dynamical process (see Sec.~\ref{sec:different-encoding-process} for details); (iii) measurement of the output state to extract information about the unknown parameter; and (iv) construction of an unbiased estimator, which provides the best estimation of the parameters (see Fig.~\ref{fig:schematic} for schematic representation of the steps). 
Optimizing these steps is, in general, a highly non-trivial task and has been addressed using powerful modern techniques, including machine learning (particularly reinforcement learning), tensor-network methods, and quantum Monte Carlo optimization. These computational tools complement analytical approaches by enabling optimization and simulation in regimes where exact analytical methods are missing, thereby playing a central role in the design of next-generation sensing protocols. We briefly review some of these developments below.

\textit{Reinforced learning (RL)}. RL~\cite{Sutton18} is a branch of machine learning (ML) in which an agent learns to perform and optimize a specific task through interactions with its environment, taking a series of actions to maximize a reward function. 
In order to improve precision, optimal control during evolution can play a significant role which can be obtained using RL. Specifically, RL has been successfully applied to general time-dependent sensing problems. For example, Xu \textit{et al.~}~\cite{Xu19} provides a reward function based on the QFI, enabling the agent to learn control strategies, while Xiao \textit{et al.}~\cite{Xiao22} design a general framework  in which the reward function is constructed from a new derivation of QFI which finally guides the learning process to achieve optimal metrological performance. Beyond reinforcement learning, supervised and unsupervised machine learning can also be directly used at the extraction stage of information from the probe. In particular, the regression methods are employed to infer an unknown parameter through repeated measurements, adopting a frequentist framework that eventually converges to the maximum likelihood Bayesian estimator~\cite{Nolan21a} when the number of training measurements is small, while for a large number of measurements, this estimator converges to Cra\'mer-Rao bound. 
A related approach considers parameter estimation as a series of classification problems, where artificial neural networks are used to efficiently implement Bayesian estimation
~\cite{Greplova17,Nolan21b}.

\textit{Tensor networks (TNs).} The tensor network~\cite{Schollwock11,Orus14,Banuls23} formalism arises fundamentally from the area law observed in the ground states of QMB systems. This approach employs the repeated Schmidt decompositions to approximate the given states of large interacting systems, especially non-integrable ones, for which analytical solutions of the eigenstates are not known.
TN formalism are especially effective for ground and low lying excited state of non-integrable Hamiltonian, thereby enabling the study of QFI near critical points, symmetry-breaking transitions and dynamical phase transitions. Since it was found that many-body criticality often enhances metrological sensitivity (as we discuss in Sec.~\ref{sec:applications})~\cite{montenegro2024review}, TN methods turn out to be an effective tool for identifying quantum sensing advantages in strongly correlated systems. Recently, techniques involving low-rank matrix product states (MPS) and matrix product operators (MPO) have been successfully applied in quantum sensing with the lossy interferometer~\cite{Cimini19,Khan25}, where the quantum state of the detected light is modeled as an MPO~\cite{jarzyna2013,Chabuda2020,Khan25}.  Furthermore, a dedicated numerical package written in Python called TNQMetro has been developed to compute fundamental quantum bounds on precision using tensor-network methods, both for finite-size systems and  for extracting the asymptotic scaling laws in large systems~\cite{Chabuda2022}.

\textit{Convex optimization — semi-definite programming~\cite{Boyd04}}. Semi-definite programming (SDP) is a subfield of convex optimization in which the objective function is linear and the constraints are linear matrix inequalities. A standard form of an SDP can be written as
\begin{align*}
    \text{Minimize:} \quad & \langle C, X \rangle \\
    \text{Subject to:} \quad & \langle A_i, X \rangle = b_i, \quad i = 1, \dots, m \\
    & X \succeq 0,
\end{align*}
where, $X \in \mathbb{S}^n$ is the optimization variable (a real symmetric matrix), $C, A_i \in \mathbb{S}^n$ are given symmetric matrices, $\langle A, B \rangle = \Tr(A^TB)$ denotes the inner product, and $X \succeq 0$ means that $X$ is positive semidefinite. Since positivity constraints naturally arises for quantum states, channels, and measurements, SDP becomes a powerful tool in quantum metrology. SDPs generalize linear programming (LPs), having a wide range of applications, including control theory, combinatorial optimization, and quantum information. The \textit{attainability} of the QCRB is of critical importance both in single- and multiparameter estimation scenarios.  In a realistic situation, small number of data samples hinders saturation of the bound. These challenges can be addressed within a fully Bayesian formulations, which replaces fixed prior distributions with reward functions. Within this context, Bavaresco \textit{et al.} utilize SDP-based methods to calibrate an optimal quantum tester that yields the best estimation strategy~\cite{Bavaresco24}. In multiparameter metrology, HCRB can also be formulated and solved via SDPs~\cite{Albarelli19}. Several software packages have been developed for this purpose in various programming languages~\cite{Agrawal18,Informatics2025May}.

\subsection{Progress in reducing assumptions in precision bounds}
\label{subsec:global-estimation} 
Most quantum parameter estimation bounds discussed so far are derived under several key assumptions. While this assumptions are mathematically convenient and often asymptotically justified, they may be restrictive in realistic setup. Let us outline these assumptions below and present the developments in which these assumptions are weakened.

(1) \textit{Unbiased estimators.} Quantum metrological bounds like the Cra\'mer-Rao bound are typically derived by optimizing over the class of locally unbiased estimators. However, verifying unbiasedness is rarely guaranteed,
especially with limited data. In practice, estimators tend to be biased, as unbiasedness is generally meaningful only in the asymptotic regime where large sample sizes allow the expectation value to be reliably compared with the true parameter~\cite{Meyer25}. In such cases, biased quantum metrological bounds including additional correction terms accounting for estimator biasness should be considered.
To overcome this, a recent study includes different bounds that incorporate estimator bias, or explicit corrections in the Cram\'er-Rao bound for finite sample sizes~\cite{Meyer25}.

To overcome this issue, an alternative framework, global quantum parameter estimation within the frequentist approach, has been proposed. 
For example, rather than having precise knowledge of the parameter of interest, only partial knowledge of its domain is available, making the approach more practical.
To illustrate this, a metrological scenario is considered in Ref.~\cite{Montenegro21} where a parameter $\theta \in \left[\theta^* - \frac{\Delta\theta}{2},\, \theta^* + \frac{\Delta\theta}{2} \right]$ is to be estimated with only its range known a priori and the prior distribution representing the information about the parameter of interest within the sensing interval $\Delta \theta$, $f(\theta)$, is known beforehand. Here $\theta^*$ denotes the true value of the parameter of interest with a control parameter present in the encoded probe, say $B$. The cost functional in such estimation tasks for a given control parameter $B$ is defined as follows:
\begin{equation}
    g(B) \coloneqq \int_{\theta^* - \frac{\Delta\theta}{2}}^{\theta^* + \frac{\Delta\theta}{2}} \frac{f(\theta)}{\mathcal{F}(\theta|B)} d\theta,
\end{equation}
where  $\mathcal{F} (\theta|B)$ denotes the QFI corresponding to the true value of the parameter of interest $\theta$ for a given control parameter $B$.
In other words, $g(B)$ denotes the maximum information achievable from the partial knowledge about the range and the prior distribution of the parameter.
The goal here is to tune the control parameter $B$ in such a way that one can achieve $g(B^*) \coloneqq \min_B g(B)$. Here, note that when $\Delta \theta \rightarrow 0$, we get local sensing, i.e., the lower bound given by Cram\'er-Rao can be achieved by the optimal measurement.
See refs~\cite{Hayashi22,Rubio23,Zhou24b,Mukhopadhyay24,Mukhopadhyay25,Mukhopadhyay2025Feb} for further works in this direction.

(2) \textit{Smoothness and regularity conditions.} The metrological bounds typically rely on certain regularity conditions, such as 
the smoothness in differentiability and integrability of differentiation and integration. Such assumptions may fail or becomes difficult to verify realistic open system dynamics, critical systems, etc., which we discuss in Sec.~\ref{sec:applications}.

(3) \textit{Existence of optimal estimators for finite samples.} In many practical scenarios, optimal estimators that saturate the bound may not exist for finite copies of the input probe state.
This issue can arise in
multiparameter estimation tasks 
where HCRB is asymptotically achievable but generally difficult to realize with limited resources. To address this limitation, several alternative lower bounds and protocols have been developed that remain valid even  when a limited number of probe states or measurement rounds are available. A later direction leads to the development of adaptive strategies (discussed in Sec.~\ref{sec:estimation-strategies}), especially relevant in quantum multiparameter estimation, in which measurements are updated sequentially based on the outcomes obtained in earlier rounds~\cite{Fujiwara11,Albarelli20}.

(4) \textit{ Independence of optimal measurements on the true parameter value.} 
In frequentist estimation, the optimal measurement that saturates the quantum bound often depends on the unknown true value of the parameter of interest. This makes the standard metrological bounds less reliable,
thereby creating an intrinsic circularity.

\subsection{Geometrical interpretation of QFI} 
\label{subsec:geometrical interpretation}
The QFI  admits a natural geometric interpretation, connecting it to the curvature and structure of the quantum state space, which is often visualized using the Bloch sphere or its higher–dimensional generalizations. The QFI quantifies how rapidly a quantum state changes under an infinitesimal variation of an encoded parameter and hence it measures
the distinguishability between neighboring quantum states and its inverse determines the ultimate precision bound for a parameter encoded in a quantum system. It implies that a larger geometric separation between nearby states provides higher achievable precision.
In this sense, parameter estimation can be understood geometrically in terms of distinguishability between quantum states, which can be interpreted as a ``distance'' on a statistical manifold. In classical statistics, this statistical distance induces a 
curved geometry in the probability simplex~\cite{Amari2007,Braunstein94}. However, by expressing probabilities in terms of probability amplitudes, this curvature can be removed. 
Quantum mechanics extends this framework by replacing classical probability distribution with density matrices on state vectors, while the statistical manifold is thus replaced by a manifold of quantum states and quantum theory naturally uses probability amplitudes.

Within this geometric framework, the distance between two quantum states is quantified by the angle between their corresponding state vectors, {$\ket{\psi}$ and $\ket{\psi+d\psi}$} defined through the Fubini-Study metric,
\begin{equation}
    ds^2=\frac{\langle d\psi|d\psi\rangle}{\langle \psi|\psi\rangle}-\frac{\langle d\psi|\psi\rangle\langle \psi|d\psi\rangle}{\langle \psi|\psi\rangle^2}.
\end{equation}
Using the normalization condition $\langle\psi|\psi\rangle=1$ and writing $\ket{d\psi}=\sum_{\eta}\ket{\partial_{\eta}\psi}\,dx_\eta$, the metric can be rewritten as
\begin{equation}
    ds^2=\sum_{\eta\nu}\frac{1}{4}\mathbb{F}_{\eta\nu}\,dx_\eta dx_\nu,
\end{equation}
where $\mathbb{F}_{\eta\nu}$ denotes the $(\eta,\nu)$ element of QFIM. This relation immediately establishes a link between the precision of parameter estimation and
distinguishability of quantum states: the greater the distance between neighboring states, the higher the achievable precision.

For mixed states, the geometric counterpart of the Fubini-Study metric is the Bures metric, a fundamental distance measure derived from the Uhlmann fidelity between density operators. In this case, the QFI can be expressed in terms of the fidelity between neighboring quantum states, $\sigma_\theta$ and $\sigma_{\theta+d\theta}$~\cite{Braunstein94} as
\begin{equation}
    \mathcal{F}(\sigma_\theta)=\frac{8}{\delta \theta^2}\left[1-F(\sigma_\theta,\sigma_{\theta+\delta\theta})\right],
\end{equation}
where $F(\rho,\tau) \coloneqq \left[\Tr[\sqrt{\sqrt{\rho}\tau \sqrt{\rho}}]\right]^2$ denotes the Uhlmann fidelity~\cite{Uhlmann76,Jozsa94}. This expression, commonly referred to as the \textit{fidelity-based QFI}~\cite{Nielsen10}, 
establishes as the infinitesimal statistical distance in quantum theory, thereby providing a physical picture of ultimate precision limit. see~\cite{Yuan17} for channels.

\textit{Quantum geometric tensor}~\cite{Provost80,Venuti07}. A unified geometric object underlying multiparameter estimation is the quantum geometric tensor. For a pure quantum state \(\ket{\Phi}\), it is defined on the projective Hilbert space as
\begin{equation}
\mathcal{G}_{\eta\nu} = \langle \partial_\eta \Phi | \partial_\nu \Phi \rangle - \langle \partial_\eta \Phi | \Phi \rangle \langle \Phi | \partial_\nu \Phi \rangle,
\end{equation}
$\eta,\nu = 1, \ldots, \text{dim} X$ with $X$ being the parameter space.
Its real and imaginary parts carry two complementary physical meanings. In particular, the real part of QGT is related to a Riemannian metric tensor, which is connected to the fidelity between two states, thereby establishing a relation with QFIM as
\begin{equation}
\mathbb{F}_{\eta\nu} = \frac{1}{4} \, \text{Re} \left( \mathcal{G}_{\eta\nu} \right).
\end{equation}
The imaginary part of the QGT is linked to the Berry curvature. Specifically, it can be written as
\begin{equation*}
\text{Im}(\mathcal{G}_{\mu\nu}) = \text{Im} \left( \langle \partial_\mu \psi | \partial_\nu \psi \rangle \right) = -\frac{1}{2} \left( \partial_\mu A_\nu - \partial_\nu A_\mu \right),
\end{equation*}
where \(A_\eta := i \langle \psi | \partial_\eta \psi \rangle\) is the {Berry connection}, and \(\Upsilon_{\eta\nu} := \partial_\eta A_\nu - \partial_\nu A_\eta\) is the \textit{Berry curvature}~\cite{Berry84}. 

Furthermore, considerable efforts have been devoted to explore the connection between the QFIM and the Berry curvature through various uncertainty relations~\cite{Guo16,Lu21}. 
In this context, a Heisenberg-type uncertainty 
relation has been derived for the regret of Fisher information which quantifies the inaccuracies of a measurement for estimating unknown parameters, relating the measurement error obtained through measurement uncertainty relation~\cite{Ozawa03,Hall04,Lu21}. For simultaneously estimating parameters ${\vec{\theta}} \coloneqq \{\theta_j\}_j$, the normalized square-root regret for a parameter $\theta_j$ is defined as $\Delta_j \coloneqq \sqrt{R_{jj}/\mathcal F_{jj}}$, where $R_{jj} \coloneqq \mathcal F_{jj}-\mathscr{F}_{jj}$ quantifies the gap between the QFI $\mathcal F_{jj}$ and the classical Fisher information $\mathscr{F}_{jj}$  achieved by a chosen measurement corresponding to the parameter $\theta_j$ and $\Delta_j \in [0,1]$. The trade-off relation between the regrets of Fisher information corresponding to two parameters $\theta_j$ and $\theta_k$ 
reads as
\begin{equation}
\label{eqn-heiseberg-type}
    \Delta_j^2 + \Delta_k^2 + 2\sqrt{1-C_{jk}^2}\,\Delta_j \Delta_k \ge C_{jk}^2,
\end{equation}
with a real number, $C_{jk} \coloneqq \frac{|\text{Im}(\mathcal Q_{jk})|}{\sqrt{\mathcal F_{jj}\mathcal \mathscr{F}_{kk}}}$, where $\mathcal Q_{jk} \coloneqq \Tr(L_jL_k\rho_{\vec{\theta}})$ is constructed from the SLDs $L_j$ corresponding to the parameter $\theta_j$ and the encoded state $\rho_{\vec{\theta}}$
(for a family of pure encoded states, this inequality is saturated; there exists a quantum measurement that achieves equality.) Note that the real part of $\mathcal Q_{jk}$ corresponds to the off-diagonal element of the QFI matrix $\mathcal{F}_{jk}$, while the imaginary part of the $\mathcal Q_{jk}$ denotes the Uhlmann matrix, which quantifies the incompatibility of the optimal measurements corresponding to different parameters to be estimated. Interestingly, for pure encoded states, $\text{Im} (\mathcal Q_{jk})$ reduces to the Berry curvature, defined before. The inequality in Eq.~\eqref{eqn-heiseberg-type} can also be tightened for mixed quantum states by using variants of $C_{jk}$. Therefore, this establishes a connection between the precision bound of multiparameter estimation and the geometry of the underlying parameter space.
 
\textit{Connection to quantum speed limit.} The quantum speed limit (QSL)~\cite{Mandelstam91,Anandan90} represents the minimum time required for a quantum system to evolve between two distinguishable states. This concept is inherently linked to the geometry of quantum state space, as the fastest evolution corresponds to a 
traversing the shortest geodesic path in the underlying quantum state space which, for example, can be measured by Fubini-Study or Bures metric. Given this geometric interpretation, QFIM naturally emerges as a tool to quantify the speed limit. Indeed, recent studies have utilized both the QFI  and QFIM to establish bounds on the quantum speed limit~\cite{Taddei13,Deffner17,Gessner18,Maleki23} for both closed and open quantum evolution.
Furthermore, M. Beau and A. del Campo~\cite{Beau17} demonstrate a connection between the QFI associated with coupling constants between system-environment, and the quantum speed limit where the multi-body Hamiltonian is considered in the presence of baths. Their results reveal an interplay between nonequilibrium dynamics, and critical sensitivity in quantum metrology.

\section{\textbf{Estimation for different encoding processes}}
\label{sec:different-encoding-process}
In quantum parameter estimation, the first step is to encode the unknown 
parameter(s) into a probe state via a quantum process. The aim here is to infer these parameters
as precisely as possible, as quantified by various metrological bounds, discussed in Sec.~\ref{sec:parameter-estimation}. Hence, encoding plays a central role in determining achievable precision and designing optimal strategies for preparing probes and measurement.
These quantum processes may arise from unitary operations, open system dynamics described by 
noisy channels, thermal processess, and more general parameter dependent quantum operations. 
We provide an overview of the different types of encoding processes that have been studied within
parameter(s) estimation theory, highlighting their distinct features, responsible for enhanced precision.

\subsection{Unitary encoding}
\label{subsec:unitary-encoding}
The most common encoding process in quantum parameter estimation involves unitary evolution, 
which imprints the unknown parameter onto a probe state. Here, the parameter to be estimated may correspond to quantities such as time, phases of a unitary operator, or parameters appearing in the generator or Hamiltonian of the system. Suppose that the parameter of interest is denoted by $\theta$, which is encoded into a quantum state through a unitary operation. The resulting encoded state can be written as
\begin{equation}
\rho(\theta) = U(\theta)\,\rho_0\,U^{\dagger}(\theta) = e^{-iH\theta}\,\rho_0\,e^{iH\theta},
\end{equation}
where $\rho_0$ represents the input probe state and $\rho(\theta)$ denotes the parameter-dependent encoded state. Here $H$ is the generator (or Hamiltonian) responsible for the unitary encoding. In such parameter-estimation tasks, the goal is to estimate the value of $\theta$ with the maximum possible precision by optimizing the probe, measurements and estimator according to the chosen metrological framework.

A significant amount of work has been devoted to local quantum parameter estimation with unitary encoding~\cite{Giovannetti04,Giovannetti06,Bhattacharyya24b,pal25}. In this setting,  the generator is often a sum of local one-body terms, resulting in the well-known shot-noise and Heisenberg scaling laws, depending on the available quantum resources. More recently, this framework has been extended to nonlocal generators, containing few-body or genuine many-body interaction terms. For a $k$-body Hamiltonian, the nonlinear dependence of the dynamics can significantly enhance the scaling of the QFI with the number of probes, in some cases surpassing the conventional Heisenberg limit and yielding super-Heisenberg behavior~\cite{Boixo07, Boixo08,Napolitano11}. Such interaction-enhanced metrology has been studied in spin systems, bosonic platforms, and nonlinear interferometric settings, both for ideal closed dynamics and in the presence of decoherence, loss, and control imperfections, which is discussed in detail in Sec.~\ref{sec:applications}.

Unitary encoding also provides an appropriate benchmark for more general and realistic metrological scenarios. These include parameter estimation in open quantum systems with dissipative or non-Markovian noise, estimation of noise strengths and channel parameters, thermometry based on thermal states, continuous-time monitored protocols, and metrology assisted by indefinite causal order or quantum switches~\cite{Chiribella13,Ebler18,Mukhopadhyay24}. Comparing such conditions to the unitary limit can be useful in determining how noise, memory effects, causal structure, and nonequilibrium dynamics affect the ultimate precision constraints. Each of these broader encoding paradigms is discussed in the following subsections.

\subsubsection{Super-Heisenberg scaling through nonlinearities}
\label{subsubsec:super-Heisenberg}
When the encoding Hamiltonian involves many-body interactions or  time dependence, the precision scaling can exceed the super-Heisenberg scaling (see also Ref.~\cite{rams2018}) can be achieved~\cite{Boixo07, Roy08,Boixo08,Choi08,Napolitano11,Pang17,Beau17,Mishra21,Yang22,he2023,Yousefjani23,Mihailescu24,Adani24,Sahoo24b,Mihailescu2024Jul,Mondal25b,cheng2025}. In this regime, the estimation error scales as either $1/N^{\beta}$ or $1/t^{\beta}$ with $\beta > 2$, where $N$ represents the number of particles involved in the estimation process and $t$ denotes the evolution time. Furthermore, many-body phase transitions can lead to exponentially enhanced quantum sensing, even when realistic constraints such as state-preparation time are taken into account~\cite{Sarkar25}.

\begin{figure*}[t]
    \centering
    \includegraphics[scale=0.28]{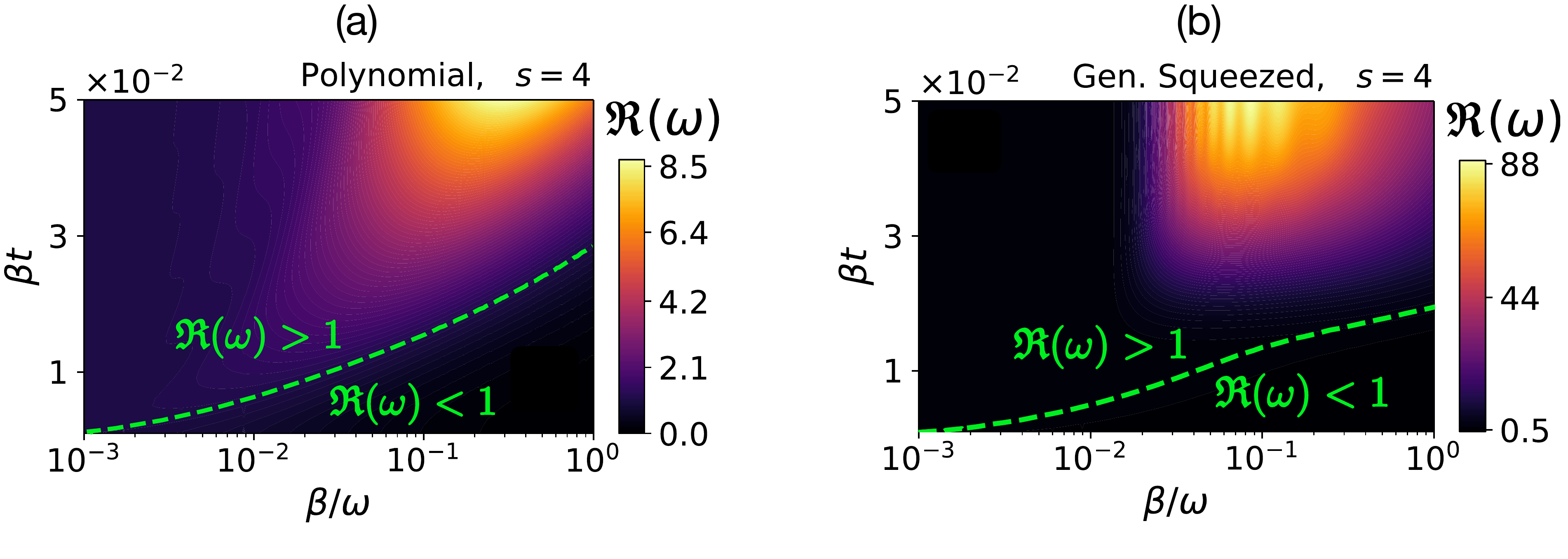} \hspace{5mm}
    \includegraphics[scale=0.26]{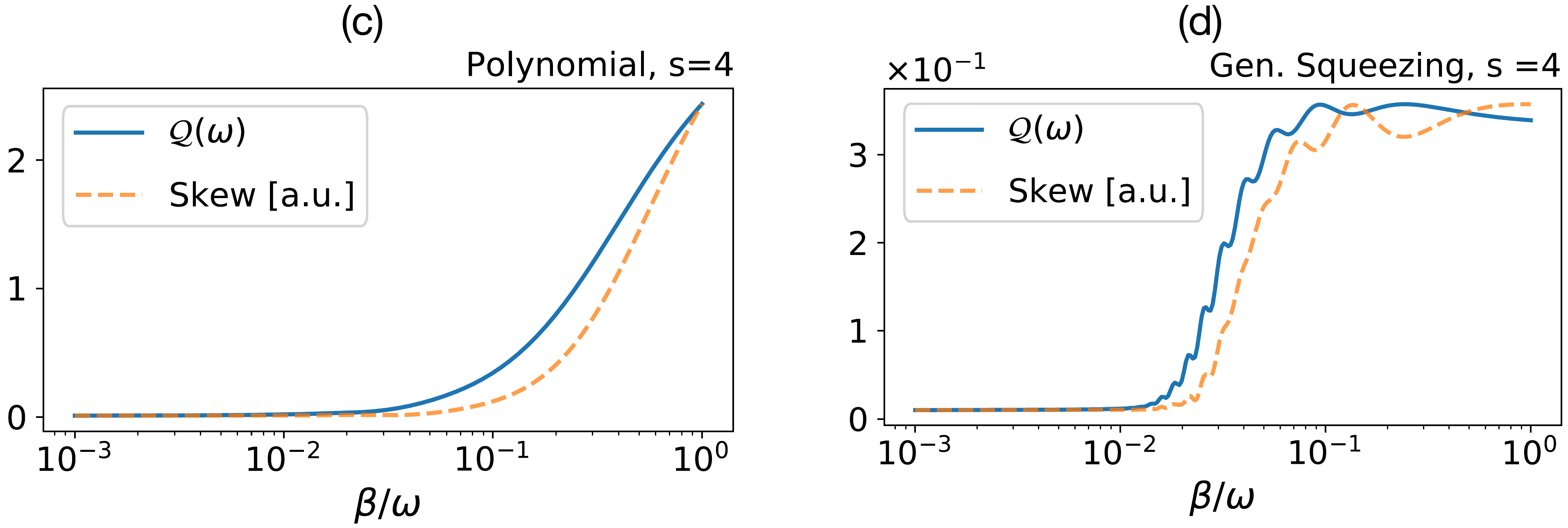}
    \caption{Nonlinear-enhanced frequency estimation~\cite{Montenegro25}.
Panels (a) and (b) show the ratio of QFIs, $\mathfrak{R}(\omega)$, as functions of the nonlinearity strength $\beta/\omega$ and time $t$ for $s=4$, corresponding to the polynomial and generalized squeezing cases, respectively. Both the figures demonstrate nonlinear enhancement in frequency sensitivity. Panels (c) and (d) depict the Wigner–Yanase skew information and the QFI as functions of $\beta$ (for $s=4$ and fixed $\beta t = 0.05$), illustrating the connection between degree of noncommutatiity and quantum sensing.
}
    \label{fig:paris_prl2025}
\end{figure*}

Recent works have demonstrated that, even without explicit many-body interactions or externally time-dependent Hamiltonians, suitably engineered higher-order nonlinearities can substantially enhance estimation precision and uncover a direct connection between quantum metrological sensitivity and quantum scrambling~\cite{Li23,Montenegro25,Xie25}. In particular, a substantial enhancement in estimation precision is reported via a nonlinear Hamiltonian used to generate dynamical states compared to the linear case. Furthermore, a relationship between this nonlinearity-induced enhancement in estimation precision and the Wigner–Yanase skew information~\cite{Wigner63,Luo03,Chen05} is established. The Hamiltonian containing the nonlinear term \(H_1\) reads
\begin{equation*}
    H = \omega H_0 + \beta H_1, \hspace{1cm} H_0 := a^\dagger a, \label{eq_hamiltonian_probe}
\end{equation*}
where $\omega$ is the unknown frequency parameter to be estimated, $\beta$ is a known coupling constant, and $a$ ($a^\dagger$) are the annihilation (creation) operators, satisfying $[a, a^\dagger] = 1$. Three kinds of nonlinearity are incorporated, namely  the polynomial case corresponding to   \( H_1= (a^\dagger + a)^s \), the generalized squeezed case given by \( H_1 = (a^{\dagger s} + a^s) \), and the generalized Kerr case as \( H_1 = a^{\dagger s} a^s \), 
where $s \in \mathbb{N}$ and $s \geq 1$~\cite{Montenegro25}. Starting from a coherent state probe, the system is evolved under the nonlinear Hamiltonian with nonvanishing coupling strength \(\beta\). To quantify the metrological advantage induced by the nonlinear term, one considers the ratio
\(  \mathfrak{R}(\omega) \coloneqq \frac{\mathcal{F}(\omega)}{\mathcal{F}_0(\omega)}\),
where $\mathcal{F}(\omega)$  and $\mathcal{F}_0(\omega)$ denote the QFI with respect to $\omega$  evaluated with ($\beta > 0$) and without ($\beta = 0$) the nonlinear contribution, respectively. Consequently, the condition \(\mathfrak{R}(\omega) >1 \) provides a clear signature that the nonlinearity enhances estimation precision relative to the linear benchmark~\cite{Montenegro25}.
To make a fair comparison between probes with and without nonlinearities, 
the QFIs are computed for evolved coherent states $|\alpha\rangle$ and $|\alpha_0\rangle$, respectively, under the condition that both probes possess the same average energy initially,
 \(\langle \alpha_0 | \omega a^\dagger a | \alpha_0 \rangle = \langle \alpha | H | \alpha \rangle\).
Under this energy constraint, enhancement is explicitly observed, for example, at order \(s\geq 3\) for all representative families of nonlinearities except the generalized Kerr nonlinearity (see Fig.~\ref{fig:paris_prl2025}). Thus, higher-order interactions can serve as effective resources for improving parameter sensitivity~\cite{Montenegro25}.

A particularly notable observation is that the maxima of the QFI and the degree of
noncommutativity, as quantified by the Wigner-Yanase skew information,\footnote{Let us define the Wigner-Yanase skew information, \(\mathcal{W}_s(A, N) = -\frac{1}{2} Tr([\sqrt{A}, N]^2)\), where \(A\) and \(N\) are positive and Hermitian operators respectively. For this problem, \(A\) is chosen to be the evolved state and \(N = a^{\dagger} a\) is the number operator.} occur at nearly the same value of the nonlinear coupling \(\beta\). This close correspondence reveals a direct interplay between metrological performance and quantum scrambling, with maximal sensitivity emerging near the regime of strongest noncommutativity generated by the nonlinear dynamics (see Fig.~\ref{fig:paris_prl2025}). 

Recent work has shown that nonlinear scrambling not only provides a genuine quantum enhancement in precision bounds, but can also enable super-Heisenberg scaling in both closed and open quantum systems~\cite{Xie25}.

\subsection{Noisy quantum metrology}
\label{subsec:noisy-channels}
Noise is unavoidable in realistic quantum systems, and hence realization of metrological protocols must incorporate its effects in fundamentally distinct contexts. One major line of study treats noise as an external disturbance acting during probe engineering, parameter encoding, or measurement. In this scenario, the aim is to determine how estimation precision degrades
in presence of decoherence, and how to overcome such losses through optimal probes, control strategies, error-correction techniques or adaptive measurements.
A second, conceptually distinct direction considers noise as the quantity to be estimated itself.
In this case, the task is to infer the physical parameters that characterize a noisy quantum process or channel, such as dephasing rates, loss coefficients, coupling strength between system and environment, bath temperature of a thermal bath, correlation time or memory parameters. 
This information is operationally useful because characterizing the noise enables improved control, mitigation, or error handling in other quantum information protocols that experience the same type of noise. In what follows, we first review the parameter estimation where the parameter(s) is(are) encoded in the presence of  noise~\cite{Huelga97,Alipur14,Yousefjani17,Peng23,Bhattacharyya24a}, whereas in the latter part of this subsection, we discuss the problem of estimating the parameters of noisy quantum channels,
referred  to as quantum channel metrology~\cite{Fujiwara01,Fujiwara03,Hayashi10,Rafal12,Rafal14,Pirandola17a,Laurenza18,Razavian19b,Zhou21,Kurdzialek2b}.

\subsubsection{Encoding in presence of noise}
\label{subsubsec:encoding-presence-of-noise}
In realistic setup, perfectly noiseless parameter encoding is rarely unattainable,  and consequently the interaction of environment with the probe state during encoding typically
leads to a noisy encoded state. A central problem in this situation is to determine how the optimal achievable precision scales in the presence 
of different categories of noise, specifically, uncorrelated and correlated noise, 
providing different metrological limits.

\textit{Uncorrelated noise.}
Uncorrelated noise acts independently on each probe particle or sensing subsystem. Typical examples include local dephasing, amplitude damping, depolarization, photon loss, and independent thermal relaxation channels. In contrast to the ideal noiseless unitary setting, where entangled probes can attain Heisenberg-limited precision, such local noise generally degrades coherence and, in the asymptotic regime, suppresses the quadratic enhancement in precision. As a result, the achievable scaling often reverts to the standard quantum limit, possibly with only a constant-factor quantum advantage, known as no-go theorem for quantum metrology when uncorrelated time-homogeneous dephasing noise acts on the system~\cite{Escher11,Knysh14,Kolodynski14,Albarelli18}.

Nonetheless, there are certain situations when SQL can be exceeded. A paradigmatic example is the estimation of a magnetic-field strength in the presence of decoherence induced by an independent bosonic environment~\cite{Matsuzaki11}. The total Hamiltonian considered, in this case, takes the form as \(H_{total} = H_{sys} + H_{env} + H_{int}\) where \(H_{sys} = \omega \sum_{l=1}^{N} \sigma_z\) with \(\omega\) the unknown field parameter and \(N\) the number of qubits, \(H_{env}\) is chosen to be a bosonic bath while \(H_{int}\) represents the interactions between the system and the bath. Starting from a GHZ probe state, by optimizing the interrogation time according to  \(s N^{-z}\) (where \(s\) is constant and \(z\) is non-negative real number), the estimation uncertainty is shown to scales as \(\Delta(\omega) \sim N^{-3/4}\), thereby surpassing the SQL scaling \(N^{-1}\), although it remains above the ideal Heisenberg scaling of \(N^{-2}\). It highlights that even under independent decoherence, properly optimized strategies can retain a nonclassical scaling advantage. For other scenarios like correlated noise having memory etc, see Refs.~\cite{Huelga97,Dorner12,Ozeri13,Jeske14,Kessler14,Arrad14,Dur14,Lu15,Brask15,Plenio16,Bai23}.
From a group-theoretic perspective, the suppression of the standard quantum limit becomes possible when the dynamics departs from the semigroup structure at short times, thereby generating time-inhomogeneous non-Markovian evolution~\cite{Smirne16}.

An immediate and  natural question is whether the standard quantum limit can still be beaten, when the noise is strictly Markovian. A decisive answer can be obtained by considering the adaptive strategies. For Hamiltonian parameter estimation with Markovian noise described by  Gorini-Kossakowski-Sudarshan-Lindblad (GKSL) master equation \cite{open_quan_book} 
\begin{eqnarray}
    \frac{d\rho}{dt} = - i \omega [H, \rho] + \mathcal{D}[L_k](\rho),
\end{eqnarray}
where \(\omega\) is the estimation parameter,  \(H\) is the Hamiltonian  and \(L_k\)'s are Lindblad noise operators, Heisenberg-limited scaling can still be restored using quantum error correction or adaptive protocols~\cite{Sekatski17,rafal17,Zhou18,Das25} 
provided $G$ and $\{L_k\}$ obey a condition, known as HNLS (Hamiltonian-not-in-Lindblad span) condition, where the Hamiltonian is $H = \omega G$.
Any Hamiltonian $H$ obeys the HNLS condition if $H$ does not lie in the linear span $\mathcal{S}$ of the operators ${\mathbb{I}, L_k, L_k^\dagger, L_k^\dagger L_j}$ for all $k, j = \{1, \dots, r\}$, where $r$ is the rank of the noise channel, i.e., the minimal number of Lindblad operators required to describe the noise. 
When the HNLS condition is violated by any Hamiltonian, then the QFI scales, at most, linearly with $t$ with $t$ being the total sensing time. In both cases, the best possible precision can be found by solving a semidefinite program.
This result reveals that Markovian noise does not universally forbid quantum enhancement; rather, the attainability of super-classical scaling depends on the geometric relation between the signal generator and the noise operators. Subsequent works developed explicit quantum error-correcting sensing codes and semidefinite-program methods to construct optimal protocols whenever the HNLS condition is satisfied (see e.g. Ref.~\cite{Albarelli22b,Wan22}).

In Ref.~\cite{Chapeau-Blondeau18}, it is found that in the typical phase estimation problem, where the GHZ state shows a $1/N^2$ scaling in the estimation error, this advantage vanishes in the presence of uncorrelated noise, i.e., the QFI asymptotically goes to zero. Instead, at some intermediate $N$, a state with nonzero partial entanglement yields the best estimation error in such case (see also Ref.~\cite{Bhattacharyya24, Arieli26}).

\textit{Correlated noise.} In realistic quantum systems, noise is seldom independent, and correlations, both spatial and temporal, are often unavoidable. This makes the study of correlated noise not only relevant but essential. Importantly, correlated noise is not purely detrimental; it can also give rise to constructive effects and plays a significant role across various areas of quantum information processing.
For example, in quantum computing, standard error-correction schemes based on independent noise models often fail to capture realistic scenarios, motivating the development of protocols that explicitly incorporate noise correlations~\cite{Aharonov06,Novais07,Preskill13}. At the same time, such correlations can be harnessed to mitigate decoherence through decoherence-free subspaces (DFS)~\cite{Lidar98,Duan98,Lidar2001a,Lidar2001b}. In quantum communication, while uncorrelated (product) noise typically removes any advantage of entangled states for classical information transmission, correlated noise can restore this advantage; for instance, in both qubit~\cite{Macchiavello02,Macchiavello04} and Bosonic Gaussian channels~\cite{Cerf05,Giovannetti05}, entanglement has been shown to enhance mutual information beyond a certain correlation threshold.
Even, more recently, correlated noise has also been exploited in experimental platforms, for example, to suppress decoherence effects in piezo-optomechanical microwave–optical quantum transducers, thereby improving their performance~\cite{Hou25}. Overall, the study of correlated noise has attracted considerable attention across diverse physical platforms and applications~\cite{Clemens04,Novais06,Buscemi10,Higginbotham18,Xia23,Harper23,Salhov24}.

Correlated noise in multipartite quantum systems can be broadly classified into three categories: (i) \emph{spatial correlated noise}, (ii) \emph{temporal correlated noise}, and (iii) \emph{spatiotemporal correlated noise}. These arise from the structure of environmental fluctuations and their influence on different subsystems across space and time.

To formalize this, consider \(N\) quantum systems interacting with an environment. Spatially uncorrelated noise corresponds to each system being coupled to an independent bath, whereas spatial correlations arise when multiple systems interact with a common environment. A convenient way to characterize such correlations is through the two-point bath correlation function~\cite{Riberi22}. Let \( B_n \) denote the bath operator associated with the environment coupled to the \( n \)-th system. Its Heisenberg evolution is given by
\[
B_n(t) = e^{i H_B t} B_n e^{-i H_B t},
\]
where \( H_B \) is the bath Hamiltonian. The two-point correlation function is defined as
\begin{align*}
C_{nm}(t_1, t_2) \equiv \langle B_n(t_2) B_m(t_1) \rangle_{\mathrm{B}} 
= \mathrm{Tr}_{\mathrm{B}} \big[ B_n(t_2) B_m(t_1) \rho_{\mathrm{B}} \big],
\end{align*}
with \( t_2 \ge t_1 \ge 0 \), and \( \rho_{\mathrm{B}} \) is the initial bath state. Assuming time-translational invariance, the correlation function depends only on the time difference,
\[
C_{nm}(t_1, t_2) = C_{nm}(t_2 - t_1) \equiv C_{nm}(t) = \langle B_n(t) B_m(0) \rangle_{\mathrm{B}}.
\]
Building on this, mathematically, the different types of correlated noises are described as follows:
\begin{itemize}
    \item \textit{Spatial correlated noise.} Noise is spatially correlated if the correlation function cannot be factorized as
    \[
    C_{nm}(t) = \delta_{nm} f_n(t),
    \]
    where \( f_n(t) = f_n(-t)^* \). Violation of this condition indicates correlations between different subsystems.

    \item \textit{Temporal correlated noise.} Noise is temporally correlated (non-Markovian) if the correlation function cannot be written as
    \[
    C_{nm}(t) = c_{nm} \delta(t),
    \]
    where \( c_{nm} = c_{mn}^* \). Deviation from the delta-function form signals the presence of memory effects.

    \item \textit{Spatiotemporal correlated noise.} Noise is spatiotemporally correlated if the correlation function deviates from the fully uncorrelated form
    \[
    C_{nm}(t) = \delta_{nm}\delta(t),
    \]
    indicating the simultaneous presence of both spatial and temporal correlations.
\end{itemize}

Having introduced different types of noise correlations, we now turn to their implications for quantum metrology. Realizing the full potential of quantum metrology requires a careful accounting of realistic noise sources. While temporally uncorrelated noise acting independently on each probe forbids any superclassical scaling, the presence of correlations can restore metrological gain.
In the context of parameter estimation under noise, a wide range of studies have explored the role of spatial~\cite{Jeske14,Layden18,Czajkowski19,Layden19,Layden20}, temporal~\cite{Chin12,Macieszczak15,Smirne16,Haase18,Smirne19,Tamascelli20,Altherr21,Yang24}, and spatiotemporal correlations~\cite{Szankowski14,Beaudoin18,Riberi22}; see also Ref.~\cite{Das26} for further developments. These works demonstrate that different forms of correlations can be exploited to enhance precision.
For example, temporal correlations in otherwise spatially uncorrelated noise can be harnessed to achieve a superclassical, Zeno-like scaling at short interrogation times~\cite{Chin12,Smirne16}. In contrast, when noise lacks temporal correlations, spatial correlations can be exploited to recover enhanced precision, either by encoding the probes into decoherence-free subspaces~\cite{Dorner12,Jeske14} or by employing quantum error-corrected sensing protocols to suppress noise while preserving the signal~\cite{Layden18}. More broadly, when both spatial and temporal correlations coexist, as in systems coupled to a common environment with a structured (colored) spectrum, environmental memory effects can help to sustain improved sensitivity over longer timescales, provided the noise can be effectively described by a classical model~\cite{Szankowski14}.

Moreover, significant experimental progress has been made toward quantum metrology in the presence of correlated noise, particularly in optomechanical platforms. In recent years, a variety of techniques have been developed to enhance the sensitivity of optomechanical sensors. For example, the use of squeezed light has enabled significant suppression of imprecision noise, leading to a $3$ dB sensitivity improvement in Advanced LIGO~\cite{Tse19}, as well as enhanced sensitivity and bandwidth in optomechanical magnetometry~\cite{Li18}. Such improvements have also been realized through several approaches, including two-tone driving~\cite{Ockeloen-Korppi16,Shomroni19}, negative-mass oscillators~\cite{Tsang12,Moller17,Lepinay21}, and intrinsic optomechanical Kerr nonlinearities~\cite{Mason19}.
More recently, joint force sensing using entangled probes across multiple optomechanical systems has been shown to reduce the effective noise floor by about $2$ dB, resulting in nearly $40\%$ enhancement in force sensitivity in both thermal-noise- and shot-noise-dominated regimes~\cite{Xia23}.

\subsubsection{Noise strength estimation}
\label{subsubsec:noise-estimation}
We mow turn to discuss the estimation of the strength of noise affecting quantum systems. Here the quantity of interest is to assess the noise parameters, which is essential to benchmark quantum hardware, identify error sources, and designing robust quantum sensing architecture. Noise can be broadly classified into two categories: (i) uncorrelated noise in which each system interact with the environment separately, and (ii) correlated noise, where distinct subsystems experience common environmental effects.

\textit{Uncorrelated noise.} Let us first focus on the estimation of noise parameter(s) of an uncorrelated quantum noisy channel in the parallel scheme, where the channel acts locally and independently on the multipartite probe, with the goal of estimating the channel's noise strength. So far, most works on estimating the noise strength of quantum channels have focused on the purification-based definition of the extended QFI~\cite{Escher11}.

Since the QFI of an encoded state $\rho_\theta$ is related to the fidelity between two nearby states $\rho_\theta$ and $\rho_{\theta + d\theta}$, where $\theta$ is the parameter to be estimated~\cite{Zhou19}, the QFI of any encoded state $\rho_\theta$ can alternatively be expressed as~\cite{Kolodynski13}
\begin{equation*}
    \mathcal{F}(\rho_\theta) = 4 \Big [ \min_{\ket{\psi_\theta}} \langle \dot{\psi_\theta} | \dot{\psi_\theta} \rangle - |\langle\dot{\psi_\theta}|\psi_\theta\rangle|^2\Big ],
\end{equation*}
where $\ket{\psi_\theta}$ denotes a purification of $\rho_\theta$, i.e., $\rho_\theta = \Tr_E[\ket{\psi_\theta} \bra{\psi_\theta}]$. It can, equivalently, be written as \(\mathcal{F}(\rho_\theta) = 4 \min_{\ket{\psi_\theta}} \langle \dot{\psi_\theta} | \dot{\psi_\theta} \rangle\)~\cite{Fujiwara08}.  The optimal QFI for the noise strength of a quantum channel \(\Lambda_\theta\), obtained by optimizing over all possible probes, referred to as channel QFI, is defined as
\begin{equation}
\mathcal{F}(\Lambda_\theta) \coloneqq \max_{\psi_{\text{in}}} \mathcal{F}(\Lambda_\theta(\psi_{\text{in}})),
\end{equation}
where $\psi_{\text{in}} \coloneqq \ket{\psi_{\text{in}}} \bra{\psi_{\text{in}}}$ denotes the input probe state, $\theta$ is the parameter characterizing the channel $\Lambda_\theta$ and \(\Lambda_\theta(\psi_{in})\) is the resulting state after the action of a channel. Now, using the purification-based expression for QFI, the channel QFI can be reformulated as~\cite{Kolodynski13}
\begin{eqnarray}
     \nonumber \mathcal{F}(\Lambda_\theta) &=& 4 \max_{\psi_{\text{in}}} \min_h \langle\dot{\tilde{\psi_\theta}}|\dot{\tilde{\psi_\theta}}\rangle\\&=&4 \max_{\psi_{\text{in}}} \min_h\langle\psi_{in}|\sum\dot{\tilde{K}}_i(\theta)^\dagger\dot{\tilde{K}}_i(\theta)|\psi_{in}\rangle
     \label{eq:pure_qfi}
\end{eqnarray}
where $\Lambda_\theta(\psi_{\text{in}}) = \Tr_E\bigg[\ket{\tilde{\psi}_\theta}\bra{\tilde{\psi}_\theta}\bigg]$, and 
$\ket{\tilde{\psi}_\theta} \coloneqq u_\theta^E U_\theta^{SE} \left( \ket{\psi_{\text{in}}} \otimes \ket{1} \right)$ with $U_\theta^{SE}$ being a global unitary acting on the system and environment generating the map $\Lambda_\theta$, $u_\theta^E$ is a unitary acting solely on the environment while $h$ is the generator of $u_\theta^E$, and $\ket{1}$ is a fixed initial state of the environment. The minimization involved in Eq.~(\ref{eq:pure_qfi}) is not easy to compute and hence it is essential to obtain an upper bound on the channel QFI. Towards this the optimal QFI for estimating the noise strength of a quantum channel $\Lambda_\theta$, obtained by optimizing over all possible input probes, referred to as the channel QFI, is defined as \cite{Fujiwara08,Kolodynski13}
\begin{equation*}
\mathcal{F}(\Lambda_\theta \otimes \mathbb{I}) \coloneqq \max_{\rho_{\text{in}}} \mathcal{F}(\Lambda_\theta \otimes \mathbb{I},(\rho_{\text{in}})).
\end{equation*}
where $\rho_{\text{in}}$ denotes the input probe acting on one joint system and auxiliary. $\mathbb{I}$ is the identity operator acting on the Hilbert space of the auxiliary. In terms of \(\tilde{K}_i^\theta\), the above definition reduces to
\begin{equation}
   \mathcal{F}(\Lambda_\theta \otimes \mathbb{I}) 
 = 4 \min_h || \sum_{i=1}^r \dot{\tilde{K_i}}(\theta)^\dagger \dot{\tilde{K_i}}(\theta) ||,
 \label{eq:core_qfi}
\end{equation}
where $\dot{\tilde{K_i}}(\theta) \coloneqq \dot{{K_i}}(\theta) -\iota \langle i |h|j\rangle K_j(\theta)$ and $K_i(\theta)$ are the Kraus operators of channel $\Lambda_\theta$, $||\cdot||$ denotes operator norm of ``$\cdot$''. $\{\ket{i}\}_{i=1}^r$ denotes the basis of environmental Hilbert space where $\ket{1}$ is one amongst them.

The channel QFI is related to the extended channel QFI as $\mathcal{F}(\Lambda_\theta) 
 \leq \mathcal{F}(\Lambda_\theta \otimes \mathbb{I})$, when a single copy of the input probe is available. This relation even holds when one considers arbitrary finite number $N$ of copies, i.e., $\mathcal{F}(\Lambda_\theta^{\otimes N}) 
 \leq \mathcal{F}((\Lambda_\theta \otimes \mathbb{I})^{\otimes N})$.
 However, it was shown that in the asymptotic limit, the exact expression of extended channel QFI is same as in Eq. (\ref{eq:core_qfi})~\cite{Kolodynski13} with an additional $\beta_K \coloneqq \sum_{i=1}^r \dot{\tilde{K_i}}(\theta)^\dagger \tilde{K_i}(\theta) =0$, i.e., $\mathcal{F}_{\text{as}}(\Lambda_\theta \otimes \mathbb{I}) = \lim_{N \to \infty} \frac{\mathcal{F} ((\Lambda_\theta \otimes \mathbb{I})^{\otimes N })}{N}
 = 4 \min_{h, \beta_K=0} || \sum_{i=1}^r \dot{\tilde{K_i}}(\theta)^\dagger \dot{\tilde{K_i}}(\theta) ||$. The asymptotic QFI is known to scale either linearly or quadratically with number of copies of a quantum channel and the scaling is achievable only in specific cases. A criteria to achieve quadratic scaling and the corresponding of asymtotic QFI of the channel is derived~\cite{Hotta05,Hotta06} and quantum error correction protocol is provided to achieve the bound~\cite{Pirandola17a,Pirandola17b,Laurenza18}.

The upper bounds of extended channel QFI in estimating noise parameter of any CPTP map have been found when any arbitrary finite number copies of input probe is available and the channel acts on the input probe according to the parallel and sequential strategies~\cite{Zhou21}.

Ref.~\cite{Mondal25} examines whether entanglement is required in optimal probes for estimating the noise strength of a wide class of local quantum encoding processes, known as vector encoding, which includes depolarizing and bit-flip channels. The authors establish a necessary criterion to find optimal probe(s) for noise parameter estimation of these encoding processes and use it to study the entanglement features of optimal probes in two representative cases: arbitrary-dimensional local depolarizing channels and local bit-flip channels. For noise estimation using two-qudit local depolarizing channels, maximally entangled probes are found to be not always optimal. Instead, as the noise strength decreases, the entanglement of the optimal probe grows in discrete steps, increasing one Schmidt rank at a time, until a sudden jump to a maximally entangled state occurs at a critical noise value that depends on the local Hilbert space dimension. Importantly, for any local depolarizing channel of arbitrary local dimension, there always exists a dimension-dependent noise regime where pure product probes achieve the best possible precision, and in the high-noise limit, fully product probes remain optimal even for multiparty probing. In contrast, for local bit-flip channels, entanglement is not a necessary metrological resource: a fixed product probe reaches the optimal precision across the entire noise range, and although certain entangled probes can attain the same precision, they offer no advantage over the optimal product strategy.

\begin{figure}
    \includegraphics[scale=0.18]{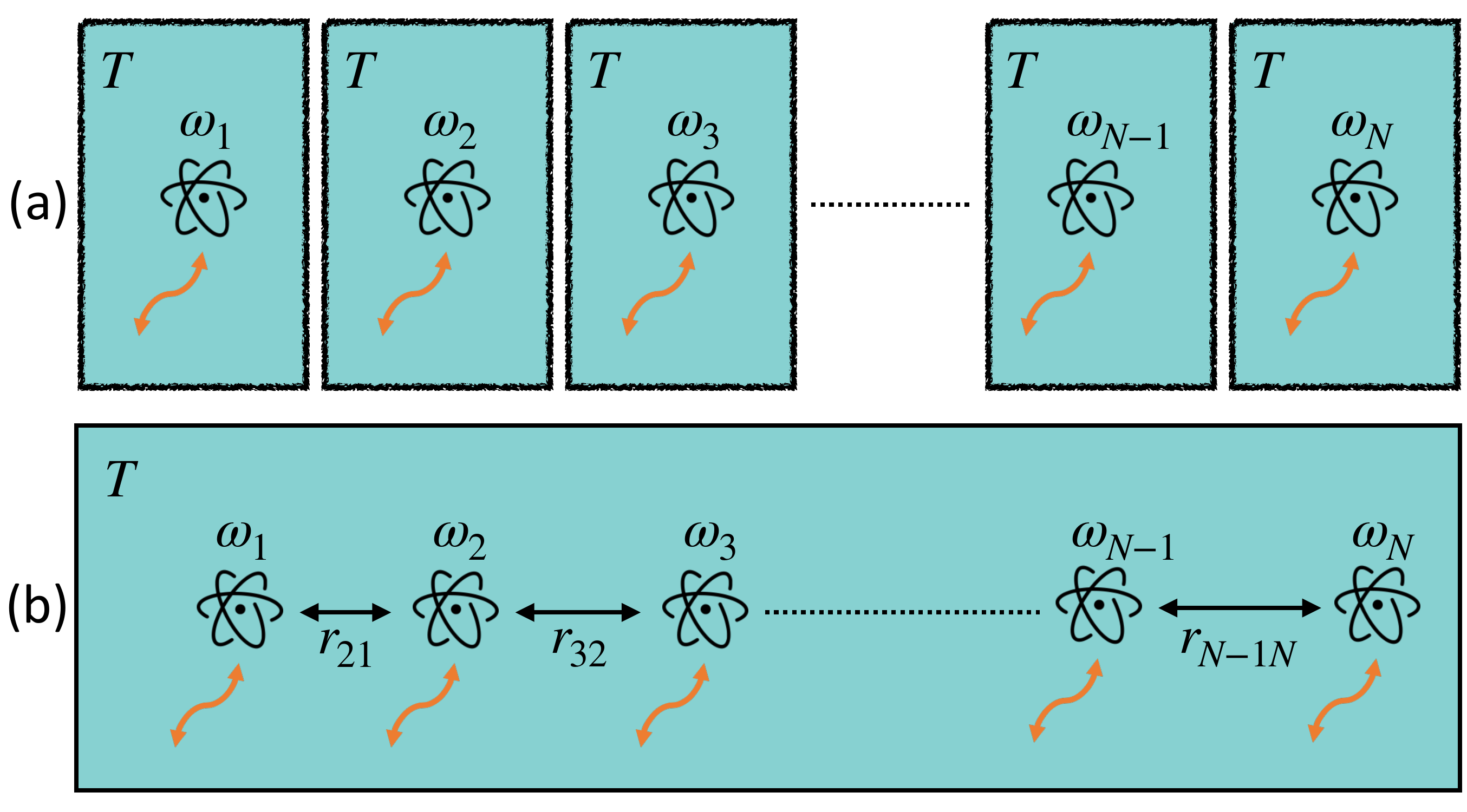}
   \caption{Schematic of thermometry protocols discussed in Ref.~\cite{Planella22}. (a) Independent baths: each thermometer couples to a separate bath, yielding additive precision. (b) Common bath: all thermometers couple to the same bath, leading to superadditive precision at low temperatures.}
\label{thermometry-in-correlated-noise}
\end{figure}

\textit{Correlated noise.} We now discuss the the estimation of correlated noise strength acting on input probes. As discussed before, such kind of noise introduces inter probe correlations that can modify the metrological performance. A paradigmatic example is provided in Ref.~\cite{Planella22} where thermometry (which is discussed in detail in the next subsection) is studied using bosonic multiparticle probes that are initially uncorrelated and non-interacting but affected by the common thermal bath (spatially correlated noise). This scenario is illustrated schematically in Fig.~\ref{thermometry-in-correlated-noise}. Although the probe remains seperable, the bath induces nontrivial correlations among systems, demonstrating environment-induced correlations, that significantly enhance precision of temperature estimation of the bath.  More generally,  one can estimate parameters present either in the system Hamiltonian or in the Lindblad operators. Within the most general adaptive framework, the attainable precession can depend on the noise correlations and the encoding stategy~\cite{Escher11}. For more works in this direction, see Refs.~\cite{Matsuzaki18,Wang24b,Dey25,Brady26}.

\begin{figure*}
\textbf{(a)} 
\includegraphics[scale=1.15]{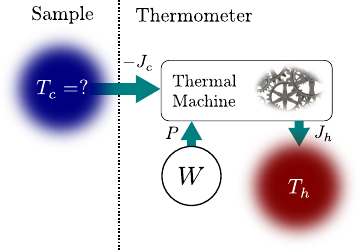} \hspace{5 mm}
\textbf{(b)}
\includegraphics[scale=1.15]{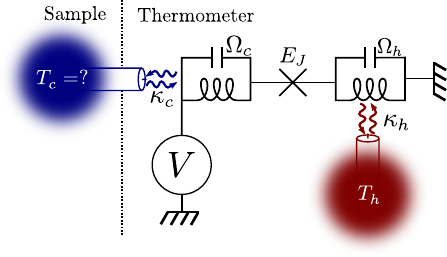} 
\caption{(a) A model-independent schematic representation of a quantum thermal machine operating as a thermometer. The temperature of the cold bath, \(T_c\), is estimated by calibrating it against a known hot-bath temperature \(T_h\). Heat flows between the cold and hot reservoirs, while the thermal machine is operated near the Carnot point, where both the heat currents and power consumption vanish, i.e., \(P = J_c = J_h = 0.\) At this operating point, the cold-bath temperature can be inferred from the Carnot efficiency relation, \( \eta = \eta_C = 1 - \frac{T_c}{T_h}. \) This protocol effectively converts a comparatively low-precision measurement of the hot-bath temperature into a high-precision estimate of the low temperature \(T_c\). Furthermore, operating the thermal machine in the refrigeration regime prevents unwanted heating of the cold bath during the measurement process. (b) Circuit-QED realization of a quantum thermal-machine thermometer. Two harmonic oscillators with frequencies $\Omega_h$ and $\Omega_c$ are individually coupled to thermal reservoirs at temperatures $T_h$ and $T_c$, respectively, while interacting with each other through a Josephson junction. An externally applied bias voltage \(V\) drives the system into a refrigeration regime, ensuring cooling of the cold reservoir while simultaneously enabling the estimation of the cold-bath temperature \(T_c\). The figure is taken from Ref.~\cite{Hofer17}.}

  \label{fig:cqed}
\end{figure*}

\subsection{Quantum thermometry}
\label{subsec:thermometry}
One of the primary tasks in parameter estimation is the accurate calibration of the temperature of a system, particularly to enhance the precision of temperature sensing in the quantum regime. Precision thermometry addresses the longstanding challenge of interpreting temperature fluctuations, since temperature is not a directly observable quantity and must instead be inferred from the statistical properties of measurable observables.

In quantum thermometry, the objective is to accurately estimate the temperature of a quantum system by coupling a quantum probe (thermometer) to a thermal reservoir, such that thermal information is encoded into the probe state. The temperature can then be inferred from measurements performed on the probe, either in the stationary equilibrium state~\cite{Stace10,correa15,Paris16,DePasquale16,Correa17,Campbell18,Latune20,Ullah25,Sarkar25b,Saha26,Sarkar26}, from stationary nonequilibrium states, or from time-evolving nonequilibrium states~\cite{Jevtic15,Kiilerich18,Cavina18,Feyles19,Zhang22,Sekatski22,Brenes23,Anto-Sztrikacs24,Ullah25c,Aiache24,Aiache25,Trombetti25,Xie26}.
When the probe reaches thermal equilibrium with the bath, it can usually be described by the Gibbs state
\begin{equation}
\rho = \frac{e^{-\beta H}}{\Tr\left[e^{-\beta H}\right]},
\end{equation}
where $\beta \coloneqq 1/(k_B T)$, $H$ is the system Hamiltonian, and $k_B$ is the Boltzmann constant. In this sense, quantum thermometry constitutes a form of \emph{dissipative parameter encoding}, in contrast to the unitary encoding schemes.
In equilibrium thermometry, energy measurements are optimal, as they saturate the QCRB independent of temperature. This implies that the QFI coincides with the energy variance of the thermometer and is therefore directly proportional to its heat capacity~\cite{correa15,Mehboudi19a}. However, in general, the optimal measurement often depends on the unknown temperature itself, making its practical implementation challenging.
Recent studies~\cite{Cavina18,Frazao24,Aiache24} have demonstrated that nonequilibrium thermometry, where the probe is measured before reaching thermal equilibrium, can offer significant advantages. Such approaches not only reduce the required measurement time but can also enhance estimation precision. In particular, while the estimation error for equilibrium probes typically diverges exponentially as $T \to 0$, nonequilibrium strategies can reduce this divergence to a polynomial scaling, although the error may still remain substantial. Therefore, many approaches have been developed to handle low-temperature quantum thermometry~\cite{Potts19,Jorgensen20}.

Building on the intuition that the QFI with respect to the temperature is proportional to the energy variance of the system and the energy eigenstates of the Hamiltonian constitute the optimal SLD basis for temperature estimation, Correa \emph{et al.} have demonstrated existance of an optimal structure of the Hamiltonian that maximizes temperature estimation precision~\cite{correa15}. Specifically, they proves that the optimal \( D \)-dimensional Hamiltonian corresponds effectively to a single qubit system, where the first excited energy level exhibits a degeneracy of \( D-1 \). This structure concentrates the thermal sensitivity around a well-defined energy gap, thereby enhancing thermometric performance. 
Notably, this optimal precision is highly sensitive to the exact value of the temperature being estimated. This strong dependence underscores the natural relevance and necessity of adopting the global estimation approach in such thermometric scenarios. To do so, a global parameter estimation framework has been employed in quantum thermometry~\cite{Mok21}, where it is shown that as the width of the temperature interval (i.e., sensing range) increases, the optimal Hamiltonian gradually shifts away from the effective two-level configuration. Beyond a certain threshold of the sensing interval, the optimal Hamiltonian becomes an effective three-level system. Further increasing the interval results in a transition to a four-level optimal Hamiltonian, and this trend continues with growing temperature uncertainty. This reveals how the optimal energy spectrum becomes increasingly complex as the available prior knowledge about the temperature becomes less precise. In addition, a notable advancement in the global quantum thermometry protocol is proposed in~\cite{Rubio21}, which relies on minimal measurement data and requires little to no prior knowledge of the system's temperature.  In contrast, Bayesian quantum thermometry~\cite{Boeyens21,Alves22,Jorgensen22,Mehboudi22,Glatthard22b} incorporates prior information about the temperature and accounts for the statistical structure of measurement outcomes, thereby extending the accessible sensitivity range and improving estimation accuracy.

Following the foundational developments in quantum thermometry, significant progress has been made in various directions.  For instance, on the practical side, quantum thermal machines have been demonstrated as effective thermometric devices for estimating ultra-low temperatures. More precisely, a conceptually different approach to quantum thermometry was proposed in \cite{Hofer17}, where a quantum thermal machine is employed as a temperature probe (see Fig.~\ref{fig:cqed}). The central idea is to couple the system of interest to a small quantum heat engine or refrigerator operating between two thermal reservoirs. When the machine operates close to the Carnot point, a simple relation between the transition frequencies of the working medium and the temperatures of the reservoirs emerges. In particular, for machines performing an Otto cycle, the condition
\[
\frac{\omega_c}{\omega_h}=\frac{T_c}{T_h}
\]
holds at the transition between the refrigeration and heat-engine regimes, where $\omega_c$ and $\omega_h$ denote the characteristic frequencies associated with the cold and hot transitions, respectively. By tuning the machine to this operating point and measuring the relevant frequencies, the unknown temperature $T_c$ of the sample can be directly inferred without requiring detailed knowledge of the system–bath coupling constants. Remarkably, the achievable precision approaches the optimal bound predicted by the QFI, making this scheme particularly suitable for thermometry in the ultra–low temperature regime.

As previously discussed, quantum advantage in thermometry can be achieved through quantum correlations. The role of quantum resources in surpassing classical limits has been studied extensively. 
For instance, 
if the thermalizing probe has an extensive thermodynamic energy, then the uncertainty in temperature estimation scales as \( \sim 1/\sqrt{N} \)~\cite{Stace10}. This scaling implies that, for a probe in a thermalized state, the SQL cannot be surpassed. To overcome this shot-noise limit and achieve Heisenberg-limited scaling, one must go beyond simple thermalization. This requires engineering structured interactions between the probe and the thermal bath, such that the probe does not fully thermalize. In such scenarios, enhanced precision beyond the SQL becomes achievable. On the other hand, optimal thermometry protocols have been studied in various settings, including probe optimization~\cite{Mukherjee19,Glatthard22b}, criticality-enhanced thermometry~\cite{Burgarth15,Plozien18,Aybar22,Montenegro22,Mihailescu24,Yang24b,Srivastava25}, and global thermometric frameworks~\cite{Mok21,Rubio21,Zhou24b,Chang24}. On the other hand, various quantum resources, including quantum coherence~\cite{Stace10,Jevtic15,Frazao24}, strong coupling~\cite{Mihailescu23,Brenes23,Rodriguez24}, quantum correlations~\cite{Gebbia20,Planella22}, squeezing~\cite{Mirkhalaf24}, discord~\cite{Sone18},
and periodic driving~\cite{Mukherjee19,Glatthard22}, have been utilized to surpass the standard quantum limit in several quantum thermometry protocols.
Even, non-Markovian dynamics have also been shown to play a crucial role in achieving quantum advantages, primarily due to environmental feedback mechanisms, and have been exploited to enhance thermometric precision in several contexts~\cite{Riberi2022Oct,Zhang22,Xu23,Rodriguez24,Aiache25}. Additionally, auxiliary-assisted thermometry has been explored for its potential to improve measurement accuracy~\cite{Kiilerich18}. Moreover,  
a collisional thermometry
framework is introduced~\cite{Seah19,Mendonca25,ghosh2025_q}, where a quantum probe is employed to infer the temperature of a thermal bath. The probe interacts sequentially with a series of $N$ auxiliary systems that correlate over time. The information about
the temperature of the bath is extracted from the state
of the auxiliary systems whose QFI is shown to scale superlinearly with $N$. Quantum many-body systems have been proposed as effective platforms for temperature estimation in non-equilibrium scenarios, highlighting their potential in advanced thermometric techniques~\cite{marzolino_pra_2014}. Despite these advancements, single-qubit thermometry remains highly practical and experimentally feasible, as demonstrated in~\cite{Jevtic15}. Quantum phases have also been found to enhance thermometric performance. For example, antiferromagnetic structured systems yield significantly higher thermometric accuracy compared to non-structured systems~\cite{guo_pra_2015}. {Moreover, gapped systems tend to be inefficient for local thermometry, whereas gapless systems are more favorable~\cite{Hovhannisyan18,Potts19}.} Interestingly, the Kondo effect has also been identified as beneficial for thermometric applications~\cite{Mihailescu23}. Whereas, the concept of interferometric power has been explored in the context of quantum thermometry, further demonstrating its applicability across a wide range of quantum systems~\cite{Yang2024May}. See Refs.~\cite{Mehboudi15,Paris16,Correa17,Campbell17,DePasquale18,Mehboudi19b,Razavian19a,Mirkhalaf21,Jorgensen20,Mehboudi22,Cenni22,Xie22,Boeyens23,Hossein23,Alves23,Yuan23,Tan24,Alves24,Brattegard24,Xie24,Zhang24b,Chang24,Abiuso24,Sharafiev25,Grattan25,Ullah25,Ullah25b,Ullah25c,Ullah25d,Mehboudi25,Pedram25,Zhang25,Hosseiny25} for further developments in this direction. Even, critical quantum thermometry offers enhanced precision due to quantum
phase transitions~\cite{Salado-Mejia21,yu2024}. 
In more detail, the effect of criticality in many-body systems on quantum thermometry is discussed later in the part, criticality-based sensing, of Sec.~\ref{sec:applications}. In addition, several experimental proposals, including quantum optical systems~\cite{Tham16,Mancino17}, Bose–Einstein condensates (BECs)~\cite{Sabin14}, cold Fermi gases~\cite{Mitchison20}, superconducting quantum circuits~\cite{Sultanov21,Lvov25}, trapped ions~\cite{deSaNeto22,Li24},  nitroge-nvacancy centers in diamond~\cite{Toyli13,Kucsko13}, 
quantum dots~\cite{Haupt14},
cavity optomechanical systems~\cite{Purdy17},
optical nanofibres~\cite{Grover15}, 
and so on~\cite{Yang11,Johnson16,Wu22} for quantum thermometry have been reported in the literature.

\subsection{Indefinite causal order in quantum metrology}
\label{subsec:ico}

The standard formulation of quantum mechanics assumes a definite causal order between successive physical processes. However, recent developments in quantum gravity suggest that causal order may be indefinite~\cite{Hardy2007}. This insight has been incorporated into quantum information theory through the formulation of higher-order processes that can not be represented in usual quantum circuit but that are still governed by quantum mechanics. A particularly important manifestation of this is
\emph{indefinite causal order} (ICO) provides a framework in which quantum channels are applied in a coherent superposition of different causal orders, rather than in a fixed sequential order~\cite{Oreshkov12,Chiribella13,Brukner14,Araujo15,Oreshkov16}. The paradigmatic example is the quantum SWITCH~\cite{Chiribella13}, which coherently controls the order of two channels $\mathcal{E}_1$ and $\mathcal{E}_2$ via an auxiliary control qubit. For an input state $\rho$ and control state  $|+\rangle_c=(|0\rangle+|1\rangle)/\sqrt{2}$, the SWITCH implements the transformation
\begin{eqnarray}
    \mathcal{S}(\rho \otimes |+\rangle\langle+|)
=
\frac{1}{2}
\sum_{i,j}
\bigg (
|0\rangle\langle0|_c \otimes K_i^{(1)}K_j^{(2)} \rho K_j^{(2)\dagger}K_i^{(1)\dagger}
\nonumber\\+
|1\rangle\langle1|_c \otimes K_j^{(2)}K_i^{(1)} \rho K_i^{(1)\dagger}K_j^{(2)\dagger}
\bigg)
+ \text{coh.}, \quad 
\end{eqnarray}
where $\{K_i^{(1)}\}$ and $\{K_j^{(2)}\}$ are Kraus operators of
$\mathcal{E}_1$ and $\mathcal{E}_2$, respectively, and the ``coh.'' terms arise from off-diagonal terms in the control basis. These interference terms vanish in any classical mixture of causal orders but survive only in the coherent superposition, leading to operational advantages. 
Remarkably, 
this mechanism can activate communication through otherwise useless channels. Specifically,
there exist situations where both definite orders $\mathcal{E}_1 \circ \mathcal{E}_2$ and $\mathcal{E}_2 \circ \mathcal{E}_1$ are individually entanglement-breaking (and hence have zero quantum capacity), yet their indefinite-order combination enables nonvanishing information transmission \cite{Ebler2018}. This enhancement originates from the noncommutativity of the Kraus operators, $[K_i^{(1)},K_j^{(2)}]\neq 0$, which generates observable interference in the control system. Experimentally, the quantum SWITCH has been realized using photonic systems~\cite{Procopio2015, Rubino2017, Goswami2018, Rubino2021, Rubino2022, Stromberg2024}, where the order of two polarization-dependent operations is coherently controlled by a path qubit. These demonstrations confirm that coherent control of causal order can mitigate certain noise effects without redundancy (as in quantum error correction) or fast control pulses (as in dynamical decoupling), highlighting ICO as a distinct quantum resource. ICO protocols have been shown to improve channel discrimination tasks ~\cite{Chiribella13,Ebler18,Chiribella2021}, quantum devices~\cite{felce2020,zhu2023}.
\begin{figure}
    \includegraphics[width=\linewidth]{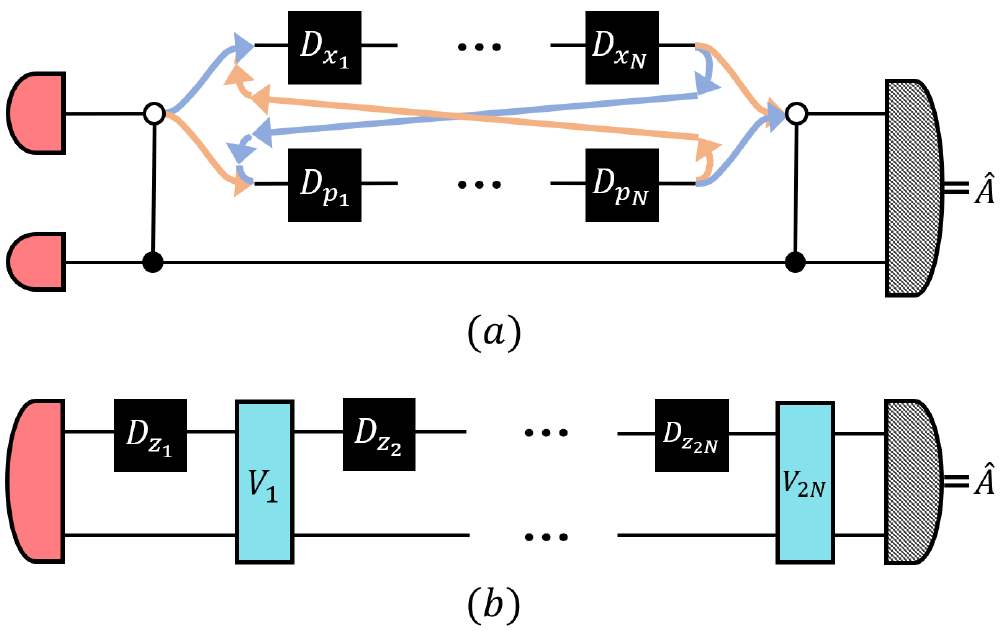}
    \caption{A schematic illustration of the key distinction between definite and indefinite causal order~\cite{Zhao20}. In the indefinite causal-order protocol considered in this work, the total position displacement operations \(D_{x_1}D_{x_2}\cdots D_{x_N}\) and momentum displacement operations \(D_{p_1}D_{p_2}\cdots D_{p_N}\) are applied in a coherent superposition of two alternative orders, with the ordering controlled by an auxiliary control qubit. Such a superposition of causal structures leads to a super-Heisenberg scaling of the quantum Fisher information, yielding a precision scaling of order \(1/N^2\)(upper panel). In contrast, for the conventional definite-order scheme, where the sequence of operations remains fixed, the achievable precision is bounded by the standard Heisenberg scaling \(1/N\) (lower panel).}
    \label{fig:ico_metrology}
\end{figure}

Recently, ICO has been developed as a promising resource for quantum metrology~\cite{Mukhopadhyay18,Frey19,Zhao20,Chapeau-Blondeau21,Francois21,Xie21,Chapeau-Blondeau22,Delgado22,Francois22,Francois23,Ban23,Goldeberg23,Liu23,Procopio23,Gu23,An24,Yuan25a,Chen25,Yuan25b,Guo25,Xu26}. Early studies demonstrate that ICO can improve thermometric precision
~\cite{Mukhopadhyay18} and enhance the estimation of noise parameters of quantum channels. Subsequently, ICO-based protocols are developed for unitary-encoding phase estimation in the presence of noise~\cite{Zhao20,Chapeau-Blondeau21,Francois21,Francois22,Francois23,Xie21,Chapeau-Blondeau22,Ban23,Goldeberg23,An24,Yuan25a,Chen25,Yuan25b,Guo25,Xu26} where the unitary and noisy channels act on the probe via a quantum switch.
For example, in order to establish that quantum SWITCH can improve the precision, Zhao ~\emph{et al.}~\cite{Zhao20} has considered a continuous variable system in which one is given $2N$ unknown quantum devices. Half of these devices imprint displacement along one phase-space quadrature while the remaining half induce displacement along the canonically conjugate quadrature. Towards inferring the product of the mean shifts associated with two conjugate directions, it is shown that the attainable estimation precision is limited by Heisenberg limit when the protocol obeys a definite sequential protocol while 
it is possible to overcome this limitation
when the order of applying the devices is coherently controlled through indefinite causal order (see Fig.~\ref{fig:ico_metrology})\footnote{Note that unlike Ref.~\cite{Zhao20}, we refer to \(N^{-1}\) as SQL and \(N^{-2}\) as Heisenberg limit in this review.}. More recently, a distinct class of ICO, based on time-flip operations, 
coherent superposition of forward and backward time evolution, is shown to achieve 
Heisenberg scaling using separable probe states~\cite{Agrawal25} and its feasibility has been further supported by an experimental simulation~\cite{Yin23,Guo25}. 
Beyond highlighting the advantages of ICO-based protocols, an important question has arisen: how do these methods compare to conventional estimation strategies, such as sequential and parallel schemes? Comparative analyses with conventional parallel, sequential and ICO-based strategies reveal that there is a hierarchy of the precision among strategies~\cite{Liu23}.
Conversely, explicit counter examples are reported where an ICO-based approach does not outperform methods employing a definite causal order, thereby challenging the universality of ICO's advantage~\cite{Mothe24,Hayashi25}. This underscores the context-dependent nature of benefits that can be obtained from ICO and emphasizes the need for continued investigation into its operational potential. Beyond ICO, related advantages in quantum metrology can also be found in protocols based on coherent superposition of quantum channels, without causal indefiniteness, thereby broadening the landscape of higher order quantum metrology~\cite{Xie21,Francois21}.

\subsection{Effect of relativistic motion and gravity in quantum sensing}
\label{subsec:relativistic-metrology}

Theoretical researches within the free field model of single-mode approximation, or within a more practical and operable theoretical framework that considers coupling of quantum systems with fixed energy gaps to quantum fields, it has been shown that quantum resources, such as  steering, entanglement, discord and coherence are disrupted by the Unruh effect, particularly when one of the detectors undergoes acceleration, and thus can influence the performance of quantum sensing devices~\cite{He18,Liu18,Wang2020b,He24,Liu24,Fan24b,Itou24}.  In gravitational-wave astronomy, parameter estimation is commonly formulated through the FIM, quantifying the precision with which source parameters can be inferred from noisy detector outputs. Vallisneri has shown that although the FIM provides a powerful framework for forecasting parameter sensitivity, its reliability requires sufficiently high signal-to-noise ratios and careful treatment of waveform degeneracies~\cite{Vallisnero2008}.  Relativistic quantum metrology further reveals that spacetime curvature itself can influence quantum-limited phase estimation. In particular, gravitational redshift modifies the accumulated phase of quantum probes evolving at different gravitational potentials, thereby affecting the QFI. The local gravitational field influence both the effective phase accumulation and the optimal measurement strategy. Recent works demonstrate that gravitational redshift can either suppress or enhance phase-estimation precision depending on the detector configuration and initial quantum correlations~\cite{Wei2025_a,Wei2025}. In particular, squeezed and entangled photonic probes propagating through curved spacetime can exhibit amplified QFI under suitable gravitational conditions, establishing gravity itself as a metrological resource.

\section{\textbf{Estimation strategies}}
\label{sec:estimation-strategies}

\begin{figure}
    \includegraphics[scale=0.6]{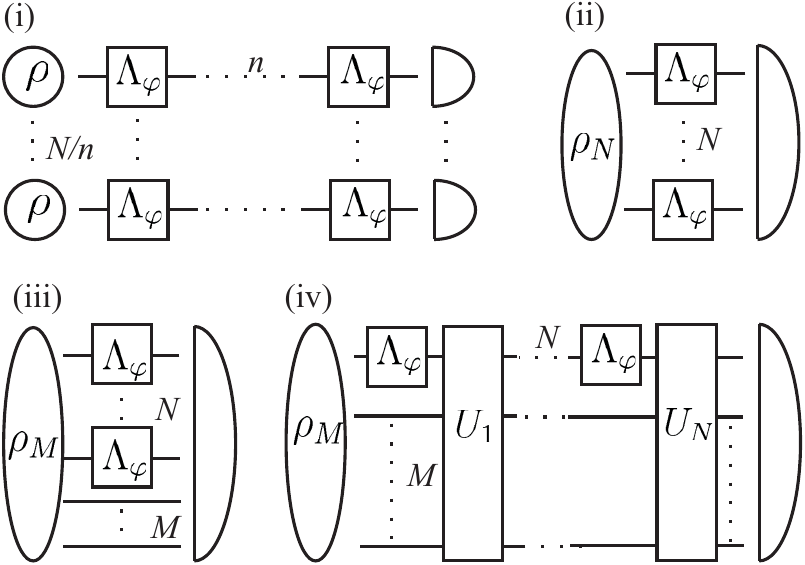}
   \caption{Schematic representation of various metrological strategies~\cite{Rafal14}. The parameter $\varphi$ to be estimated is encoded through the quantum channel $\Lambda_\varphi$. \textit{(i) Sequential scheme (SS):} an input state $\rho$ undergoes repeated local applications of $\Lambda_\varphi$ in sequence, without employing entanglement. \textit{(ii) Parallel scheme (PS):} an entangled multipartite probe $\rho_N$ is subjected to multiple local actions of $\Lambda_\varphi$ acting in parallel. \textit{(iii) Passive ancilla scheme (PAS):} the probe system is supplemented with noiseless ancillas that are entangled with the probes but do not directly undergo parameter encoding. \textit{(iv) Active ancilla-assisted scheme (AAAS):} intermediate unitary operations $U_i$ are applied between successive channel uses, enabling adaptive probe-ancilla interactions during the estimation process.}
\label{fig:rafal2014prl}
\end{figure}

In parameter estimation, one of the central tasks is to infer an unknown  parameter \(\varphi\) encoded through an operation within an input probe. 
In the frequentist approach (as discussed in Sec.~\ref{sec:parameter-estimation}),  the precision of any unbiased estimator, is bounded by the quantum Cram\'er-Rao inequality, with the QFI which quantifies the maximum extractable information about the parameter \(\varphi\). For example, when the encoding process is a unitary operation in the absence of noise, this parameter may represent a frequency, a magnetic field, time, or more generally, a parameter of the generator Hamiltonian. 
Over the past few decades, several encoding architectures, whether unitary or noisy, have been considered. Let us discuss these strategies which form a hierarchy towards general metrological protocols (see Fig.~\ref{fig:rafal2014prl} for a schematic illustration of different metrological strategies).

\begin{enumerate}
    \item \textit{Sequential scheme $(\mathrm{SS})$:} The \(N\) uses of the encoding map are applied sequentially in \(n\) rounds on \(N/n\) independent probes. This protocol is natural in interferometric settings.

    \item \textit{Parallel scheme $(\mathrm{PS})$:} A globally entangled \(N\)-partite probe state undergoes \(N\) encoding operations in parallel. This scheme allows multipartite entanglement to contribute to enhanced sensing.

    \item \textit{Passive auxiliary scheme $(\mathrm{PAS})$:} \(N\) probes are entangled with \(M\) noiseless auxiliarys, and the entangled probe state passes through the \(N\) encoding maps in parallel. This is a more general setup than the PS.

    \item \textit{Active auxiliary-assisted scheme $(\mathrm{AAAS})$:} The application of \(N\) encoding maps is interleaved with arbitrary unitaries \(U_i\) that enable interaction between the probes and auxiliaries. This is the most general setting,  also known as adaptive scheme briefly discussed in Sec.~\ref{subsec:noisy-channels}, encompassing all the above schemes as special cases.

    \item \textit{Causal superposition strategies $(\mathrm{CSS})$:} 
    Quantum metrology can be extended beyond definite causal order as discussed in Sec.~\ref{subsec:ico} in which
    channels may be probed in a superposition of different causal orders, thereby providing qualitatively new metrological resource.
\end{enumerate}

Let us denote the encoding map by \(\Lambda_\varphi\) and the parameter to be estimated by $\varphi$. The input probe, let's say \(\rho\), is a single-probe state used in the sequential scheme, while \(\rho_N\) and \(\rho_M\) denote the global input probes in the parallel and auxiliary-assisted schemes, respectively. In the active auxiliary-assisted strategy, let \(U_i\) denote arbitrary global unitaries that act jointly on all probes and auxiliarys. Let us denote the maximum QFI achievable under sequential, parallel, passive auxiliary, and active auxiliary-assisted schemes, $\mathcal{F}^\mathrm{SS}, \mathcal{F}^\mathrm{PS}, \mathcal{F}^\mathrm{PAS},$ and $ \mathcal{F}^\mathrm{AAAS}$, respectively, which 
can be defined as follows:
\begin{align}
&\mathcal{F}^\mathrm{SS} \coloneqq \max_{\rho, n} \mathcal{F}\left\{[\Lambda^n_\varphi(\rho)]^{\otimes N/n}\right\}, \label{seq-eq} \\
&\mathcal{F}^\mathrm{PS} \coloneqq \max_{\rho_N} \mathcal{F} \left[\Lambda_\varphi^{\otimes N}(\rho_N)\right], \label{par-eq} \\
&\mathcal{F}^\mathrm{PAS} \coloneqq \max_{\rho_M} F\left[(\Lambda_\varphi^{\otimes N} \otimes \mathbb{I}^{\otimes M})(\rho_M)\right],   \label{pass-anc-eq} \\
&\mathcal{F}^\mathrm{AAAS} \coloneqq \max_{\rho_M, \{U_i\}} \mathcal{F}\left[U_N \Lambda_{\varphi} \cdots U_1 \Lambda_{\varphi}(\rho_M)\right]\label{act-anc-eq} .
\end{align}

For a unitary encoding process, an universality emerges: all encoding strategies, sequential, parallel, passive auxiliary, and active auxiliary-assisted schemes, achieve the same asymptotically optimal scaling of the precision for a single-parameter, i.e., \(\delta\varphi \sim 1/N\)~\cite{Childs2000,Giovannetti06}. This implies that the inclusion of auxiliarys, whether passively or actively, does not improve the asymptotic scaling behavior for estimating a single parameter.

When the encoding process involves an \textit{uncorrelated} noisy channels~\cite{Rafal14,Zhou21}, the equivalence between different strategies no longer hold. In particular, for certain types of noise, like amplitude damping noise, the parallel encoding scheme can even yield a higher precision than the passive auxiliary scheme asymptotically. However, the active auxiliary-assisted scheme does not offer any asymptotic advantage over the passive auxiliary scheme. Furthermore, the causal superposition protocol also provides no asymptotic improvement over either the active auxiliary-assisted or the passive auxiliary schemes asymptotically since asymptotic achievable QFI are equal in all three schemes.
Specifically, when the 
``Hamiltonian-not-in-Kraus-span''\footnote{Suppose $H = \iota \sum K_i^\dagger K_i$ and $S = \text{Span}_H \{ K_i^\dagger K_j \forall i,j\}$, where $\text{Span} \{\cdot \}$ denotes all Hermitian operators that are linear combinations of operators in the argument. The ``Hamiltonian-not-in-Kraus-span'' (HNKS) condition states that $H \notin S$. Note that the HNKS condition can reduce to the ``Hamiltonian-not-in-Lindblad-span'' condition, discussed in Sec.~\ref{subsec:noisy-channels}~\cite{Zhou18}.} (HNKS) condition is violated, both passive and active auxiliary-assisted schemes are asymptotically restricted to the precision scaling of \(O(\sqrt{N})\), corresponding to the standard quantum limit. In contrast, When the HNKS
condition holds true,  the detrimental effect of noise can be fully mitigated, in terms of scaling and both passive and the active auxiliary-assisted schemes 
recover the Heisenberg scaling of \(O(N^2)\) which is also achievable for both schemes.

\subsection{Distributed quantum sensing}
\label{subsec:distributed-sensing}
The formalism of multiparameter quantum estimation naturally extends to scenarios involving spatially distributed networks of quantum sensors, giving rise to the framework of distributed quantum sensing~\cite{Komar14,Ge18,Eldredge18,Zhuang18,Proctor18,Volkoff18,Gatto19,Qian19,Xia19,Guo2019,Rubio20b,Sekatski20,oh2020,Wolk20,Gramegna21,Qian21,Zhang21,Bringewatt21,Morelli22,Sun22,Kasai22,Shettell22b,Hamann22,Oh22,Malia22,Kwon22,Shettell22,Moore23,Pelayo23,Ehrenberg23,Pelayo23,Hollendonner23,Kasai24,Wang24,Zhang24,malitessta2023,Ge24,Xu24,Ho24,Nehra24,Zang24,Ueda25,Agarwal25}. In this setting, the goal is to enable multiple sensors to collectively exploit shared entangled states in order to enhance the precision of estimating global properties of the system. Distributed quantum sensing has captured significant attention in recent years and has been developed extensively both theoretically and experimentally. It has also been proposed for novel applications, such as privacy-preserving quantum tasks and probing quantum theory in curved spacetime.

Numerous studies have shown that the Heisenber scaling (HS) can be achieved by using quantum resources,
particularly entanglement, in distributed quantum sensing~\cite{Guo2019,Liu21,Zhao21,Kim24,Hong25}. For example, multi-mode squeezed entangled states can achieve the HS in continuous variable systems. Additionally, mode-entangled spin-squeezed atomic states and mode- and particle-entangled (MePe) states enable the HS in discrete-variable systems.

\begin{figure}
\centering
    \includegraphics[scale=0.22]{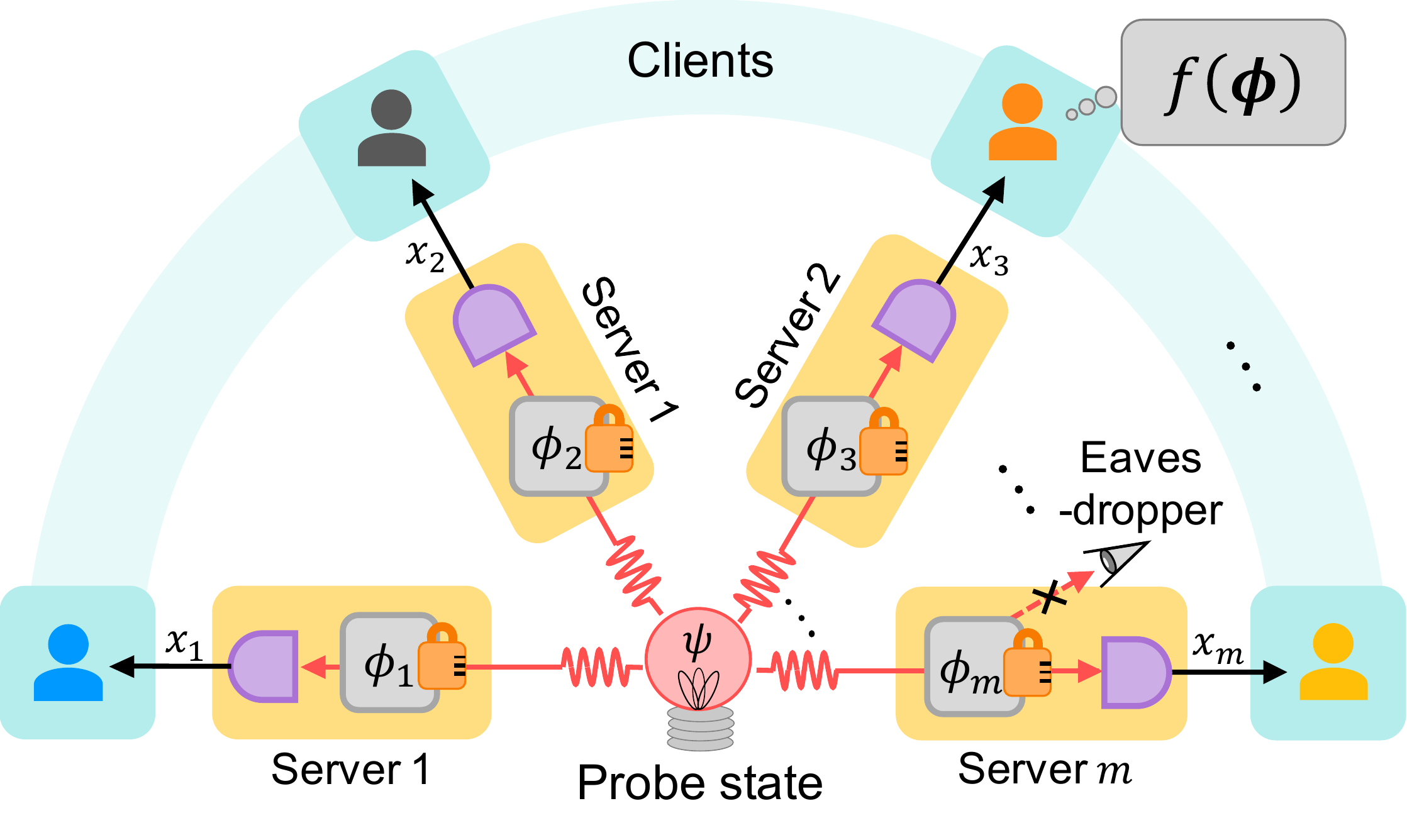}
   \caption{Illustration of a private distributed quantum sensing network~\cite{Namkung26}.
   A shared quantum probe state is encoded with spatially distributed parameters $\phi$ in such a way that information about the individual parameters remains inaccessible to untrusted servers.  Performing local measurements on
the encoded state, the servers communicate the outcomes $\{x_1, \ldots, x_m\}$ to the clients. The aim of the clients is to infer a global linear function $f(\phi) = \omega^T \phi$.}
\label{fig:distributed-sensing}
\end{figure}

On the other hand, privacy and security in distributed quantum sensing have recently attracted considerable attention. In such protocols, multiple spatially separated sensors cooperate through shared quantum resources to estimate global parameters, while simultaneously facing challenges associated with information leakage, malicious attacks, and protection of locally encoded sensing data. It has been shown that distributed sensing can be performed while hiding the identity of the sensing node itself, thereby incorporating anonymity into quantum metrology protocols~\cite{Kasai22}. Cryptographic techniques have also been integrated into quantum metrology in order to guarantee secure parameter estimation, protect sensing information from leakage, and provide robustness against malicious or dishonest parties~\cite{Takeuchi19,Yin2020,Okane21,Shettell22,Moore23,Ho24,Bizzarri25,Hassani25,Jong25,Junior25,Alushi26,Namkung26,Kianvash26}. For instance, operational notions of privacy in distributed quantum sensing protocols involving untrusted servers have recently been investigated and it has been shown experimentally that 
privacy-preserving distributed sensing can simultaneously achieve Heisenberg-limited precision while using fewer photons than the number of estimated parameters~\cite{Namkung26}. The corresponding schematic diagram is shown in Fig.~\ref{fig:distributed-sensing}. Furthermore, privacy-preserving network parameter estimation schemes have also been developed, where distributed sensors collaboratively estimate global quantities while preventing unauthorized access to locally encoded parameters~\cite{Shettell22b}. Similarly, multipartite entangled states are found that simultaneously provide robustness against noise and privacy protection of locally encoded sensing data. In particular, these states allow the extraction of global information relevant for parameter estimation while preventing unauthorized parties or intermediate nodes from accessing local parameters encoded at individual sensors~\cite{Bugalho25}. 

\subsection{Protocols for optimal state preparation and noise mitigation}
\label{subsec:state-preparation}

One of the major challenges in quantum estimation protocols is the preparation of probe states that can exploit quantum advantages. Since the attainable precession depends on the encoding of the unknown parameter on the probe state, the choice and preparation of the initial state which remain experimentally accessible is crucial for metrological protocol. In particular, achieving precession beyond SQL generally requires generation of highly nonclassical multiparty states. Among these states, Greenberger-Horne-Zeilinger state serves as an important resources as local phase encoding on a GHZ state enables Heisenberg scalings~\cite{Pezze18} which is written as
\begin{equation}
    |GHZ\rangle=\frac{1}{\sqrt{2}}(\ket{0}^{\otimes N}+\ket{1}^{\otimes N}),
\end{equation}
where \(\ket{0}(\ket{1})\) is the eigenstate of the \(z\)-component of Pauli matrix and \(N\) is the number of parties. Moreover, the corresponding symmetric logarithmic derivative basis used for extracting information is well understood and can be implemented using current state-of-the-art technology. In continuous-variable  systems, analogous quantum enhancement can be achieved using \(\mathrm{N00N}\) states. Therefore, probe state preparation remains a crucial aspect of designing efficient and experimentally viable quantum estimation protocols, which we review here.

\subsubsection{Greenberger-Horne-Zeilinger (GHZ)-like state-preparation}
\label{subsubsec:GHZ-like-state-preparation}

In order to achieve quantum enhanced parameter estimation, a wide variety of state engineering protocol has been developed. Examples include gate-based quantum criticalities~\cite{Mihailescu25b}, adiabatic protocols~\cite{Venuti07,Mihailescu24} and measurement based schemes~\cite{Albarelli2017}. More recently, with machine learning (ML) techniques, particularly reinforcement learning, are employed for optimal preparation of these useful states~\cite{bukov_prx_2018}, which can be incorporated in quantum sensing devices. Among the various approaches, one of the most promising is the preparation of GHZ-like states, which exhibit quantum advantages. This specific type of state preparation requires entangling operations between qubits~\cite{chao_science_2019,choi_prl_2014}. For example, a GHZ-like state can be generated using the collective-spin one-axis twisting (OAT) model, \(H_{OAT} = \chi J_z^2\), where \(J_z = \frac{1}{2} \sum_j \sigma_j^z\) and \(\chi\) denotes the interaction strength~\cite{ueda_pra_1993}. This model dynamically generates spin squeezing and specific evolution time \(\chi t=\pi/2\), produces GHZ states. 

A sufficient condition to detect multipartite entanglement based on spin squeezing is proposed which can be used in experiments~\cite{sorenson_prl_2001,Pezze09}. The OAT-based protocols can be disturbed by decoherence due to its long evolution time. To overcome it, either three-body collective \(XYZ\) interactions or Floquet driving of an OAT Hamiltonian can be used to generate GHZ-like state \footnote{ The state is not a perfect GHZ state although the probability distribution on the eigenstate of the dynamical state is symmetric for \(\pm m\) with \(J_z\ket{m}=m\ket{m}\).}  in a very short time which can also provide HL in metrology~\cite{zhang2024,Zeng2025Jun}

More recently, an optimal protocol based on gradient ascent pulse engineering (GRAPE) for dynamical creation of a two-dimensional Rydberg lattices~\cite{Porotti2023}. This approach referred to as adiabatic echo protocol, illustrates that GRAPE-type optimal control methods can be a robust route to generate multipartite quantum states in experimentally relevant systems. 

\emph{Domino dynamics in encoding of state.} Another experimentally attractive route for generating metrologically appropriate multipartite entangled state is provided by the domino dynamics~\cite{yoshinaga_pra_2021}. The protocol considers on-dimensional chain of \(N\)-qubits subject to external magnetic field of strength \(B\) and coupled through a nearest-neighbor (NN) Ising interactions described by the Hamiltonian, given as
\begin{equation}
H = B \sum_{j=1}^{N} \sigma_z^{(j)} + J \sum_{j=1}^{N-1} \sigma_z^{(j)} \sigma_z^{(j+1)},
\end{equation}
where $\sigma_z^{(j)}$ is the Pauli-$z$ operator for the $j$-th qubit. Starting from \(\ket{+}^{\otimes N} = \bigotimes_{j=1}^{N} \frac{1}{\sqrt{2}} (\ket{0}_j + \ket{1}_j)\), the system evolves for an optimized duration \(t_\text{opt}\). During the dynamics, the Ising interaction introduces a conditional phase correlations between neighboring qubits. Subsequently, the entanglement spreads from nearest neighbors to longer ranges in a wave-like manner, similar to how a toppling domino causes others to fall (which is called \emph{Domino effect}). After $t = t_\text{opt}$, the evolved state $\ket{\psi(t_\text{opt})}$ exhibits global entanglement, can produce GHZ-like state with a collective phase \(\phi\) encoding the parameter to be estimated. To extract the phase information a global unitary transformation is applied to the system followed by a subsequent projective measurement in the computational basis. From the resulting measurement statics, the encoded information can be extracted with a precession, approaching Heisenberg limit, \(\delta B\sim 1/Nt\). Instead of NN interactions, one can extend this protocol through long-range Ising interaction in which the interaction strength follows a power law decay~\cite{monika2025}. In this case, the long-range (LR) interaction-based quantum sensors can provide higher precession in estimating magnetic field compared to its NN counterparts. A key advantage of this protocol is that it requires neither dynamically switched interactions nor complicated control sequences, naturally rely on Ising interactions available in currently available technologies, and it can generate entangled probe, which is robust against decoherence and imperfections.

The advantages of domino dynamics is as follows: (i) Requires no dynamically switched interactions — only passive evolution under a fixed Hamiltonian. (ii) Naturally exploits always-on Ising interactions available in many physical platforms. (iii) Enables scalable entanglement generation in linear time, favorable for metrological tasks. Also the protocol remains effective even in the presence of certain types of decoherence (e.g., dephasing) and adaptable to platforms where dynamic control of $J$ is not feasible.

\subsubsection{\(\mathrm{N00N}\)-like state preparation}
\label{subsubsec:NOON-like-state-preparation}
\(\mathrm{N00N}\) states are extremely valuable resource in quantum metrology due to their ability to achieve the Heisenberg limit in the phase estimation. A \(\mathrm{N00N}\) state is a quantum superposition of \(N\) photons distributed between two paths, described as
\begin{equation}
    \ket{\mathrm{N00N}}=\frac{1}{\sqrt{2}}(\ket{\mathrm{N0}}+\ket{\mathrm{0N}}),
\end{equation}
where \(\mathrm{N}\) represents the number of photons occupying either of the two interferometer paths.  This highly nonclassical state exhibits maximal entanglement across the two paths, which can measure a phase shift with much higher precision than classical or coherent light. In this regard, a scheme is proposed in Ref.~\cite{kok2002} which employs only linear optical elements, such as beam splitters and phase shifters, together with single-photon sources and photodetectors to produce these highly entangled states conditionally. This approach is particularly significant because it avoids the need for nonlinear optical interactions and the nondetection conditions which are challenging to implement experimentally. Several other proposals in photonic setup to produce \(\mathrm{N00N}\) states include applications of intensity-symmetric multiport beam splitters and conditional measurement~\cite{pryde2003}, linear optics with feed-forward~\cite{Dowling08} and Young's double slit interference experiment~\cite{cao2023}. \(\mathrm{N00N}\) states upto \(\mathrm{N}=5\) can also be produced experimentally by spontaneous parametric down conversion~\cite{Mitchell2004} and by mixing quantum and classical light~\cite{Afek2010}. (for other physical platform, see Refs.~\cite{Merkel2010,cable2011,Kamide2017,Walther2004,nielsen2007,chen2010,su2017,zhang2018,qi2020,cao2023}) 

\subsubsection{Nonequilibrium state preparation}
\label{subsubsec:Nonequilibrium state}
Beyond static entangled probes such as GHZ and N00N states discussed above, nonequilibrium dynamics provides an important route for preparing and stabilizing metrologically useful states
that may not be accessible within the equilibrium physics. In the following we briefly discuss the strategies for preparing such nonequilibrium states. In Sec.~VII.E.2, we shall discuss these systems from a different perspective, showing time-dependent many-body evolution itself becomes the metrological resource.

\emph{Floquet-engineered probes.} The central motivation behind Floquet driving~\cite{Eckardt17} is to dynamically engineer and control quantum systems using periodic time-dependent fields. This enables access to physical behaviors and phases of matter that are inaccessible in static systems. By periodically modulating the system, we can create effective (Floquet) Hamiltonians that differ significantly from the original static Hamiltonian. This allows simulation of complex or exotic systems, such as topological insulators (even when the undriven system is trivial), synthetic gauge fields to name a few~\cite{messer2018}. In quantum sensing, Floquet driving is used to enhance sensitivity, resolution, and selectivity by engineering time-dependent control protocols. The key motivation is to exploit periodic modulation to amplify or isolate specific quantum signals and suppress noise, allowing for more precise measurements of weak external fields or interactions.

Floquet engineering has been experimentally realized across several quantum platforms, establishing periodic driving as an useful technique for  nonequilibrium state preparation. In ultracold atomic gases, lattice shaking and periodic modulation have been used to generate artificial gauge fields, density-dependent tunneling, and topology, e.g., Floquet realization of the Haldane model with ultracold fermions and anomalous Floquet topological systems in optical lattices~\cite{Jotzu14}. Other important works in different physical systems include experimental demonstrations of a Floquet topological insulator in photonic systems~\cite{Rechtsman13}, realization of Floquet dynamics of interacting spin chains in trapped-ion systems~\cite{Zhang17}, demonstration of Floquet time-crystalline order in driven disordered dipolar many-body systems at room temperature via NV centers in diamond~\cite{Choi17}, and more recently, simulation of  programmable digital Floquet engineering in superconducting-qubit processors~\cite{Zhang22b}. Floquet mechanisms are not restricted to a single physical architecture, but provide a broadly applicable strategy for accessing nonequilibrium states and dynamical phases that can further be harnessed for performing various metrological tasks.

\emph{Time-crystalline probes.} Time crystals have recently been developed as a promising platform for quantum sensing due to their robust nonequilibrium order and long-lived temporal correlations~\cite{Sacha2017}. In periodically driven many-body systems, discrete time crystals (DTCs) spontaneously break the discrete time-translation symmetry of the Floquet drive, producing a subharmonic response characterized by oscillations with period $nT$ instead of the driving period $T$~\cite{Yao17,Else2020}. Such rigidity of the oscillations against perturbations makes time crystals attractive candidates for precision metrology. A typical Floquet spin system is described by the unitary evolution operator
\begin{equation}
U(T)=e^{-iH_2T_2}e^{-iH_1T_1},
\end{equation}
where $H_1$ generates a global spin rotation and $H_2$ accounts for interacting spin dynamics. In the time-crystalline phase the system exhibits stable subharmonic oscillations in observables, such as the magnetization, which provide a robust dynamical reference signal. When an external parameter $\lambda$ (for example a weak magnetic or electric field) perturbs the system, the oscillation phase and amplitude become sensitive to $\lambda$. 

Experimentally, discrete time crystals were first observed in 2017 in two independent platforms: trapped-ion spin chains~\cite{Zhang17} and disordered ensembles of nitrogen-vacancy centers in diamond~\cite{Choi17, Beatrez2023}. These experiments demonstrated persistent subharmonic oscillations under periodic driving, confirming the realization of time-crystalline order in interacting quantum systems. Subsequent experimental progress has extended time-crystal physics to a wide variety of platforms including magnon condensates, ultracold atoms in optical cavities, and programmable quantum processors \cite{Smits2018,Kessler2022,Mi2022}. In particular, experiments on superconducting and trapped-ion quantum processors have demonstrated controllable Floquet dynamics and long-lived time-crystalline oscillations in multi-qubit systems, paving the way for potential sensing applications based on nonequilibrium many-body phases. Together, these developments establish time crystals as a powerful resource for quantum metrology, combining dynamical stability, long coherence times, and enhanced sensitivity in driven quantum systems. Using such properties an experiment of quantum sensing on DTC has been performed recently by Moon \emph{et al. }~\cite{Moon2024}. 

Nonequilibrium probe preparation is not limited to unitary periodic driving. In open quantum systems, engineered dissipation provides a complementary route in which the environment itself is shaped to stabilize metrologically useful steady states, entangled states, squeezed states, or long-lived dynamical order. This motivates the reservoir-engineering strategies discussed in Sec.~\ref{subsec:reservoir-engineering}.

\subsection{Dynamical decoupling}
\label{subsec:dynamical-decoupling}
Dynamical decoupling (DD)~\cite{Viola99a,Viola99b} is a quantum control technique used to mitigate decoherence in quantum systems by applying sequences of rapid, precisely timed control pulses. These pulses are designed to average out the effects of environmental noise that would otherwise disrupt the quantum state's coherence. By periodically flipping the state of the system, DD effectively cancels out the unwanted interactions between the system and its surroundings, allowing quantum information to be preserved for longer periods. Therefore, DD becomes a crucial tool for enhancing the stability and reliability of quantum technologies, used in quantum computing and quantum sensing. For instance, DD sequence in quantum sensing to mitigate the environmental effect. DD sequences have been applied in several magnetometry setup based on Nitrogen-vacency (NV) center  to increase sensing time~\cite{Taylor08,delange11,Goldstein2011}.

In this respect, DD techniques are employed to enhance quantum parameter estimation in the presence of environmental noise~\cite{Tan13}. In particular, for an $N$-qubit system interacting with a bosonic reservoir, an exact analytical expression for QFI is derived by using a transfer matrix formalism and time-dependent Kraus operators. Their results show that applying a sequence of instantaneous $\pi$-pulses, characteristic of DD protocols, can significantly suppress decoherence and can recover Heisenberg limit ($\propto 1/N^2$) in precision. This study provides a theoretical foundation connecting quantum control and quantum metrology, and offers guidance for experimental implementations in noisy quantum systems. Since then, DD-based sensing schemes have been explored in diverse areas of quantum sensing, including Floquet spectroscopy~\cite{Lang15}, laser frequency noise~\cite{Zhang21},  a continuous phased dynamical decoupling (CPDD)~\cite{Louzon25}. For more proposals of optimized metrology see Refs.~\cite{Tan14,Sekatski16,Feng24}. Further, noise aware optimized DD method proposed recently can also be useful to improve precession, especially when noise is non-Markovian~\cite{White2025}. In a different context, recent studies also show that pretrained multistage Hamiltonian learning model, can be applied to estimate the parameters of larger Hamiltonian through incorporation of DD techniques~\cite{kang2025}.

\subsection{Quantum error correction-aided methods}
\label{subsec:QEC}
Quantum mechanics sets ultimate limits in precision through the Heisenberg limit for which the estimation uncertainty scales
as $N^{-2}$, where $N$ denotes the total number of
probes used in an experiment. In the absence of noise, such scaling can be attained,
for instance, using the Greenberger-Horne-Zeilinger state in atomic systems or the \(\mathrm{N00N}\) state in photonic settings. In a realistic scenario,
environmental interactions with systems substantially reduces this quantum enhancement, restricting precision to $N^{-1}$ (SQL)
or ‘shot noise’ in quantum optics. Over the past few
decades, significant progress has been made in identifying strategies to retain quantum enhancement under noise, employing a variety of theoretical and
experimental tools, including dynamical decoupling.

QEC has established as a crucial tool for achieving large-scale fault-tolerant quantum computation, as it provides the clearest route for successfully running deep algorithms with a quantum advantage. A wide variety of QEC schemes have been proposed, ranging from repetition codes to concatenated codes, topological codes, and bosonic codes, to name a few. 
QEC codes store the logical information in an entangled state of many physical qubits. Error detection and correction are enabled through syndrome measurements, multi-qubit measurements that
extract information about errors without disturbing the encoded quantum information. Depending on the specific code, syndrome measurements can generally reveal whether an error has occurred, as well as its location and type. This capability has sparked the adaptation of
QEC to several other domains of quantum technologies, such as quantum communication and quantum sensing. In the context of quantum sensing, QEC has been proposed for removing unwanted environmental noise, thereby protecting the probe 
~\cite{Ozeri13,Dur14,Kessler14,Arrad14,Unden16,Reiter17,Zhou18,Martinez-Garcia19,Tan19,Layden18,Layden19,Zhou20,Zhuang20,Gorecki20,Shettell21,Kubica21,Zhou22,Rojkov22,Omanakuttan2024,Lin24,Zhou24,Kwon25a,Kwon25b,Kurzyna25,Mann25,Arieli26}.

In the task of estimating a signal Hamiltonian under general Markovian noise, the HL can be achieved if and only if the signal Hamiltonian lies outside the noise span, a condition known as the Hamiltonian-not-in-Lindblad-span criterion~\cite{rafal17,Zhou18}, or the Hamiltonian-not-in-Kraus-span  criterion~\cite{Zhou21}, discussed in Sec.~\ref{sec:estimation-strategies}.
In both the situations, there can be QEC code that can achieve the optimal estimation precision~\cite{Rafal14,Zhou20,Wan22}. 

To illustrate the basic mechanism, let us consider the estimation of phase parameter $\omega$, in the presence of perpendicular noise (noise signal is perpendicular to Hamiltonian)~\cite{Zhou18}. The probe and the auxiliary are initially prepared in the logical code state as
\begin{equation}
|\psi(0)\rangle
= \frac{1}{\sqrt{2}} \left(|+\rangle \otimes |0\rangle + |-\rangle \otimes |1\rangle \right).
\end{equation}
Neglecting the noise, the evolved state takes the form as
\begin{equation}
|\psi(t)\rangle = \frac{1}{\sqrt{2}}
\left(e^{-i \omega t/2} |+\rangle \otimes |0\rangle
+e^{+i \omega t/2}|-\rangle \otimes |1\rangle\right).
\end{equation}
If a quantum jump occurs at time $t$, the state is transformed to
\begin{equation}
|\psi'(t)\rangle
= \frac{1}{\sqrt{2}}
\left( e^{-i \omega t/2} |-\rangle \otimes |0\rangle
+ e^{+i \omega t/2} |+\rangle \otimes |1\rangle \right).
\end{equation}

These jumps can be detected by performing a two-outcome measurement that projects either onto the code space $\mathrm{span}\{\,|+\rangle \otimes |0\rangle,\; |-\rangle \otimes |1\rangle \}$
or onto the orthogonal (error) subspace $\mathrm{span}\{\,|-\rangle \otimes |0\rangle,\; |+\rangle \otimes |1\rangle \}$ and hence errors can be corrected through an appropriate recovery operations.
As long as error detection and correction happen sufficiently fast as compared to to the noise timescale, the logical qubit evolution can behave like ideal situation, thereby attaining HL.

Although QEC codes provide optimal strategies in principle, they typically rely on the availability of noiseless auxiliary systems having the same dimension as the probe~\cite{Zhou18} and on the arbitrarily fast and accurate execution of the protocol.  These requirements
present major obstacles to the practical implementation of the QEC-based schemes. 
For example, in some physically relevant scenarios, in estimating Pauli-Z signal under bit-flip noise, where  all qubit operations are influenced by noise,
quantum repetition codes are used to restore HL~\cite{Kessler14,Arrad14,Dur14,Unden16} 
(see also refs~\cite{Layden19,Peng20,Zhou24} for more general auxiliary-free EC codes).
Furthermore, most EC-based
quantum metrological protocols assume an idealized setting in which noise acts only during the signal accumulation stage, while all quantum error correction (QEC) operations, including state preparation, syndrome extraction, recovery operations, and final measurements, are treated as noiseless. This assumption significantly simplifies the analysis but is unrealistic in practical implementations, where every physical operation is itself subject to errors. To address this limitation, 
fault-tolerant quantum metrology is proposed, where all qubit operations are subject to noise~\cite{Sahu2026}.

\subsection{Noise in measurements}
\label{subsec:Noisy-measurement}
Besides noise affecting state preparation and the parameter encoding process in quantum metrology, another crucial source that can degrade the precision arises from the imperfection in the measurement. Any imperfection affecting measurement apparatus which are typically used to extract information about the parameters can directly reduce the amount of accessible information. Let us discuss some recent works which addresses this issue in a general setup.

To assess the robustness of estimation precision under imperfect measurements, a quantity based on Fisher information, called 
the \textit{Fisher information measurement noise susceptibility (FI MeNoS)} is introduced for single parameter estimation~\cite{Kurdzialek23a}. This quantity characterizes the maximum relative degradation of FI due to small, general measurement disturbances.
Suppose an ideal POVM $\bm{M}$ is perturbed by a noise as $\bm{\tilde{M}} = (1 - \epsilon)\bm{M} + \epsilon \bm{N}$, where $\bm{N}$ is an undesired measurement and $\epsilon \ll 1$ is the disturbance probability. The relative loss in FI is quantified as
\begin{equation}
\label{eqn:noise-in-measurement-1}
\chi[\bm{M}, \bm{N}] = \lim_{\epsilon \rightarrow 0} \frac{\mathscr{F} [\bm{M}] - \mathscr{F} [(1-\epsilon)\bm{M} + \epsilon \bm{N}]}{\epsilon \cdot \mathscr{F}[\bm{M}]},
\end{equation}
which captures the relative \emph{decrease} of FI due to infinitesimal noise $\bm{N}$. Accordingly, the effective FI in the Cram\'er-Rao bound becomes $\mathscr{F}[\bm{M}] \left(1 - \epsilon \chi[\bm{M}, \bm{N}]\right).$

To capture the worst-case effect of noise, a quantity, \textbf{FI MeNoS}, is defined as
\begin{equation}
\label{eqn:noise-in-measurement-2}
\chi[\bm{M}] = \max_{\bm{N} \in \mathcal{M}} \chi[\bm{M}, \bm{N}],
\end{equation}
where maximization is performed over POVM $\bm{N}$, thereby eliminating its dependence on $\bm{N}$.
It quantifies how \emph{susceptible} the measurement $\bm{M}$ is to arbitrary small disturbances. A larger value of $\chi[\bm{M}]$ implies a greater precision loss.

\begin{figure*}
    \includegraphics[scale=0.3]{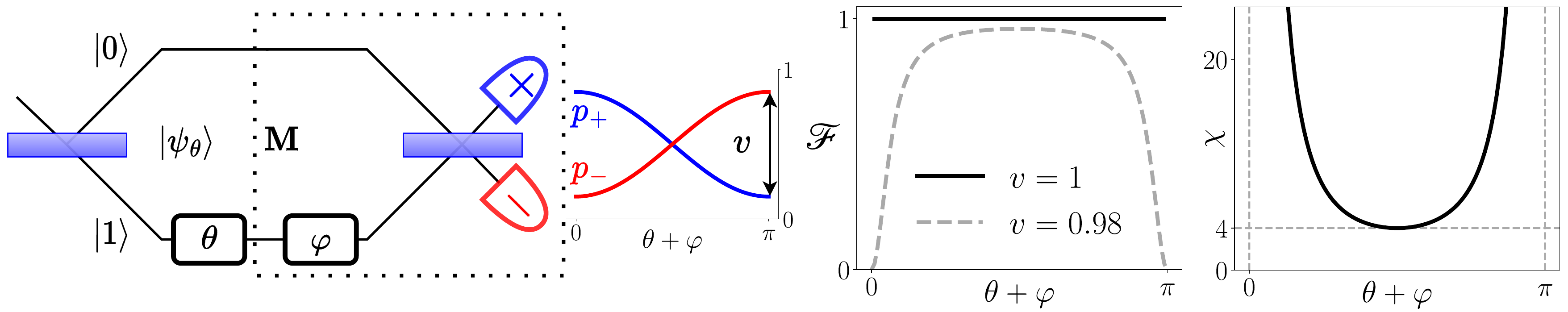}
   \caption{Measurement noise susceptibility in quantum estimation~\cite{Kurdzialek23a}. An ideal measurement $\bm{M}$ is perturbed by measurement noise, modeled as $\bm{\tilde{M}}=(1-\epsilon)\bm{M}+\epsilon\bm{N}$, where $\bm{N}$ denotes an undesired measurement and $\epsilon\ll1$ characterizes the disturbance probability. The effect of measurement imperfection is quantified by the visibility $v$. The left panel shows a schematic of phase estimation using a Mach–Zehnder interferometer, where $\theta$ is the unknown parameter to be estimated and $\varphi$ is an additional controllable phase. Alongside, the click probabilities of the upper and lower detectors, $p_{+}$ and $p_{-}$, are shown as functions of $\theta+\varphi$ for different visibilities. The middle and right panels display the FI, $\mathscr{F}$, and FI MeNoS, $\chi$, respectively. At $\theta+\varphi=\pi/2$, both the FI and FI MeNoS attain their maximum values for both perfect and imperfect visibility, illustrating the interplay between estimation precision and robustness against noisy measurements.}
\label{fig:noise-in-measurement}
\end{figure*}

The explicit formula of $\chi[\bm{M}]$ can also be found. Noting that $\mathscr{F}[\bm{M}] = 
\sum \Tr(\rho_\theta M_i) l_i^2$, with 
$l_i \coloneqq \frac{\textrm{Tr} (\dot \rho_\theta M_i)}{\textrm{Tr} ( \rho_\theta M_i)},$ where $M_i$ denotes the element of measurement $M$ and $\rho_\theta$ denotes the encoded state. With this notation, Eq.~\eqref{eqn:noise-in-measurement-1} reduces to $\chi[\bm{M}, \bm{N}] = 1+ \mathscr{F}[\bm{M}]^{-1} G[\bm{N}]$ where $G[\bm{N}] = \Tr(A_i N_i)$, $A_i = l_i^2 \rho_\theta - 2 l_i \dot{\rho}_\theta$, and the optimization reduces to the maximization of $G[\bm{N}]$. Let the logarithmic derivatives be ordered as $l_1 \le l_2 \le \dots \le l_K$ and in the worst case scenario, noise can affect smallest and largest logarithmic derivatives. This allows one to perform the maximization of Eq.~\eqref{eqn:noise-in-measurement-2}, leading to
\begin{equation}
\label{eqn:noise-in-measurement-3}
\chi[\bm{M}] = 1 + \frac{1}{2 \mathscr{F}[\bm{M}]} \left( l_1^2 + l_K^2 + \|A_1 - A_K\|_1 \right),
\end{equation}
where $\| A \|_1 = \mathrm{Tr}\left(\sqrt{A A^\dagger}\right)$ is the trace norm. Furthermore, the phase-estimation problem in a Mach-Zehnder interferometer with a single-photon input has been investigated in this setting~\cite{Kurdzialek23a}. The effect of measurement imperfections is quantified by the visibility $v$. 
The FI, $\mathscr{F}$, is independent of the controllable phase $\varphi$ for perfect visibility ($v=1$), a condition that is generally unattainable in practice. 
For imperfect visibility (e.g., $v=0.98$), the FI attains its maximum at $\theta+\varphi=\pi/2$ and vanishes for $\theta+\varphi \in \{0,\pi\}$. Interestingly, the FI MeNoS, $\chi$, is minimized at $\theta+\varphi=\pi/2$, implying that the precision of estimating $\theta$ becomes least susceptible to general measurement noise at this operating point, even in the presence of small perturbations to the ideal measurement~\cite{Kurdzialek23a} (see Fig.~\ref{fig:noise-in-measurement}).

Recent studies generalizes the concept of 
\textbf{FI MeNoS} to the multiparameter quantum estimation setting by using semidefinite programming~\cite{Albarelli24}. For more works in this direction, see Refs.~\cite{Len22,Zhou23b}.

\subsection{Reservoir engineering}
\label{subsec:reservoir-engineering}

 Metrologically useful quantum resources, such as entanglement and other forms of the quantum correlation, in general, are known to be adversely affected in the presence of noise. Reservoir engineering~\cite{Poyatos96,Plenio02,Kraus08,Schirmer10,Rai10,Krauter11,Diehl11,Berman12,Ghosh19,Ray23} refers to the deliberate design and control of system-environment interactions to derive a quantum system toward desired states or dynamics. It has emerged as a powerful tool for preparing steady entangled states~\cite{Krauter11,Lin13,Shankar13} and spin-squeezed states~\cite{Cirac93,Kronwald13,Didier14,Wollman15,Kienzler15,Lecocq15,Pirkkalainen15,Lei16,Dassonneville21}, universal quantum computation~\cite{Verstraete09,Kastoryano13},  and protecting coherence against decoherence~\cite{Murch12}. Beyond these examples, engineered dissipation has also been used to prepare metrologically useful nonequilibrium many-body states in cold atoms and Bose gases~\cite{Diehl11b,Otterbach14,Caspar16,Caspar16b}, Gaussian and cat-like entangled states~\cite{Ma17,Mamaev18,Zippilli21}, and, more recently, long-range entangled structures generated through controlled symmetry selection or dephasing-assisted dynamics~\cite{Dutta24,Saha2024}. In the context of quantum metrology, this engineering can also play an important role since many of states it generates are known to provide high precision by beating SQL~\cite{Tan17,Zhang21a,Li23a,Agarwal2025,Labarca25}. This dissipating approach can be beneficial for building large-scaling quantum sensor where keeping entanglement over long time is a major challenge, thereby providing an effective method to design noise-resilient quantum metrology.

\section{\textbf{QFI as a diagnostic of quantum resources, many-body phenomena, and processes}}
\label{sec:usefulness-of-QFI}

{QFI, beyond its central role in quantum parameter estimation, has found wide-range of applications across different areas of quantum information theory, quantum many-body (QMB) phenomena and quantum technologies.   In particular, QFI serves as an important witness of entanglement in QMB settings, where deriving experimentally accessible entanglement criteria is often highly challenging. In addition, QFI has emerged as a powerful  tool for identifying various types of quantum phase transitions in QMB systems.  Further, it has been recognized that QFI is  connected to quantum thermodynamics, and also to the dynamics of open quantum systems.

\subsection{State-space structure}
\label{subsec:state-space-structure}
\textit{Connecting entanglement with QFI.} Entanglement~\cite{Horodecki09, Otfried09,Das16} of bipartite and multipartite states is a key quantum resource that enables advantages over classical approaches in quantum information processing tasks~\cite{Ekert91,Bennett92,Bennett93,Gisin02,Briegel09}. One can classify entangled states from the perspective of quantum technological applications, specifically, by dividing entanglement classes based on additional physical or algebraic properties relevant to the task at hand. Here, focusing on QFI, we review works that use QFI to formulate various entanglement criteria and thereby connect QFI to entanglement measures. 

Let us consider a parameter estimation scenario in which a parameter \( \theta \) is encoded  on an input state \( \hat{\rho}_{\mathrm{inp}} \) by a unitary, such that
\[ \hat{\rho}_{\mathrm{out}}(\theta) {\coloneqq e^{i \theta \hat{J}} \hat{\rho}_{\mathrm{inp}} e^{-i \theta \hat{J}}},
\] where \( \hat{J}\) is the generator of the unitary. For instance, in the context of estimation of an unknown phase $ \theta$, \( \hat{J} \equiv\hat{J}_{\alpha} = \sum_{l=1}^N \hat{\sigma}^{(l)}_{\alpha} \) is the collective spin operator formed from the Pauli operators \( \hat{\sigma}^{(l)}_{\alpha} \) acting on the \( l \)-th qubit. The corresponding QFI quantifies how sensitively the state changes under the action of the generator \( \hat{J} \). It thus depends on both the input state and the generator, i.e., $\mathcal{F} \equiv \mathcal{F}(\hat{\rho}_{\mathrm{inp}},\hat{J})$. A sufficient detection criterion for entanglement in a \( N \)-qubit state is simply followed from the QFI ~\cite{Pezze09}.Let us define a quantity $\chi$ as
$\chi^2 \equiv \frac{N}{\mathcal{F}_\mathrm{Q} [\hat{\rho}_{\mathrm{inp}}, \hat{J}]}$. The QCRB for estimating the phase \( \theta \) in this setting is given by
\begin{equation*}
\Delta \theta_{\mathrm{QCR}} := \frac{1}{ \sqrt{\mathcal{F} [\hat{\rho}_{\mathrm{inp}}, \hat{J}]} } = \frac{\chi}{\sqrt{N}}.
\end{equation*}
When the estimation precision surpasses the shot-noise limit, which is attainable with classical (separable) states, we call it sub-shot-noise sensitivity. If the QFI satisfies \( \chi < 1 \) for any given input state \( \hat{\rho}_{\mathrm{inp}} \) and generator \( \hat{J} \), the state is multipartite entangled and is capable of enabling parameter estimation with sensitivity beyond the classical limit, i.e., it is useful for sub-shot-noise metrology. 
Conversely, there exist multipartite entangled states for which \( \chi \geq 1 \); such states, despite being entangled, do not outperform separable states in phase sensitivity. Thus, while \( \chi < 1 \) is only a sufficient condition for entanglement, it is both necessary and sufficient for achieving sub-shot-noise sensitivity. In this way, the condition \( \chi < 1 \) identifies a subset of entangled states that are \emph{useful} for quantum-enhanced metrology. QFI thus serves as a clean metrological entanglement witnesses.

In addition, classification of entangled states is proposed via the, so called, \textit{metrological gain}~\cite{Pezze09,toth2012,hyllus2012}. Let us consider the most typical quantum metrology scenario where the parameter to be estimated is encoded via a unitary generated by a Hamiltonian \(\mathcal{H}\). In this context, a quantum state \(\varrho\) is termed \textit{metrologically useful} if it enables a  sensitivity surpassing that achievable by any separable state, i.e.,
\begin{equation}
\mathcal{F} [\varrho, \mathcal{H}] > \max_{\varrho_{\mathrm{sep}}} \mathcal{F} [\varrho_{\mathrm{sep}}, \mathcal{H}] \equiv \mathcal{F} ^{(\mathrm{sep})}(\mathcal{H}),
\end{equation}
where \(\mathcal{F} [\varrho, \mathcal{H}]\) denotes the QFI  corresponding to the probe state \(\varrho\) and generator \(\mathcal{H}\), while $\mathcal{F}^{(\mathrm{sep})}(\mathcal{H})$  represents the maximum QFI attainable by all separable states under the same encoding Hamiltonian.
To quantify this advantage, the metrological gain is defined as
\begin{equation}
g_{\mathcal{H}}(\varrho) \coloneqq \frac{\mathcal{F} [\varrho, \mathcal{H}]}{\mathcal{F}^{(\mathrm{sep})}(\mathcal{H})}.
\end{equation}
A particularly relevant figure of merit is the maximum metrological gain over all \textit{local Hamiltonians} (i.e., sums of single-particle Hamiltonians), defined as
\begin{equation}
g(\varrho) \coloneqq \max_{\text{local } \mathcal{H}} g_{\mathcal{H}}(\varrho).
\end{equation}
For any state $\varrho$, if \(g(\varrho) > 1\), then \(\varrho\) is guaranteed to be metrologically useful, meaning it offers an estimation precision beyond the limit imposed by separable states, thereby confirming its entanglement.

Beyond witnessing the presence of entanglement, QFI can further be used for probing the finer structure of the entanglement in the multiparty setting, by investigating the hierarches in the multiparty entanglement, for example through the bounds that track the entanglement depth~\cite{toth2012,hyllus2012}. It turns out that for a $N$-qubit $k$-producible states, the QFI is upper bounded by
\begin{eqnarray}
    \mathcal{F} [\hat{\rho}_{\mathrm{inp}}, \hat{J}] \le s k^2+r^2,
    \label{toth}
\end{eqnarray}
where $s=\left\lfloor \frac{N}{k} \right\rfloor$ denotes the largest integer smaller than or equal to $\frac{N}{k}$, and $r=N-sk$. Therefore, the state, for which the QFI exceeds the bound in~\eqref{toth}, must contain at least $(k+1)$-partite entanglement. QFI thus not only detects presence of entanglement in a state, but detect entanglement depth, as well.

Although the QFI provides a useful operational bridge between metrological performance and the structure of quantum correlations, e.g., by witnessing entanglement depth, as discussed above, the relation between entanglement and metrological usefulness is, however, subtle and task-dependent. It has been demonstrated that entanglement is an important resource in several metrological tasks and enables surpassing the classical bound generated by separable states~\cite{Giovannetti06,Giovannetti11}. Furthermore, non-distillable class of entangled states, known as bound entangled states, have been shown to be metrologically useful~\cite{Czekaj15,Toth18}. However, presence of entanglement by itself does not guarantee quantum-enhanced metrology. For example, considering  two-qubit flip-flop Hamiltonian (anisotropic Heisenberg model) coupled with a bath,  if the curvature of entanglement (CoE), i.e., the second derivative of concurrence, of the evolved state equals to the negative of the QFI corresponding to the coupling constant of the interaction Hamiltonian, product measurements are shown to be sufficient to saturate the QCRB, although at all other times entangled measurements become optimal~\cite{Saleem25}. Strikingly, this result holds irrespective of whether the initial state is separable or entangled. In principle, there may exist highly entangled states that fail to offer any metrological advantage~\cite{Hyllus10}.
Also, certain bipartite entangled states, while unable to outperform separable states in a linear interferometer, can still exhibit metrological advantages when multiple copies are used or when an auxiliary is introduced~\cite{Toth20}. These examples collectively demonstrate that metrological usefulness is governed not merely by the presence or amount of entanglement in the input state, but also by the underlying structure of the entanglement, the nature of the parameter-encoding process, and the choice of measurement strategy.

However, the operational significance of the QFI is not limited to its connection with entanglement alone.
{From the broader perspective of quantum resource theories, several distinct forms of quantum resources have been shown to enhance metrological performance and can be quantitatively characterized through QFI. For example, beyond multipartite entanglement, more intricate many-body quantum correlations,such as many-body nonlocality, have been identified as operational resources for quantum-enhanced sensing~\cite{niezgoda2021}. More recently, non-stabilizerness (or magic) \cite{Gottesman97,Gottesman98,Bravyi05,Bravyi12,Veitch14,Howard17}, a central resource in fault-tolerant quantum computation, turns out to provide metrological advantages in suitable parameter-estimation tasks~\cite{tanaus2026}. Other examples include quantum coherence \cite{Streltsov17}, which forms the fundamental basis of phase sensitivity and interferometric precision~\cite{Streltsov17,Ahnefeld25}; quantum asymmetry with respect to the encoding generator, which determines the ability of a state to encode phase information and thereby directly governs interferometric sensitivity~\cite{Zhang17,Albarelli22}; and non-Gaussianity~\cite{Weedbrook12}, which can also enhance estimation precision in continuous-variable and bosonic metrological protocols beyond what is achievable with Gaussian resources alone~\cite{Hanamura21,Rahman25} (see also Sec.~\ref{subsec:CV-metrology}). Thus, QFI provides a unifying operational framework connecting metrological performance to a broad spectrum of quantum resources, far beyond entanglement alone.

A more fundamental question that naturally arises in this context is: {\it Can  Fisher information (both quantum and classical) identify whether a quantum state is resourceful or not?} Remarkably, the answer is affirmative~\cite{Tan21}. In particular,  both FI and QFI can faithfully distinguish resourceful quantum states from free states, independent of the specific quantum resource under consideration. More generally, for any well-defined quantum resource theory, every resourceful state can always be associated with a suitable metrological task in which it achieves a strictly better estimation performance than all free states lacking that resource. This establishes Fisher information as a universal operational witness of quantum resources and provides a fundamental metrological interpretation of resourcefulness in quantum theory.}

\subsection{QFI as an indicator of many-body quantum phenomena} 
\label{subsec:QFI-indicator}

The critical behavior of a quantum many-body (QMB) system manifests most prominently near a continuous quantum phase transition (QPT), where the ground state undergoes a qualitative change with the variation of the control parameter~\cite{Sachdev11}. Such criticality is characterized by a bunch of key properties, such as gap-closing in the thermodynamic limit, long-range correlations, and universal scaling behavior. In the case of Landau-type transitions, symmetry breaking and the emergence of an order parameter also play a central roles. Quantum phase transitions can belong either to the conventional Landau paradigm, as in second-order symmetry-breaking transitions, or to broader classes beyond it, such as localization-driven and topological transitions. Moreover, the critical behavior near the QPT is  often classified into different universality classes; for instance, the phase transition in an Ising chain falls under the broader category of the Ising universality class (\(\mathbb{Z}_2\) universality class). Therefore, identifying the critical behavior and the associated phases of a system is a fundamental task in many-body physics. While various indicators of QPTs have been proposed in the literature, in this review, we primarily focus on the role of QFI in revealing the phases of the QMB systems.

Recent advances in quantum information geometry have illuminated universal features of quantum phase transitions through three landmark studies. A unifying geometric perspective that is primarily focused on the second-order quantum phase transitions has emerged through several seminal works. Venuti and Zanardi~\cite{zanardi_2} introduced a scaling theory of the quantum geometric tensor (QGT), showing that its singular behavior near criticality captures essential features of quantum phase transitions, including both fidelity susceptibility and Berry curvature. Shortly thereafter, Zanardi, Giorda, and Cozzini~\cite{zanardi_2} formalized this geometry by equipping the space of Hamiltonian parameters with a Riemannian metric derived from the QGT, showing that its divergence directly signals quantum critical points. Building on these geometric foundations, Zanardi, Paris and Venuti~\cite{Zanardi08} then highlighted the operational consequences by interpreting criticality as a resource for quantum estimation: they showed that near a quantum critical point, the QFI becomes super-extensive, yielding enhanced precision in parameter estimation achievable by the optimal Cramér–Rao bound. Together, these works establish a unified framework in which the geometry of the ground-state manifold via the QGT and its associated metric both characterizes quantum phase transitions and quantifies their metrological utility.

In order to utilize the cooperative phenomenon in several aspects of quantum information as well as quantum technology, one requires the knowledge of the multiparty correlation structure in between the systems. In this direction, the evaluation of multiparty entanglement through QFI has already been  discussed in Sec.~\ref{subsec:state-space-structure}. Exploration in this direction, a novel method is proposed to detect multipartite entanglement in quantum many-body systems by relating the QFI to experimentally measurable dynamic susceptibilities\footnote{Dynamic susceptibility: It is defined as $\chi = \iota \int_0^\infty dt e^{\iota \omega t} \Tr(\rho [O(t),O])$, where $\hat{O}(t) = e^{\iota H t} \hat{O} e^{- \iota H t}$.}~\cite{hauke2016}. The key insight is that the QFI, a fundamental quantity for characterizing entanglement depth, can be lower-bounded by an integral over the system's dynamic susceptibility weighted by a thermal factor. This connection uses standard linear response techniques and enables entanglement detection without full state tomography. By applying this method on the transverse-field Ising model, it can be shown that near quantum phase transitions, the QFI peaks, revealing the presence of highly multipartite entangled states. Their approach is experimentally feasible in platforms such as cold atoms, trapped ions, and solid-state systems, offering a powerful tool for probing entanglement in complex quantum systems.

\subsubsection{Phase transitions - second and first order, topological transition, localization transition}
\label{subsubsec:phase-transition}

QFI has emerged as a powerful tool for probing many-body quantum systems, especially in the context of phase transitions and entanglement dynamics. In such cases, an operator-based criteria in distinguishing multiparty entanglement (ME) can raise the curtain between the different phases of a Hamiltonian. Such ME is studied by the QFI of an observable \(\hat{O}\) based on in Eq.~\eqref{toth}. For a pure state, the QFI associated with an observable \(\hat{O}\) simplifies to a form proportional to its variance as
\begin{equation}
\mathcal{F} [\hat{O}] = 4\left(\langle \hat{O}^2 \rangle - \langle \hat{O} \rangle^2\right).
\end{equation}
As demonstrated in Refs.~\cite{toth2012,hyllus2012}, when the observable takes the form
\begin{equation}
\hat{O}[\{\mathbf{n}_j\}] = \frac{1}{2}\sum_j \mathbf{n}_j \cdot \hat{\boldsymbol{\sigma}}_j,
\end{equation}
with \(\mathbf{n}_j = (n_j^x, n_j^y, n_j^z)\) being unit vectors, the QFI provides rigorous lower bounds on multipartite entanglement.  Specifically, if the QFI density \(f_Q = \mathcal{F}/L\) exceeds a divisor \(k\) of the system size \(L\), the state is guaranteed to be at least \((k+1)\)-partite entangled as described by~\eqref{toth}. The tightest bound is obtained by maximizing \(\mathcal{F}\) over the choice of \(\{\mathbf{n}_j\}\), which is mathematically equivalent to finding the ground state of a classical Hamiltonian
\begin{equation}
H_{\mathrm{cl}} = -\mathcal{F} [\hat{O}[\{\mathbf{n}_j\}]]
\end{equation}
with vector-spin variables \(\mathbf{n}_j\) coupled through connected two-body correlators, given as
\begin{equation}
C_{i,j}^{\alpha,\beta} = \langle \hat{\sigma}_i^\alpha \hat{\sigma}_j^\beta \rangle 
- \langle \hat{\sigma}_i^\alpha \rangle \langle \hat{\sigma}_j^\beta \rangle .
\end{equation}
Substituting the form of \(\hat{O}\), the QFI becomes
\begin{equation}
\mathcal{F} [\hat{O}[\{\mathbf{n}_j\}]] = 
\sum_{\alpha,\beta=x,y,z} \sum_{i,j} n_i^\alpha\, C_{i,j}^{\alpha,\beta}\, n_j^\beta.
\end{equation}

Using the above mentioned protocol, 
multipartite entanglement, quantified through the QFI, is capable of 
revealing the regular-to-ergodic transition in the Dicke model~\cite{gietka2019}.
In particular, the multipartite entanglement is developed in the Dicke model as the dynamics crosses from a nonthermal, memory-preserving regular regime to an ergodic one. Precisely, the ergodic regime can generates metrologically useful entanglement much more rapidly than the regular one. Extending this line of inquiry, QFI is employed to explore the detection of higher-order many-body interactions that leads to faster entanglement generation~\cite{cieifmmonde2024}. For a \(k\)-body interacting  Ising Hamiltonian, (Lipkin–Meshkov–Glick (LMG) model), the maximal QFI in a product state is ordered for a fixed interaction order \(k\). Since, QFI can be measure experimentally, such study provides a viable method to verify many-body interaction present in the Hamiltonian.

A coherent line of development has emerged to identify topological phases in qquantum many-body systems with the help of ME, quantified by the QFI.  In one-dimensional symmetry-protected topological phases, a dual-lattice formulation of the scaling of the QFI with respect to nonlocal dual spin generators sharply distinguishes different topological sectors and even resolves phases with higher winding numbers~\cite{zhang2018_a}. This idea can subsequently be generalized to genuine two-dimensional intrinsic topological order in the perturbed toric-code model, where the QFI density can be expressed in terms of gauge-invariant Wilson loops over square regions \cite{zhang2022_a}. The scaling of these loop operators provides a direct ME witness of the $\mathbb{Z}_2$ topological phase transition and remains applicable out of equilibrium, enabling the study of thermalization, disorder-assisted stabilization, and the absence of finite-temperature topological order in the thermodynamic limit. Closely related studies on disordered topological wires further establish that the superextensive or Heisenberg-like scaling of the QFI remains robust even in the presence of impurities~\cite{pezze2022}, thereby demonstrating that ME offers not only a universal diagnostic of topological quantum matter but also a practically measurable resource for robust quantum-enhanced metrology~\cite{Gabbrielli2018,Gabbrielli2019,yin2019,su2020}. 

\textit{Dynamical quantum phase transition.} Dynamical quantum phase transitions (DQPTs)~\cite{Heyl13,Heyl_review_2018}, which draws parallel ideas borrowed from adiabatic quantum phase transitions(QPTs), are characterized by non-analyticities in the time evolution of a quantum many-body system, when the ground state is quenched across quantum critical point, by abruptly or gradually changing the system parameters. Several studies~\cite{Guan21,Zhou23,Munoz-Arias23,Chen25a,Mumford25}, have investigated the detection of these dynamical quantum phase transitions through the QFI. For example, when the transverse-field Ising model quenched from the paramagnetic phase to the ferromagnetic one the QFI density diverges logarithmically with system size at the critical times, thereby signaling the occurrence of a DQPT~\cite{Chen25a}. Importantly, this behavior is robust against both non-integrability and dissipation. On the other hand,  the long-time average of QFI can also detect both type-A and type-B DQPT in spinor Bose–Einstein condensates as well as in the LMG models~\cite{Mumford25}.

\subsubsection{QFI to detect scrambling, fragmentation, measurement-induced phase transition}
\label{subsubsec:QFI-detector}
\textit{Many-body scar and QFI.} Quantum many-body scars are rare eigenstates embedded within otherwise thermal spectra that do not obey the eigenstate thermalization hypothesis. They retain memory of initial conditions and show nonthermal dynamics, but their entanglement structure is still not fully understood. One of the related question that naturally emerges is if scar states can provide a resource for quantum-enhanced metrology. Such question has been addressed in 2022 by Desaules \textit{et al.} by uncovering a striking entanglement structure in exact scar states: when these states arise from an $\mathfrak{su}(2)$ spectrum-generating algebra, their QFI scales super‑extensively with system size, indicating genuine multipartite entanglement across the entire system~\cite{desaules2022}. Through analytical derivations supported by scaling arguments and quantum statistical estimates, these nonthermal eigenstates exhibit entanglement properties distinct from those of generic thermal states, paving the way to harness them for enhanced quantum metrology. This work bridges the field of quantum many-body scars, a phenomenon known for anomalous ergodicity breaking, with the operational potential of entanglement in precision measurements, thus linking three key areas: algebraic structures in nonergodic dynamics, extreme multipartite entanglement, and quantum-enhanced metrological applications.

\emph{Measurement-induced phase transition via QFI.} Measurement-induced phase transitions (MIPTs)~\cite{Skinner19,Fisher23}, a key framework to explore the competition between unitary dynamics and measurement processes in quantum systems. Typically, studies focus on a simple, symmetry-free architecture known as the brick-wall circuit, where layers of unitary operations are alternated with random local measurements. While the unitaries generate entanglement across the system, the measurements tend to suppress it. Remarkably, beyond a critical measurement rate, the system undergoes a transition from a volume-law to an area-law scaling of entanglement entropy—which marks the onset of the MIPT. Moreover, studies on MIPTs also include traditional QMB systems, such as spin chains and fermionic lattices.  Although most investigations are centered on bipartite entanglement, growing attention is now devoted to understanding the fate of multipartite entanglement in such monitored circuits and many-body Hamiltonian systems. In this context, the QFI offers a deep insights into the multipartite entanglement structure (as illustrated in Sec.~\ref{subsubsec:phase-transition}) and quantum criticality underlying these measurement-induced phenomena. 

In this regard, a laser-driven atomic ensemble undergoing collective decay under continuous monitoring is shown to reveal a MIPT that circumvents the traditional postselection barrier~\cite{Passarelli2024_a}. By tuning the drive strength, a transition in the scaling of entanglement entropy, from subextensive to extensive, can be identified, coinciding with the superradiance threshold in the trajectory-averaged dynamics. This work demonstrates how monitoring protocols alone can generate nontrivial dynamical phases in open quantum systems. Building on this, the competition between collective generalized measurements and interaction-induced scrambling in spin-$1/2$ ensembles is analyzed by Poggi \emph{et al.} \cite{Poggi2024}. The trajectory-level study uncovers three distinct regimes in the scaling of the QFI: weak and strong measurements both yield Heisenberg-limited multipartite entanglement, whereas intermediate monitoring suppresses it, producing sub-Heisenberg-limited states. This reveals a nonmonotonic dependence of entanglement generation on measurement strength, linking monitored spin ensembles to measurement-induced transitions in hybrid circuit models. On a continuously monitored quantum Ising chain, the postselected no-click trajectory reproduces the same phase boundaries for bipartite entanglement entropy and multipartite QFI, while the inclusion of quantum jumps unveils richer phenomenology~\cite{Paviglianiti2023}: a third regime with logarithmic entanglement entropy but bounded multipartiteness. This highlights how QFI captures subtler aspects of entanglement structure beyond what bipartite measures can detect. Furthermore, in a monitored quantum circuits, it is shown that conventional unstructured random circuits fail to produce divergent multipartite entanglement even at criticality~\cite{Lira2025}. Remarkably, the structured two-site measurement schemes with protection mechanisms can stabilize multipartite entangled phases. Taken together, these works position the QFI as a unifying diagnostic tool across atomic ensembles, spin chains, and monitored quantum circuits, revealing how measurement protocols, system dynamics, and circuit structure conspire to either suppress or enhance ME.

\subsection{QFI as a detector of non-Markovianity}
QFI flow provides a powerful information-theoretic characterization of open system dynamics and can act as an important witness for non-Markovianity. For a parameter-dependent quantum state $\rho_\theta(t)$ evolving under a noisy quantum channel, the QFI flow is defined as the rate of change of the QFI,
\begin{equation}
\mathcal{I}(t)
=
\frac{d}{dt}\mathcal{F}[\rho_\theta(t)].
\end{equation}
For Markovian dynamics described by completely positive divisible channels, the QFI monotonically decreases under the action of noise, i.e., 
\begin{equation}
\frac{d}{dt}\mathcal{F}[\rho_\theta(t)]
\le 0,
\end{equation}
which reflects the irreversible loss of distinguishability between nearby quantum states. In contrast, a temporary increase of QFI, \(\mathcal{I}(t)>0\), signals information backflow from the environment to the system and therefore serves as a witness of non-Markovian dynamics \cite{Lu2010,Scandi2025}. In this framework, the environment-induced recovery of distinguishability directly enhances the achievable estimation precision, linking non-Markovian memory effects to quantum metrological advantage.

More generally, for multiparameter estimation with parameter vector $\boldsymbol{\theta}=(\theta_1,\theta_2,\dots)$, the relevant quantity is the QFIM,  as discussed in Sec.~\ref{subsubsec:quantum-multiple-parameter}, and presented in Eq.~(\ref{eq:qei_multi}). The multiparameter QFI flow in this case can be extended as the time derivative of the QFIM~\cite{Xing2021},
\begin{equation}
\mathcal{I}_{\mu\nu}(t)
=
\frac{d}{dt}
[\mathbb{F}(t)]_{\mu\nu}.
\end{equation}
Like in the single parameter case, the positivity of certain eigenvalues or trace measures of $\mathcal{I}_{\mu\nu}(t)$ provides a sensitive detector of non-Markovian information backflow in multidimensional parameter spaces \cite{Xing2021}. These works further demonstrated that multiparameter non-Markovianity can exhibit richer structures than the single-parameter case because correlations between different estimation channels contribute additional geometric information encoded in the off-diagonal elements of the QFIM. A subsequent analysis clarified several subtleties regarding the interpretation of QFI flow under general dynamical maps and emphasized the dependence of the witness on the chosen encoding strategy and parameterization~\cite{Vatasescu2020}. A several works has strengthen the connection between QFI and non-Markovianity~\cite{song2015,dhar2015,LU2020,paolo2023,Dakir2025}, thereby establishing a connection between open-system dynamics, information backflow, and quantum-enhanced metrology.

\section{\textbf{Quantum-metrological protocols and applications}}
\label{sec:applications}
Having discussed the foundational bounds, encoding mechanisms, and resource-diagnostic role of QFI, we now turn to concrete quantum-metrological protocols and their applications. We begin with interferometric sensing, which is historically the central paradigm of phase estimation. Following this, we proceed to discuss other major metrological schemes, including atomic clocks, continuous-variable quantum metrology, atom-photon and cavity-QED-based  sensors, many-body quantum sensors, quantum imaging and illumination, and related information-theoretic sensing protocols.

\subsection {Interferometric quantum sensing}
\label{subsec:inferometric-metrology}
We discuss quantum sensing based on the interferometric principle. While photonic interferometers provide the canonical setting for optical phase estimation, matter-wave and atom interferometers extend the same metrological idea to massive quantum probes for precision measurements of acceleration, gravity, rotations, and fundamental constants.

\subsubsection{Photonic interferometry}
Interferometry-based quantum sensing constitutes one of the most powerful paradigms for precision measurements, where a physical parameter is encoded into the relative phase accumulated between two coherent quantum paths. In a generic interferometric protocol, a probe state is first prepared and then split into two paths that evolve differently under the influence of an external parameter $\lambda$. After recombination, the accumulated phase difference is converted into measurable population differences. The resulting interference signal can typically be written as
\begin{equation}
P = \frac{1}{2}\left[1+\cos(\phi)\right],
\end{equation}
where $\phi$ denotes the phase difference acquired between the two paths. When the phase depends on the parameter of interest, $\phi=\phi(\lambda)$, estimating $\phi$ allows one to infer $\lambda$ with high precision. 
Among various interferometric setups, the Mach-Zehnder interferometer (MZI) is one of the most widely studied optical interferometers in this field. In quantum metrology, the achievable precision is fundamentally bounded by the quantum Cramér–Rao inequality. For an interferometric phase estimation problem, the QFI is related to the phase encoding generator $G$ as \(\mathcal{F} = 4(\Delta G)^2.\) In the case of the uncorrelated probes, the phase sensitivity is restricted to the standard quantum limit $\Delta[\phi]^2 \sim 1/{n}$, where $n$ is the number of probes.
However, several theoretical proposals have shown that by exploiting nonclassical resources, such as entanglement or spin-squeezed states, the precision of phase estimation can surpass the standard quantum limit and even approach the Heisenberg limit, which scales as $1/n^2$. Furthermore, parity measurements of photon number have been proposed as an effective detection strategy, demonstrating that the Heisenberg limit can be achieved with maximally entangled states in a lossless MZI~\cite{Gerry10}. Significant works have been devoted to generating input states with stronger nonclassical features in MZI-based systems~\cite{Campos03,Schlosshauer05,Dowling08,Anisimov10,Gerry10,Giovannetti11,Pang17,Guo18,Wang19,Kumar23,Dalidet26}. Interferometry therefore forms the basis of numerous quantum sensing technologies, including optical interferometers used in gravitational-wave detection, Ramsey interferometers in atomic clocks, and matter-wave interferometers for measurements of inertial forces and gravitational fields. These techniques rely on coherent interference of quantum states and provide some of the most precise measurement tools currently available.

Interferometric techniques become one of the most mature platforms for quantum-enhanced sensing and metrology, with applications ranging from precision spectroscopy to gravitational-wave detection. Optical interferometers have long served as a fundamental tool for phase-sensitive measurements, and quantum resources, such as squeezed and entangled states, have been proposed to enhance their sensitivity beyond the standard quantum limit~\cite{Caves1981,Giovannetti04,Giovannetti11}. Within this interferometric setting,  squeezing reduces quantum fluctuations in the measured quadrature or collective variable, thereby lowering the phase noise below the SQL. We return to the Gaussian and CV description of such squeezed probes in Sec.~\ref{subsec:CV-metrology}, while spin-squeezed many-body and cavity-mediated realizations are discussed in Secs.~\ref{subsec:atom-photon} and~\ref{subsec:many-body-sensors}.
A landmark experimental realization of quantum-enhanced interferometry is demonstrated using squeezed light in gravitational-wave detectors, significantly improving the sensitivity of large-scale interferometers such as LIGO~\cite{Aasi13}.

\subsubsection{Matter-wave and atom interferometry}
Atomic interferometry focuses on the estimation of extremely small phase shifts with the highest possible precession allowed by quantum mechanics. In particular, atomic interferometry exploits the wave nature of atoms to coherently split, manipulate, and recombine matter waves, allowing the measurement of phase shifts induced by external fields with extremely high precision.  Light-pulse atom interferometers currently employed for precision sensing of a wide range of physical quantities, including magnetic fields, accelerations, rotations, gravitational fields, and gravity gradients~\cite{Kasevich1991,Peters2001}, while recent developments using large momentum transfer techniques have further improved their sensitivity~\cite{Chiow2011}. Historically, the simplest form of atom interferometry can be related to Young’s double-slit experiment, where a beam splits into two paths by passing through a double slit and the resulting interference pattern is observed on a screen placed further along the beam path. Over the years, several types of atom interferometers have been developed, including the Mach–Zehnder interferometer~\cite{Mach1892,Kasevich1992}, Raman interferometer~\cite{Kasevich1991}, Talbot–Lau interferometer~\cite{Lau1948,clauser1994}, Ramsey–Bordé interferometer~\cite{ramsey1950,Bord1989}, and Sagnac interferometer~\cite{Gustavson1997}, among others. Matter-wave interferometers based on ultracold atoms provide another powerful platform for precision sensing.

These diverse implementations highlight the versatility of atom interferometry as a universal platform for precision measurement. Atom interferometry has found widespread application in precision sensing and fundamental physics. State-of-the-art experiments have enabled high-precision measurements of gravitational acceleration and gravity gradients~\cite{Peters1999,Snadden1998}, tests of the equivalence principle~\cite{Asenbaum2020}, and determinations of the Newtonian gravitational constant~\cite{Rosi2014}. Atom interferometers have also been employed to measure the fine-structure constant~\cite{Parker2018} and to probe inertial effects such as rotations via the Sagnac effect~\cite{Gustavson1997}. More recently, in order to probe for physics beyond the Standard Model, including searches for dark matter and gravitational waves~\cite{Graham2013,Dimopoulos2008}, atom interferometers have been employed.

In parallel, significant advances have been achieved through the incorporation of many-body quantum effects and nonclassical resources.   Interferometric sensing, in particular, has been widely addressed in Bose-Einstein condensate (BEC), a highly coherent sources of matter waves, where nonlinear interactions enable the generation of spin-squeezed and entangled states that can surpass the standard quantum limit~\cite{Gross2010,Riedel2010}, while optical lattices and trapped-atom platforms enable precise control over atomic motion and interactions~\cite{Bloch08}.  In addition, interferometric schemes using trapped ions~\cite{Leibfried2004}, and photonic platforms~\cite{Nagata07} have demonstrated quantum-enhanced phase estimation with nonclassical probe states. Furthermore, the use of entanglement and spin-squeezed states allows atom interferometers to surpass the standard quantum limit and approach Heisenberg-limited sensitivities~\cite{Pezze18}. These developments have led to the emergence of quantum-enhanced interferometry, including nonlinear schemes such as \(SU(1,1)\) interferometers~\cite{Yurke1986}, which exploit parametric amplification and quantum correlations to further improve sensitivity. 

\emph{\(SU(1,1)\) interferometers.} Unlike Mach-Zender interferometer, \(SU(1,1)\) a different class of interferometers which can play important role in achieving high sensitivity by using fewer optical elements than the \(\mathrm{SU}(2)\) interferometers called as \(SU(1,1)\) interferometer~\cite{Yurke1986} where the role of the first beam splitter is played by an active nonlinear element that generates correlated particle pairs via parametric amplification. This process is described by the unitary operator
\( \hat{U}_{\mathrm{PA}}(r) = \exp\left[-i r \left(\hat{b}_1^\dagger \hat{b}_2^\dagger + \hat{b}_1 \hat{b}_2 \right)\right], \) where $\hat{b}_1$ and $\hat{b}_2$ denote the bosonic annihilation operators corresponding to the two interferometric arms (referred to as the side modes).  Assuming that the side modes are initially in the vacuum state, the action of $\hat{U}_{\mathrm{PA}}(r)$ produces a two-mode squeezed vacuum state, which can be a coherent superposition of twin-Fock states, characterized by an average particle number in the side modes given by \(N_s = 2 \sinh^2 r.\) These particles are typically generated from a highly populated reservoir mode (the pump), which is assumed to remain undepleted, i.e., its occupation is much larger than $N_s$. After an interrogation time, a phase shift $\phi/2$ is imprinted onto each of the two side modes. Subsequently, a second parametric amplification stage effectively inverts the initial transformation which can be implemented by introducing a relative phase shift of $\pi/2$ in the pump mode, corresponding to the transformation $r \rightarrow -r$. Finally, a measurement of the total particle number in the side modes, \(\hat{N}_s = \hat{b}_1^\dagger \hat{b}_1 + \hat{b}_2^\dagger \hat{b}_2,\) provides information about the accumulated phase $\phi$. Evaluated at the optimal operating point $\phi = 0$, the corresponding phase sensitivity is given by~\cite{Szigeti2017,Szigeti2021}
\begin{equation}
\Delta \phi_{\mathrm{SU}(1,1)} = 
\left.
\frac{\sqrt{\mathrm{Var}(\hat{N}_s)}}{\left| \partial \langle \hat{N}_s \rangle / \partial \phi \right|}
\right|_{\phi = 0}
= \frac{1}{\sqrt{N_s (N_s + 2)}},
\end{equation}
which exhibits Heisenberg limit with respect to the number of particles in the side modes. Experimentally, \(\mathrm{SU}(1,1)\) interferometers is implemented in spinor Bose-Einstein condensate and optical cavity QED platforms have been demonstrated to enable sensing~\cite{linnemann2016,Burd2019,Szigeti2021,Szigeti2017,Parker2018,Mao2023,Kunkel2022,Liu2021_c,Linnemann2017,Hayes2018,Barberena2024,Liu2023,Jiao2024,Kruse2016,Lewis-Swan2020,Du2022,li2016,kreifmmode2023,chen2025}.

These developments collectively establish interferometry as a central paradigm for quantum sensing, enabling precision measurements across optical, atomic, and photonic quantum platforms.

\subsection {Atomic clocks and frequency metrology}
\label{subsec:clock-and-interferometry}

Atomic clocks rely on a well-defined transition between two long-lived internal atomic states, called clock states, typically denoted by $\ket{g}$ and  $\ket{e}$~\cite{Diddams04,Poli13,Ludlow15}. Depending upon the physical platform, these states may belong to a single trapped ion, an array of trapped ions, or an ensemble of neutral atoms confined in optical magnetic trap. The energy difference between the clock states provides a standard for frequency, and therefore time. In practice, the two clock states are coherently coupled by an external electromagnetic field so that a superposition of them can be prepared and later interrogated. Depending on the atomic species and the specific clock transition, this coupling may arise through an electric-dipole transition, a weakly allowed transition, or higher-order processes such as magnetic-dipole or electric-quadrupole couplings. The transition frequency is then extracted from the phase accumulated during the subsequent time evolution of the atomic state ~\cite{Liu20c,Olivares25}. In the atomic clock, the goal is to estimate the transition frequency of an atom as precisely as possible. There are three techniques commonly used in atomic clock operation: 
i) the \textit{Rabi} method~\cite{Rabi54}, where the atoms or ions interact with a single electromagnetic-field pulse;
ii) the \textit{Ramsey} scheme~\cite{Ramsey50}, in which two Rabi-like pulses are separated by a much longer field-free interval; 
iii) the \textit{coherent population trapping (CPT)}
phenomenon~\cite{Vanier05}, where two phase-coherent fields couple the clock states to a common excited state.

\textit{Rabi method~\cite{wineland1994,bollinger1996,pezze2019}:} A two-level atom, initially prepared in the state $\ket{g}$, interacts with an oscillating electric field with frequency $\omega$ for a time $t$. At resonance, i.e., when $\omega$ is equal to the atomic transition frequency denoted by $\omega_A$, the atom can be found in the excited state $\ket{e}$ with unit probability provided the interaction time is chosen as $t = \pi/\Omega_0$, where \(\Omega_0\)  is the on-resonance Rabi frequency. However, in the presence of a detuning $\Delta \omega = \omega_A - \omega \neq 0$, the pulse brings the atom to its excited state with non-zero probability but less than unity. If the interaction time is set to $t = \pi/\Omega_0$, the probability in finding the atom in excited state is given by,  
\begin{equation*}
\label{eq:Rabi:pe}
p^e(\omega) = \frac{1}{\left[\Theta_{\Omega_0}(\Delta \omega)\right]^2}
\sin^2 \left[
\frac{\pi}{2}\, \Theta_{\Omega_0}(\Delta \omega)
\right]\,,
\end{equation*}
with
\begin{equation*}
\label{Theta}
\Theta_{\Omega_0}(\Delta \omega) = 
\sqrt{1 + \left(\frac{\Delta \omega}{\Omega_0}\right)^2 }\,.
\end{equation*}
The precision in the estimation of $\omega_A$ corresponding to $p^e(\omega)$, obtained from Rabi interferometry, can be calculated through evaluating at $\omega \approx \omega_A$ and  the QFI is given as~\cite{Olivares25} 
\begin{equation*}
\label{Fri:expansion}
\mathcal{F}^{\text{Rabi}} (\omega) \approx \frac{1}{\Omega_{0}^2} \left[
4 - \left( 8- \frac{3}{4}\, \pi^2 \right) ( \omega - \omega_A )^2
\right]\,.
\end{equation*}

\textit{Ramsey method:} {An alternative approach for probing atomic resonances is the Ramsey method of separated oscillatory fields, in which the atoms interact sequentially with two phase-coherent laser or microwave pulses separated by a free-evolution interval, commonly referred to as the \emph{dark time}~\cite{Ramsey90}. In contrast to the Rabi  scheme, where the atom interacts continuously with a single driving field, the Ramsey protocol splits the interrogation into two coherent interactions. More specifically, after the interaction with a first pulse, the atom undergoes a free evolution for a finite time $T$, and is then subjected to a second pulse before the final measurement. 
If the interaction time is set to $\tau = \pi/(2\Omega_0)$ for both the pulses, corresponding to $\pi/2$-pulses, one can compute the probability of finding the atom in the excited state after the full sequence. Note that in the limiting case, $T \rightarrow 0$, two pulses effectively merge into a single continuous interaction, and the standard Rabi transition probability is recovered.  The maximum of the QFI is still attained at $\omega = \omega_A$ just like in the Rabi scheme.} If $\omega \approx \omega_A$, the QFI associated with the Ramsey scheme can be expanded around the resonance frequency as
\begin{align*}
\mathcal{F}^{\text{Ramsey}} &(\omega) \approx \frac{1}{\Omega_{0}^2} \Bigg\{
4\left(1 + \frac{\pi}{4}\, \kappa\right)^2 \nonumber\\[1ex]
&- \left[ 8- \frac{3}{4}\, \pi^2
+ (10 - 3 \pi) \frac{\pi}{2}\, \kappa \right] ( \omega - \omega_A )^2
\Bigg\}\,,
\end{align*}
where, for the sake of simplicity, set $T = \kappa \tau$ with $\tau = \pi/(2 \Omega_0)$~\cite{Olivares25,Ludlow15,pezze2019}. Note that for $\kappa = 0$, corresponds to a vanishing dark interval between two pulses, i.e.,  $ \mathcal{F}^{\text{Ramsey}} (\omega)$ coincides with $\mathcal{F}^{\text{Rabi}} (\omega)$.
Remarkably, if the free evolution time $T = \kappa \tau$ is much longer than the interaction time $\tau$,
namely, $\kappa \gg 1$, $\mathcal{F}^{\text{Ramsey}}$ for $\omega \approx \omega_A$ reads as
\begin{align*}
\mathcal{F}^{\text{Ramsey}} &(\omega) \approx \frac{4}{\Omega_{0}^2}
\left(1 + \frac{\pi}{4}\, \kappa\right)^2\,,
\end{align*}
that is independent of $\omega$~\cite{Olivares25}, thereby highlighting the metrological sensitivity on accumulated probe. It is also worth noting that, in the correspondence of the maximum
($\omega = \omega_A$), the ratio between the QFIs corresponding to the Ramsey
and Rabi methods scales as
\begin{equation}
\frac{\mathcal{F}^{\text{Ramsey}} (\omega_A)}{\mathcal{F}^{\text{Rabi}}(\omega_A)} =
\left(1 + \frac{\pi}{4}\, \kappa\right)^2\,,
\end{equation}
which demonstrate the benefit of the Ramsey Method over Rabi scheme.

\textit{Coherent population trapping phenomenon (CPT):} A third important paradigm in atomic frequency metrology is coherent population trapping (CPT), a quantum interference phenomenon with three level \(\Lambda\)-type transitions in which two metastable states (which may correspond to the two levels defining a clock transition) are coupled to a common excited state by two quasi-resonant, phase-coherent laser fields~\cite{Vanier05,Arimondo96}. The trapping occurs when the frequency difference of the two laser fields, $\omega_{L1}-\omega_{L2}$, matches the atomic transition frequency $\omega_A$. Under this condition, the atomic population is driven into a coherent superposition of the two clock states, known as a dark state. In this state, the atoms no longer absorb photons from the laser fields, resulting in a strong reduction of the fluorescence signal and the appearance of a narrow dark resonance. Thus CPT becomes a crucial ingredient to build complete atomic clocks, and quantum sensors~\cite{Vanier05}. It has been shown that, for CPT-based schemes, FI coincides with the QFI for all values of the detuning $\Delta \omega = \omega_A - \omega$~\cite{Olivares25}. In contrast, for Rabi and Ramsey methods, the QFI exceeds FI only for nonzero detuning~\cite{Olivares25}.

{Beyond the interrogation schemes discussed above, such as  Rabi, Ramsey, and CPT spectroscopy,  optical clocks and superradiant lasers provide a complementary route to frequency metrology~\cite{Rabi1938,Ramsey50,Arimondo96,Vanier05,Ludlow15}. 
In these systems, an ensemble of atoms emits collectively into an optical cavity, and in the superradiant regime, narrow atomic transition dominates over the cavity resonance~\cite{Meiser10,Bohnet12,Norcia16a,Norcia16b}.
This suppresses sensitivity to cavity frequency or cavity length noise, which is one of the primary limitations of conventional laser-based clock operation. Recent work on fully collective superradiant lasing has proposed a continuous-wave active-clock architecture in which repumping is also implemented collectively through an auxiliary cavity, causing ultranarrow superradiant emission and even operating regimes with vanishing sensitivity to cavity-length vibrations~\cite{Reilly26}.
These developments place superradiant lasing at the interface of optical-clock physics, collective atom-cavity dynamics, and quantum-enhanced frequency sensing.}

{Among various applications, atomic clocks have enabled a broad range of applications with extremely precise measurements, including estimating the age of the Universe with an uncertainty below 100 ms~\cite{Huntemann16,Marti18,Mann18}. This level of precision establishes the significany role of atomic clocks as one of the most successful implementtaions of quantum metrology.}

\subsection{Continuous variable metrology}  
\label{subsec:CV-metrology}
Continuous variable (CV) metrology provides the framework for treating coherent states, squeezed states, Gaussian states, and quadrature measurements. It provides the formal structure to the squeezed-light interferometric protocols, and also displacement, force, multiparameter sensing tasks via CV. The area of optical systems is one of the most well understood and matured field of research -  thanks to the decades of theoretical and experimental research efforts. Gaussian states constitute the central resource for CV quantum metrology and quantum optics. Their importance arises because coherent, squeezed, and thermal states are all Gaussian and are experimentally accessible in cavity-QED and photonic platforms. Within this broad field, CV systems constitute an important branch in which substantial advances have been made toward implementing a wide range of quantum information tasks~\cite{braunstein2005,ferraro2005,Adesso2014,Serafini2023}. Their roles in quantum sensing has garnered much attention in recent times. As the name suggests, in CV systems  the relevant information-carrying observables, such as position and momentum quadratures, have continuous spectra. 

In many optical realizations of CV systems, each mode of the electromagnetic field is modeled as a bosonic harmonic oscillator. For an \(\mathcal{N}\)-mode noninteracting system, the Hamiltonian can be written as \(\hat{H} = \sum_{j=1}^{\mathcal{N}} \hat{H}_{j},\) where \(\hat{H}_{j} = \hbar \omega_{j} \left( \hat{a}_{j}^{\dagger}\hat{a}_{j} + \tfrac{1}{2} \right)\) and \(\omega_{j}\) denotes the frequency of the \(j\)th mode. Here, \(\hat{a}_{j}\) (\(\hat{a}_{j}^{\dagger}\)) represents the annihilation (creation) operator for mode \(j\), satisfying the bosonic commutation relation \([\hat{a}_{j}, \hat{a}_{j}^{\dagger}] = 1\). The position and momentum quadrature operators, defined as 
\begin{eqnarray}
    \hat{x}_{j} = \tfrac{\hat{a}_{j}^{\dagger} + \hat{a}_{j}}{\sqrt{2}}, \quad \hat{p}_{j} = \tfrac{\hat{a}_{j} - \hat{a}_{j}^{\dagger}}{\iota\sqrt{2}},
\end{eqnarray}
leading to the complete set of canonical operators grouped in vector form as 
\begin{eqnarray}
    \hat{R} = (\hat{x}_{1}, \hat{p}_{1}, \ldots, \hat{x}_{\mathcal{N}}, \hat{p}_{\mathcal{N}})^{T}.
\label{eq:displacement}
\end{eqnarray}
These operators satisfy the commutation relations
\begin{eqnarray}
[\hat{R}_{k}, \hat{R}_{l}] = \iota \mathcal{M}_{kl}, \quad
\text{where} \quad \mathcal{M} = \bigoplus_{j=1}^{\mathcal{N}} \Omega_{j}
    \label{eq:comutation}
\end{eqnarray}
denotes the \(\mathcal{N}\)-mode symplectic form. Here, \(\mathcal{M}\) is the \(\mathcal{N}\)-mode symplectic form and \(\Omega_j\) is denoted by 
\begin{eqnarray}
\Omega_j&=&
\begin{pmatrix}
0 & -1 \\
1 & 0
\end{pmatrix}.
    \label{eq:Omega}
\end{eqnarray}
Now we are interested in Gaussian states \(\rho\), which include all ground and thermal states of a second-order (quadratic) Hamiltonian. Such states can be completely described by their first and second moments, known as the displacement vector \(\mathbf{d}\) and covariance matrix \({\Xi}\), in the phase space, where \(d_{k} = \langle \hat{R}_{k} \rangle_{\rho},\)
and \(\Xi_{kl} = \tfrac{1}{2} \langle \hat{R}_{k}\hat{R}_{l} + \hat{R}_{l}\hat{R}_{k} \rangle_{\rho} - \langle \hat{R}_{k} \rangle_{\rho} \langle \hat{R}_{l} \rangle_{\rho}.\) Here, \({\Xi}\) is a real, symmetric, and positive-definite matrix whose elements correspond to the two-point correlation functions between the \(2\mathcal{N}\) canonical variables. 
The displacement vector and covariance matrix can also be redefined in terms of the moments of the creation and annihilation operators of different modes\cite{Adesso2014}.
This greatly simplifies the calculation of the means and variances of \(\hat{N}_{j}\), which are essential for analytical treatment. In the phase-space formalism, the evolution of the state can be represented by a symplectic matrix \({S}\), under which the first and second moments of the state transform as
\begin{equation}
    {d} \rightarrow {d}' = {S}{d}, 
    \qquad
    {\Xi} \rightarrow {\Xi}' = {S}{\Xi}{S}^{T}.
    \label{eq:symplectic_evolution}
\end{equation}

Continuous-variable quantum metrology studies the estimation of an unknown parameter $\chi$ encoded in bosonic modes whose observables are the field quadratures. A probe state $\rho_0$ evolves under a unitary encoding $U_\chi=\exp(-i\chi \hat G)$ such that $\rho_\chi = U_\chi \rho_0 U_\chi^\dagger$, where the generator $\hat G$ is typically linear or quadratic in the mode operators for displacement or phase estimation. The QFI  reduces to $\mathcal F = 4\,\mathrm{Var}(\hat G)$ for pure states~\cite{Braunstein94,Giovannetti11} which directly determine the ultimate precession via quantum Cramer-Rao bound~\cite{Cramer46,Rao1992}.
For Gaussian probes, the state is fully characterized by the displacement vector and covariance matrix. CV quantum metrology exploits bosonic modes and quadrature measurements to achieve high-precision parameter estimation in optical and atomic platforms. In these systems, nonclassical Gaussian resources such as squeezed states play a central role by redistributing quantum noise in phase space and enabling sensitivities beyond the standard quantum limit, forming the basis of modern quantum sensing protocols with continuous degrees of freedom~\cite{Lawrie2019}. 

In CV-metrology parameter encoding corresponds to a symplectic transformation $S\in Sp(2N,\mathbb{R})$ acting as $\Xi\rightarrow S\Xi S^{ T}$while the displacement vector transforms as \(d\to Sd\)~\cite{Weedbrook2012,Adesso2014}. Since Gaussian states remain Gaussian under linear optical transformation, quadratic Hamiltonian evolution. This symplectic description can be quantum senor. In particular,  a complete treatment of single-mode Gaussian parameter estimation is developed where the exact QFI expressions for arbitrary mixed Gaussian states is derived~\cite{Pinel13}. These results unify estimation of (i) phase shifts, (ii) displacement amplitudes, (iii) squeezing parameters, (iv) optical losses, (v) purity and (vi) temperature. For an arbitrary Gaussian state, the QFI can be written entirely in terms of first and second moments as~\cite{Safrnek2015,Monras2013,Jiang2014} 
\begin{equation} 
\mathcal F = \frac{1}{2} \mathrm{Tr} \left[ (\Xi^{-1}\partial_\chi \Xi)^2 \right] + (\partial_\chi \mathbf{d})^T \sigma^{-1} (\partial_\chi \mathbf{d}), \label{eq:gaussian_qfi} 
\end{equation}
with \(\chi\) being the estimated parameters. For example, in phase estimation using a squeezed vacuum state with mean photon number $\bar{n}$, the QFI scales as \(\mathcal F = 8\bar{n}(\bar{n}+1),\) which exhibits Heisenberg-like enhancement~\cite{Giovannetti11} over coherent-state probes where \(\mathcal F \sim \bar{n}\),  demonstrating the metrological advantage provided by squeezing and nonclassical fluctuations. Monras~\cite{Monras2013} has further developed a general framework for phase-space formalism of Gaussian-state metrology, while \v{S}afr\'anek \emph{et al.}~\cite{Safrnek2015} have derived compact exact formulas for multimode Gaussian systems (see also \cite{rigovacca2017,marian2016,patra2026}).  In such cases, maximum QFI can be achieved via particular Gaussian measurements depending upon squeezing, displacement operations~\cite{Oh2019}. On the other hand, entangled Gaussian states enhance sensitivity beyond the shot-noise limit by redistributing fluctuations in phase space, while non-Gaussian resources can further increase the QFI through higher-order moments~\cite{Kwon2019}. Recent developments connect CV metrology with resource theories of nonclassicality, quantum correlations and geometric approaches to parameter estimation, unifying covariance-matrix methods with information-geometric bounds and multiparameter estimation frameworks~\cite{Sidhu2020,Fadel2025}. Experimentally, homodyne detection measure of rotated quadratures $\hat x_\phi=\hat x\cos\phi+\hat p\sin\phi$ has been conducted, enabling near-optimal readout of the encoded parameter in modern squeezed-light interferometric platforms~\cite{Aasi13,Grote2013,Li2014_a,mcintyre2024}.

Beyond single-mode sensing, multimode Gaussian states enable distributed and multiparameter quantum metrology.Generally, analytical expressions for the full QFIM and the compatibility conditions required for simultaneous optimal estimation which are derived in Ref.~\cite{rosanna2018}, establishing a general framework for Gaussian multiparameter metrology. In particular, the attainability of the quantum Cram\'er-Rao bound depends crucially on the commutativity of the optimal symmetric logarithmic derivative operators, making measurement compatibility a central issue absent in single-parameter estimation. A major practical advance is introduced through the concept of multiparameter squeezing \cite{Gessner2020}, which extends conventional spin squeezing to sensor networks estimating several phases, gradients, or spatially distributed fields simultaneously. Instead of minimizing fluctuations of a single collective observable, the relevant resource becomes the covariance matrix of multiple generators, and quantum enhancement is witnessed through
\begin{equation}
\xi_{\mathrm{multi}}^2
=
\frac{\bar{n}\,\mathrm{Tr}\left[\Xi^{-1}\right]}
{\mathrm{Tr}\left[\mathcal F\right]},
\end{equation}
where $\Xi$ is the covariance matrix of local generators. The condition $\xi_{\mathrm{multi}}^2<1$ signals genuine multiparameter quantum enhancement beyond the shot-noise limit.  

Theoretical developments in CV metrology have led to analytic tools for evaluating metrological performance of Gaussian states through fidelity-based methods and covariance-matrix approaches, which allow closed-form bounds on achievable precision even in multimode settings and noisy channels~\cite{Banchi2015}. These frameworks focus on direct application in realistic interferometric platforms, including gravitational wave detectors and large-scale optical sensors, where quantum correlations and measurement backaction collectively determine the ultimate sensitivity limits of CV measurements~\cite{Danilishin2012}. Recent works explore new directions such as distributed CV sensing and variational optimization of measurement strategies, demonstrating that squeezed resources and linear optical networks can approach Heisenberg limit for global parameter estimation tasks~\cite{ge2025,Nielsen2025}. In parallel, integrated photonic implementations and engineered squeezing sources are emerging as scalable platforms for continuous-variable metrology, enabling compact architectures for quantum-enhanced sensing and real-time parameter estimation~\cite{IntegratedSqueezing2025}. Together, these advances highlight how Gaussian-state engineering, symplectic dynamics, and optimized quadrature measurements form a unified framework for  CV quantum metrology.
Recently, there have been new theoretical and experimental developments  on distributed quantum phase and displacement sensing on CV systems with single and several number of modes, which is practically useful in multiparameter estimation ~\cite{Gatto19,Guo2019,oh2020,Sekatski20,zhuang2018,malitessta2023,ge2025,wang2020,liang2020,gramegna2021,alushi2024, ivanov2020,Chattopadhyay2025}.

\subsection{Atom-photon and cavity-QED sensors} 
\label{subsec:atom-photon}

Quantum optics and cavity quantum electrodynamics (cavity-QED) study the coherent interaction between quantized electromagnetic modes confined in a resonator and quantum emitters such as atoms, ions, artificial atoms, or spin ensembles. For sensing, the cavity-QED plays two complementary roles: (1) it enhances light-matter coupling and readout, (2) it can mediate effective interactions that can generate nonclassical collective states such as spin-squeezed states. Hence, atom-photon interactions in cavity-QED provide a highly versatile and controllable platform for quantum sensing.

In such systems, the coherent interaction between a quantized cavity mode and a two-level atom leads to enhanced sensitivity to external perturbations. Within the rotating-wave approximation, the dynamics is governed by the Jaynes-Cummings Hamiltonian,
\begin{equation}
H = \omega_a a^\dagger a + \omega_\sigma \sigma^\dagger \sigma
+ g \left( a^\dagger \sigma + a \sigma^\dagger \right),
\label{eq:jayens_cumming}
\end{equation}
where the coupling strength $g$ determines the rate of coherent energy exchange between atomic and photonic degrees of freedom. Small variations in system parameters, such as the cavity frequency $\omega_a$, atomic transition frequency $\omega_\sigma$, or the coupling $g$, are imprinted onto the quantum state of the coupled system through its time evolution. The resulting strong dependence of observables and QFI on these parameters enables high-precision estimation beyond classical limits~\cite{lu2022,Alushi2025}. Moreover, cavity QED platforms allow for the incorporation of dissipation, driving, and effective non-Hermitian dynamics, offering additional avenues for sensitivity enhancement via criticality, strong fluctuations, or spectral degeneracies~\cite{Beaulieu25,larson2024,Yu2025,LI2025_b}. These features make atom-photon cavity QED systems, promising candidates for quantum sensors of weak fields, frequencies, and forces. The light-matter interaction Hamiltonian in Eq.~(\ref{eq:jayens_cumming}) serves as the minimal building block for a broad family of cavity-QED models that play a central role in quantum sensing and critical metrology. By retaining the counter-rotating terms neglected under the rotating-wave approximation, one obtains the quantum Rabi model, which captures the full atom-photon interaction in the ultrastrong-coupling regime~\cite{rabi1936,rabi1937,braak2011}. Extending this single-atom description to $N$ collectively coupled two-level systems leads to the Dicke model,~\cite{dicke1954,hepp1973,wang_hioe_1973}. In the thermodynamic limit, this model undergoes a zero-temperature superradiant quantum phase transition at the critical coupling \(g_c=\frac{1}{2}\sqrt{\omega_a\omega_0}\), accompanied by the closing of the excitation gap, \(\Delta \sim |g-g_c|^{z\nu}\),  which underlies the divergent susceptibility and enhanced QFI near criticality \cite{emary_brandes_2003}. Within the rotating-wave approximation, the corresponding many-body extension yields the Tavis-Cummings model, while the single-atom limit reduces to the Jaynes-Cummings model~\cite{jaynes1963,tavis1968}. Although the Tavis-Cummings transition belongs to a distinct symmetry class due to excitation-number conservation, it also exhibits zero-temperature criticality and gap closing in the thermodynamic limit. These generalized cavity-QED Hamiltonians therefore provide a unifying framework in which light-matter hybridization, quantum criticality, and vanishing spectral gaps can be directly exploited for quantum-enhanced sensing of coupling strengths, frequencies, and external perturbations. 

Such presence of criticality and several distinguished physical phenomena in the cavity QED system has been explored for designing several quantum sensing protocols. It is shown that using a critical point of finite component phase transition, critical quantum optical probes can achieve a quadratic precision scaling with 
system size in a frequency-estimation protocol with the aid of Kerr nonlinearlity~\cite{garbe2020}. 
Further, criticality-induced quantum sensors are built by exploiting the ground states of the Jaynes-Cumming~\cite{chu2021},  
 Tavis-Cumming~\cite{zhu2024b}
 and the Dicke models~\cite{Gietka2022}.
 Other examples include cavity magnonic systems~\cite{ying2025}) and Rabi model with auxiliary systems in which globalized critical quantum metrology are executed to combat decoherence~\cite{Wang2024,Chen25b}.

 
Beyond equilibrium scenario, cavity QED platforms can also be used to obtained nonequilibrium sensing, especially in the presence of dissipation, which can effectively be described by non-Hermitian dynamics. In particular, exceptional point (EP)-enhanced sensing is  experimentally demonstrated in coupled optical microcavities, where tuning to an EP produces an anomalously large frequency response to nanoscale perturbations~\cite{Chen17} as also discussed in non-Hermitian quantum sensing (Sec. \ref{subsubsec:QMBcriticalsensors}). Independently, cavity-mediated generation and readout of entangled atomic ensembles has been used to realize metrological spin squeezing well below the standard quantum limit, enabling direct phase-sensitivity improvements in atomic clocks and inertial sensors \cite{Hosten2016, Wu2019}.  More recently, hybrid cavity QED readout of solid-state spins (NV ensembles) has demonstrated cavity-enhanced magnetic sensitivity that pushes sensitivities toward the PT regime, illustrating the applicability of cavity readout beyond optics to microwave and solid-state platforms~\cite{Wang2024}.  Finally, entanglement-enhanced atom interferometry and related cavity-assisted protocols have shown experimentally that cavity QED can both prepare and deploy nonclassical states for real sensing tasks, bridging the gap between state engineering and sensor operation \cite{Greve2022}.  Together, these demonstrations establish cavity QED as a versatile toolbox for quantum sensing: optical microcavities realize engineered spectral responses, high-finesse atomic cavities enable deterministic spin squeezing and quantum non-demolition readout, and hybrid microwave cavities extend these capabilities to solid-state spin ensembles and superconducting platforms, all with concrete experimental proofs of enhanced sensitivity.  In this sense, cavity-QED sensors do not only estimate cavity or atom-photon parameters; they also provide a mechanism for preparing and reading out nonclassical collective states, especially spin-squeezed ensembles, that improve atomic clocks and interferometers. The emphasis in this subsection is on light-matter coupling as a sensing mechanism and as a route toward cavity-enhanced state preparation and readout. Broadly, the concept of many-body sensors based on criticality, topology, localization, and collective correlations are discussed in the following section.

\subsection {Quantum many-body  (QMB) sensors}
\label{subsec:many-body-sensors}

In Sec.~\ref{sec:usefulness-of-QFI}, QFI was primarily employed as a diagnostic tool, revealing the structure of quantum states and signatures of many-body phenomena. In the present section, the perspective is reversed: the same quantum resources and physical phenomena are treated as tools for achieving enhanced parameter estimation. Quantum sensing has rapidly emerged as the most promising technology with potential implementation in several engineered quantum materials that are specifically designed for simulating various kinds of exotic QMB phenomena. Remarkable experimental progress over the past few decades has made such platforms increasingly accessible for the realization of many-body sensing protocols. Many-body systems offer distinct advantage, as collective quantum effects and correlations can be harnessed for enhancing the sensing performance in comparison to the uncorrelated probes, and thus help with designing a new class of robust sensing devices. The prime effort of the community is to examine various available platforms, such as ultacold atoms, trapped ions, atoms in cavity, that can be exploited for building quantum sensors tailored to specific metrological tasks.  Quantum sensing in many-body systems can be categorized in several broad directions, including criticality-based symmetry-breaking, topological and localization-delocalization based sensors.

\subsubsection{Critical quantum sensors: second-order, first-order, topological, localization} 
\label{subsubsec:QMBcriticalsensors}

Theoretical studies in many-body quantum sensing have primarily focused on quantum phase transitions, which occur in closed quantum systems at zero temperature. In particular, near criticality, the ground state undergoes a qualitative change driven by quantum fluctuations. In such settings, probe system is prepared near the quantum criticality, which is typically characterized by gap-to-gapless transition in the thermodynamic limit and long-range correlations leading to collective quantum effects.  The noncommutative nature of Hamiltonian terms triggers these transitions. A criticality based sensing is relatively recent emerging idea, starting with the works of Zanardi \emph{et al.}~\cite{Venuti07} where a scaling theory of QGT shows its singular behavior near criticality through fidelity susceptibility and Berry curvature as discussed in Sec.~\ref{subsec:QFI-indicator}. More specifically, let us consider a system with Hamiltonian, \(H=H_1+\chi H_2\) with \([H_1,H_2]\ne 0\) that undergoes a phase transition at $\chi = \chi_c$ with variation of the parameter \(\chi\). In the criticality based adiabatic quantum sensing, the ground state of \(H\) is typically used as the probe state for estimating the parameter \(\chi\). Near the criticality point \(\chi=\chi_c\), the QFI scales non-linearly with the system size \(N\), i.e., QFI \(\propto N^\beta\) with \(\beta>1\). See also Refs.~\cite{shi2024,hotter2024,cheng2025} in this regard.

\emph{Enhancement of sensing using ground state of Hamiltonian.} Considering the scenario of \(H=H_1+\chi H_2\) with \([H_1,H_2]\ne 0\), let the ground state of the Hamiltonian is given as \(\ket{\Psi(\chi)}\) which contain the information of the parameter \(\chi\). The maximum information that can be harnessed from the state is given by QFI,
\begin{equation}
   \mathcal{F}(\ket{\Phi(\chi)})=4(\langle \partial_\chi\Phi(\chi)|\partial_\chi\Phi(\chi)\rangle-|\langle\partial_\chi\Phi(\chi)|\Phi(\chi)\rangle|^2),
\end{equation}
which is a specific form of Berry curvature denoting the fidelity deviation due to small changes of the parameter of the Hamiltonian. Specifically, the Berry curvature is given as 
\begin{equation}
   \Lambda_{\chi\nu}=(\langle \partial_\chi\Phi|\partial_\nu\Phi\rangle-\langle\partial_\chi\Phi|\Phi\rangle\langle\partial_\nu\Phi|\Phi\rangle), 
\end{equation}
with \(\mathcal{F}(\ket{\Phi(\chi)})=4\Lambda_{\chi\chi}\). It has been established~\cite{Venuti07} that at the criticality, this geometric tensor scales extensively with the system size. In the following, we briefly discuss the proof~\cite{Zanardi08}.

Consider a family of Hamiltonians \(H(\lambda)=H_1+\chi H_2,\) with \(\ket{0}\) being its ground state with energy \(E_0\). The geometric tensor associated with the parameter \(\chi\) can be shown as
\begin{equation}
\Lambda \equiv \Lambda_{\chi\chi}
=
\sum_{n\neq 0}
\frac{|\langle 0|H_2|n\rangle|^2}{(E_n-E_0)^2}.
\label{eq:chi_def}
\end{equation}
Using the spectral representation,
Eq.~\eqref{eq:chi_def} can be written as
\begin{equation}
\Lambda
=
\int_0^{\infty} d\tau \, \tau \,
\langle H_2(\tau)H_2(0)\rangle_c,
\label{eq:chi_corr}
\end{equation}
where \(H_2(\tau)=e^{\tau H}H_2e^{-\tau H}\) and the subscript \(c\) denotes the connected correlator. Assuming that the perturbation \(V\) is extensive and can be written as
\begin{equation}
H_2=\sum_{i=1}^{N} h^i_2,
\end{equation}
where \(h^i_2\) is a local operator. Then \(\langle H_2(\tau)H_2(0)\rangle = \sum_{i,j} \langle h^i_2(\tau)h^j_2(0)\rangle.\) In the thermodynamic limit, the sums can be replaced by integrals,
\begin{equation}
\Lambda
\sim
\int_0^\infty d\tau \, \tau
\int_0^{N} dx
\int_0^{N} dx'\,
\langle h_2(x,\tau)h_2(x',0)\rangle.
\label{eq:chi_integral}
\end{equation}

At a quantum critical point described by the conformal field theory (CFT)-based universality class in \(d=1\), dynamical exponent \(z=1\)~\cite{Sachdev11}. Then the scale invariance implies that the two-point function of a local operator with scaling dimension \(\Delta_v\) behaves as
\begin{equation}
\langle h_2(x,\tau)h_2(0,0)\rangle
\sim
\frac{1}{(x^2+\tau^2)^{\Delta_v}}. {}
\label{eq:scaling_corr}
\end{equation}
Substituting Eq.~\eqref{eq:scaling_corr} into Eq.~\eqref{eq:chi_integral}
and introducing the relative coordinate \(r=x-x'\),
the integral over the center-of-mass coordinate produces a factor \(L\), yielding
\begin{equation}
\Lambda
\sim
N
\int_0^\infty d\tau \, \tau
\int_0^{N} dr \,
\frac{1}{(r^2+\tau^2)^{\Delta_v}}.
\label{eq:chi_scaling_int}
\end{equation}
Under the scale transformation \(r\to b r\), \(\tau\to b\tau\), the integral in Eq.~\eqref{eq:chi_scaling_int} scales as \(\chi \sim N \, b^{3-2\Delta_v}.\) Choosing \(b\sim N\), one finds
\begin{equation}
\Lambda \sim N^{4-2\Delta_v}.
\label{eq:general_scaling}
\end{equation}
For a relevant perturbation driving the quantum phase transition, the scaling dimension satisfies \(\Delta_v = d+z-\frac{1}{\nu}.\) In one dimension (\(d=1\)) with \(z=1\) and correlation-length exponent \(\nu=1\), one has \(\Delta_v=1\). Substituting into Eq.~\eqref{eq:general_scaling} yields
\begin{equation}
\Lambda \sim N^2.
\end{equation}

The scaling of Fisher information at a continuous quantum phase transition can be directly connected to the divergence of the correlation length. If the estimated parameter is the relevant coupling driving the transition, for which $\zeta=|\lambda-\lambda_c|^{-\nu}$, where $\lambda_c$ is the critical point and the associated scaling exponent $\nu$ dictates the rate of divergence. Then finite-size scaling implies that the Fisher information scales as $\mathcal{F}=N^{2/\nu}$, i.e., as $\mathcal{F}=N^{2}$, when $\nu=1$.

This results highlights the role of criticality in quantum sensing and provides an universal result which governs the parameter estimation in a single universal class of Hamiltonian. For example, one can consider Ising classes, for which the model is given by 
\begin{equation}
    H= {h\sum_{i=1}^{N}\sigma_z^i}+\frac{J}{4}\left[\sum_{i=1}^{N-1}(1+\gamma)\sigma_x^i\sigma_x^{i+1}+(1-\gamma)\sigma_y^i\sigma_y^{i+1}\right],
\end{equation}
where \(h\) is the strength of the magnetic field and \(J\) is the coupling between the spins. For convenience, it's customary to introduce the dimensionless parameter $\chi = h/J$, and then the Hamiltonian can be expressed in the general form of \(H=H_1+\chi H_2\) where \(H_2\) represents local magnetic field, while \(H_1\) represents the interaction between the spins. This model belongs to the Ising class and can be solved analytically via a standard Jordan-Wigner
transformation~\cite{LSM_main,barouch_pra_1970,barouch_pra_1971,santoro_ising_beginners_2020}, which maps spin operators to fermionic ones,
\begin{equation}
\hat{c}_i =
\left(\prod_{j=1}^{i-1} -\hat{\sigma}^z_j\right)
\hat{\sigma}^-_i,
\qquad
\hat{c}^\dagger_i =
\left(\prod_{j=1}^{i-1} -\hat{\sigma}^z_j\right)
\hat{\sigma}^+_i,
\end{equation}
where $\hat{\sigma}^{\pm} = (\hat{\sigma}^x \pm i\hat{\sigma}^y)/2$.
The QFI  associated with the parameter
$\chi$ has been computed analytically for the ground state~\cite{Venuti07,zanardi_2}, and is shown to scale quadratically with
the system size as
\begin{equation}
\mathcal{F}(\ket{\Phi(\chi)}) \sim N^2,
\end{equation}
in the vicinity of the critical point $h/J \to \pm 1$. The critical point is marked by the closing of the energy gap, and near to it, the system becomes extremely sensitive to even a tiny variations in the underlying Hamiltonian, resulting in an enhanced sensitivity. 

Criticality-assisted metrology has additionally been investigated in several ultracold-atom platforms~\cite{Skotiniotis15,Montenegro21,Szigeti2021,Geiger2020,Xu2025}, as well as in cavity-QED systems~\cite{He2023_b,Su2024}, highlighting the feasibility of these ideas in current experimental architectures. Broadly, many-body quantum critical sensing exploits the enhanced distinguishability of many-body quantum states in the vicinity of a phase transition to estimate a control parameter with high precision. This enhancement is commonly quantified through the QFI, the fidelity susceptibility, or closely related quantum-geometric quantities~\cite{Venuti07,zanardi_2,Zanardi08,inverizzi2008,frerot2018,rams2018,montenegro2024review,Agarwal2025,Alushi2025}.

In spin-chain platforms, the paradigmatic examples are transverse-field Ising and anisotropic XY models, where the parameter to be sensed is typically the transverse field, the anisotropy, or a dimensionless coupling such as 
$J/h$~\cite{inverizzi2008,ye2016,Liu2016_a,frerot2018}. Moreover, investigation with control-parameter uncertainties near criticality reveals a trade-off between robustness and quantum-enhanced sensitivity in realistic critical sensors~\cite{Mihailescu25}. These studies show that the QFI may become super-extensive near the quantum critical point, allowing sensitivities beyond the standard shot-noise scaling. Extensions to less idealized settings, including minimally accessible anisotropic spin chains, modular construction that introduces periodic coupling defects into the chain, and boundary-driven nonequilibrium chains, further indicate that critical enhancement can survive under restricted measurements, or real-time feedback (adaptive protocols) open-system conditions~\cite{marzolino2017,Adani24,Mukhopadhyay24,salvatori2014,pezze2019}.

For Bose-Hubbard (BH) and related models, the relevant transition is usually the superfluid to Mott-insulator transition, with the hopping-to-interaction ratio, the chemical potential, or the onsite interaction serving as the control parameter~\cite{Carrasquilla2013,Lacki2014,Lacki2016}. Fidelity-susceptibility and QFI-based diagnostics can then use near the transition, that allows precise estimation of the control parameter driving the system across the phase boundary. Related cold-atom proposals also demonstrate how multipartite entanglement and QFI can be accessed through experimentally measurable dynamical susceptibilities~\cite{hauke2016}. There are several protocols within the framework of the BH model for  generating metrologically useful multiparty quantum states, such as GHZ-like and N00N states, that can be exploited in the estimation of the parameters, like phase and potential~\cite{sharma2011,Watanabe2012,eichler2014,Compagno2017,Yannouleas2019,Booker2020,Shao2020,koscielski2023,wittmann2023,Pelayo23}. BH systems have been proposed for enhance sensing via spin sqeezing that involve  fast-adiabatic-like preparation and optimal control theory~\cite{juli2012,lapert2012,yuste2013,marcin2020,Dziurawiec2023}. Moreover, recent studies suggest BH systems as an promising platform for temperature sensing~\cite{carcy2021,suthar2021,suther2022,Pohl2022}. 

For Fermi-Hubbard (FH) and their generalized versions in the fermionic lattice systems~\cite{Arovas2022,nat2013,Tasaki1998}, the same principle applies and the estimated parameters are typically interaction strength, filling-dependent control parameter, magnetic field, or coupling that drives a fermionic critical point. Recent works on quantum-critical metals and strange-metal regimes further indicate that QFI and related susceptibilities can reveal enhanced multipartite correlations, suggesting a route toward metrological protocols based on strongly correlated fermions~\cite{Sajna2020,Mamaev2020,fang2025,balut2025,costa2021,Laurell22,Kaczmarek23,Mirani24,Bippus25}. Oplical lattice based quantum simulators of the FH model can be used for enhancing sensing of atomic clocks via spin squeezing techniques involving one-axis twisting and two-axis counter-twisting~\cite{He19,Yanes22}. Moreover, FH systems have been shown to have useful applications in thermometry~\cite{Wu03,Hartke2020,Shen23,Pasqualetti24}.

Finally, for systems governed by a CFT or universal critical description, the emphasis is on scaling rather than on microscopic details. The parameter to be estimated is the relevant perturbation measuring the distance from criticality, such as a mass term, magnetic field, or coupling. In this setting, the scaling of the QFI or fidelity susceptibility is controlled by universal data, including scaling dimensions, correlation-length exponents, and finite-size scaling forms~\cite{Venuti07,zanardi_2,Gu2010}. Thus, the metrological enhancement reflects the universality class of the transition rather than the microscopic details. Beyond the Ising universality class, a wide range of other spin models have been investigated, where critical scaling has also been shown to enhance parameter estimation and topological phase transitions~\cite{lambert2020,sarkar2022}. In contrast, for multi-parameter quantum estimation, the role of criticality remains largely unexplored. However, recent results indicate that multicritical points may enable super-Heisenberg precision scaling~\cite{Mondal25b}. Apart from quantum magnetometry, quantum criticality play role in enhancing phase thermometry~\cite{yu2024}. Interestingly, there also exist scenarios in which criticality does not directly improve precision; nevertheless, appropriately engineered adiabatic protocols can suppress excitations, thereby accelerating the evolution and giving rise to exponential super-Heisenberg scaling~\cite{cheng2025}. 
\begin{figure}
 \includegraphics[width=\linewidth]{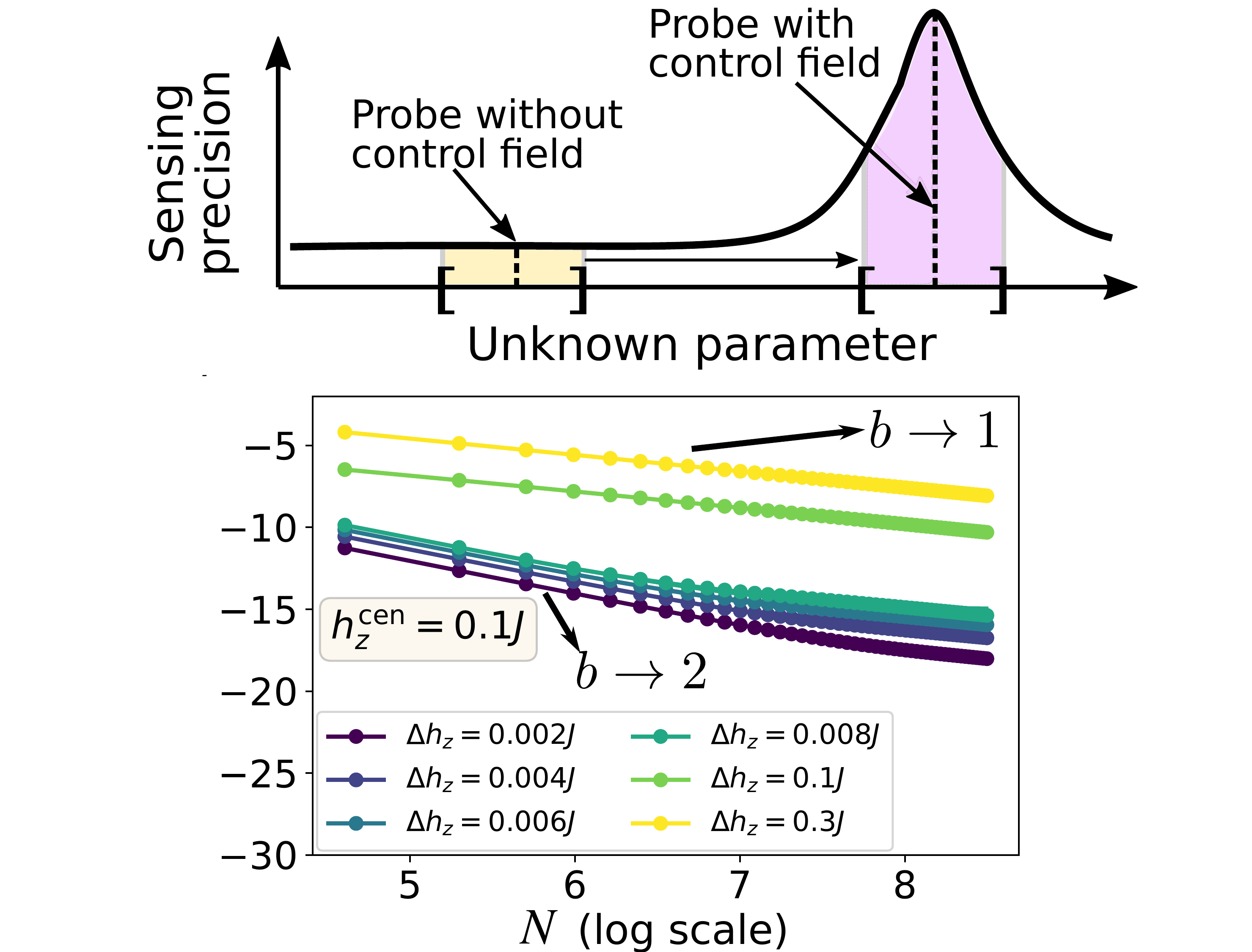}
  \caption{ The upper panel schematically illustrates the central idea of Ref.~\cite{Montenegro21}, where an external control field is tuned to shift the sensing region of the phase diagram toward the vicinity of a quantum critical point. This strategy enables enhanced parameter sensitivity even when the exact value of the parameter to be estimated is unknown and only a finite sensing interval is available \emph{a priori}. The lower panel presents the scaling behavior of \(g(B_z^\ast)\) with the system size \(N\) for the transverse-field Ising model, where \(h_z\) denotes the parameter of interest and \(B_z\) acts as the control field. The numerical results are fitted using the scaling form \(aN^{-b}+c\) for different sensing intervals \(\Delta h_z\), demonstrating how the optimal control field adapts with system size to maintain near-critical enhanced sensitivity.}
  \label{fig:montengro_2021}
\end{figure}
In addition, as an application of the global parameter estimation framework discussed in Sec.~\ref{subsec:technique}~\cite{Montenegro21}, the field strength of a transverse-field Ising model with $N$ sites is estimated by controlling another field strength. The system is described by the Hamiltonian
\begin{equation*}
\label{TI}
H_{\mathrm{TI}} = \sum_{l=1}^N J \sigma_x^l \sigma_x^{l+1} - \sum_{l=1}^N (B_z + h_z) \sigma_z^l,
\end{equation*}
where $J$ denotes the coupling constant, while $h_z$ and $B_z$ represent the parameters associated with the external transverse magnetic field. The transverse-field Ising model is known to exhibit a quantum phase transition, and maximum sensitivity can be achieved around the critical point. To estimate the single parameter $h_z$ in the Hamiltonian, the control field $B_z$ is used for which only partial prior information is known in the form of a known range $\Delta h_z$, which is far from the critical region. By suitably tuning the control parameter $B_z$, the sensing range $\Delta h_z$ can be shifted to lie almost symmetrically around the critical point, leading to the quantum-enhanced sensitivity, even when the true value of $h_z$ is far from the critical point and only the parameter's range is known (see Fig.~\ref{fig:montengro_2021} for the detail description).  Furthermore, this analysis of global multiparameter estimation is extended by incorporating a longitudinal field into the transverse-field Ising Hamiltonian and demonstrates that a similar advantage can be achieved in this case as well~\cite{Montenegro21}.

Another interesting application of critical behavior at phase transitions appears in quantum thermometry, where criticality is exploited to enhance temperature estimation~\cite{Mehboudi19a,Aybar22}. Most criticality-assisted metrology approaches use quantum critical systems as probes for estimating Hamiltonian parameters by preparing the system near its quantum critical point and performing measurements~\cite{Zanardi08,frerot2018}. Because the susceptibility diverges close to criticality, the sensitivity to parameter variations can be significantly enhanced~\cite{Zanardi08,frerot2018}. However, these methods suffer from drawbacks such as critical slowing down, which leads to diverging state-preparation times, and the requirement of global multiparticle measurements, which are especially very difficult to implement in strongly interacting QMB systems.
To overcome these drawbacks, a noninvasive thermometry scheme has recently been proposed~\cite{yu2024}. In this approach, an additional spin probe is embedded in the sample and interacts uniformly with a controlled subset of neighboring spins. Through this interaction, the probe undergoes temperature-dependent dephasing and effectively acts as a phase-sensitive probe. The temperature of the sample can then be inferred by measuring the time evolution of the probe coherence. Importantly, this noninvasive strategy allows repeated measurements on the same thermal sample, thereby alleviating the problem of long equilibration times required for preparing the system near criticality.
Using this framework, noninvasive quantum thermometry~\cite{yu2024} has been investigated for a two-dimensional Ising model~\cite{Friedli17} at zero field near its critical phase transition point, which can be described through several physical platforms including arrays of Rydberg atoms~\cite{Browaeys16,Gross17,Bernien17,Barredo18}, nitrogen-vacancy-center systems~\cite{Jelezko04a,Jelezko04b,Robledo11,Abobeih18,Degen21,Gulka21}, and spin-polarized Fermi gases~\cite{Cetina15,Cetina16}. Further, the temperature sensitivity can be critically enhanced near the phase transition, with the achievable precision depending on the probe–system coupling range and the interrogation time~\cite{yu2024}.

\emph{Topological phases as quantum sensor.} A coherent progression in topological quantum metrology has recently been investigated via both adiabatic and dynamical approaches. The first key step was the demonstration that topological phase transitions in free-fermionic systems, such as the SSH chain and Chern insulators, can yield quantum-enhanced sensing despite the absence of symmetry breaking and long-range quantum correlations~\cite{sarkar2022}. In these systems, the singular enhancement of the QFI near the phase boundary identifies bulk gap closing as the essential source of metrological gain. This static criticality-based perspective is subsequently extended to nonequilibrium dynamics through topological quantum walks with localized defects, where the evolution time itself acts as a sensing resource~\cite{tong2025}. By exploiting the propagation dynamics in the topologically nontrivial regime, the estimation precision of the defect strength can be shown to approach the Heisenberg limit while remaining highly robust against disorder. Most recently, this line of work has been further generalized through adiabatic control of topological edge states~\cite{he2025}. In this protocol, sharply localized zero modes are adiabatically transformed into delocalized states, yielding a QFI scaling $\mathcal F\sim N^{2p}$ (with \(N\) being the lattice size) determined by the order $p$ of band touching. Furthermore, when \(L\) number of multipartite entangled edge states are employed, the scaling is amplified to $\mathcal F\sim L^2N^{2p}$, thereby unifying edge-state robustness, higher-order criticality, entanglement, and adiabatic control into a scalable topological sensing framework. For example, in the simultaneous multiparameter estimation of the pairing term and the onsite potential in the one-dimensional topological Kitaev model, multicriticality enables super-Heisenberg scaling $(\sim N^6)$ within a narrow parameter regime, arising from the coexistence of symmetry breaking and gap closing~\cite{Mondal25b}.

\emph{Localization-delocalization based quantum sensor.} A unified localization--delocalization metrological framework has recently been developed in which Stark, quasiperiodic, kinetic grading and hybrid localization transitions act as highly sensitive resources for weak-field sensing. It is shown that disorder-free Stark localization can support super-Heisenberg QFI  scaling in the weak-field extended regime, \(\mathcal{F}(F)\sim N^\alpha, \alpha>2\), with the single-particle probes exhibiting $\alpha\simeq 6$ in the weak-field limit, while thermal probes reduce the scaling toward the Heisenberg limit~\cite{he2023,Yousefjani23}. This static eigenstate-based sensing paradigm is subsequently generalized through experimentally accessible operator-based approaches, where the operator Fisher information (OFI),
\begin{equation}
\mathcal{F}_O(F)=
\frac{\left(\partial_F\langle O\rangle\right)^2}{(\Delta O)^2},
\qquad 
\mathcal{F}_O\le \mathcal{F},
\end{equation}
is shown to reproduce the universal scaling of the full QFI using realistic observables such as charge-density-wave imbalance, density modulation, and wave-packet width across both Stark and quasiperiodic localization transitions.
It can be extended in the dynamical realm~\cite{Sahoo24b,Sahoo25}. Another conceptual advancement is addressed via nonequilibrium localization metrology, where Bloch-oscillation and dynamical Stark protocols is shown to reveal the universal scaling \(\mathcal{F}(t)\sim t^2N^\beta,\) demonstrating simultaneous quadratic enhancement in time and superextensive scaling in system size.  In this situation, the dynamical quantum advantage survives even in the presence of moderate dephasing and admits an effective non-Hermitian description for noisy Stark probes~\cite{Sarkar2025}. More recently, the sensing capacity of Stark probes has been further enhanced by nonlinear gradient encoding, where the nonlinearity itself acts as an additional critical resource in both single-particle and interacting systems~\cite{Yousefjani25}. Further, the interplay between multiple localization-inducing criticalities, such as Stark and Aubry-Andr\'e-Harper potentials, has been shown to yield even stronger scaling than pure Stark probes, while analogous localization-enhanced sensing signatures have now been extended to interacting BECs in tilted optical lattices through directly measurable cloud-width and fidelity observables \cite{debnath2026}. Another recent work shows that an infinitesimal kinetic grading can drive a localization transition with a diverging localization length and universal finite-size as well as Kibble–Zurek scaling, thereby identifying a sharp critical resource in an inhomogeneous lattice system. For sensing, the central message is that tuning near this localization criticality can strongly amplify parameter sensitivity through the enhanced ground-state response, making graded-hopping systems a promising platform for critical quantum metrology~\cite{Debnath25b}. Together, these works establish localization physics as one of the most versatile routes toward experimentally viable quantum-enhanced sensing. 

\emph {Metrological resource for critical sensors.} The metrological resource underlying a critical sensor depends on the nature of the transition. In conventional Landau-type second-order quantum phase transitions, enhanced sensitivity is usually associated with a closing of excitation gap together with the onset of an order parameter and symmetry breaking. Localization-delocalization transitions, by contrast, do not generally involve Landau symmetry breaking; their enhanced QFI or fidelity susceptibility is often accompanied by the closing of relevant energy gap~\cite{he2023,Sahoo24b}. Similarly, topological transitions are characterized not by a local order parameter but by a change in the topological structure of the state or band, usually enabled by gap closing~\cite{sarkar2022}. Ref.~\cite{Mondal25b} has, however, recently reported that such gap-closing is always not necessary, and symmetry breaking can be an independent resource resulting in Heisenberg limit without invoking gap closing in the one-dimensional Kitaev model.

\emph{Effect of \(k\)-body interaction in quantum sensing.} Although two-body interactions remain a central focus of majority of the works, the scaling of the QFI can be increased
by taking higher-body couplings into effect, e.g., the work by Boxio \emph{et.al.}~\cite{Boixo07}, which considers the encoding of the parameter through a \(k\)-body Hamiltonian. Beyond HL, a super-Heisenberg scaling emerges in a parametrized Hamiltonian including \(k\)-body nonlinear interaction, which scales as \(\sim N^k\) with \(k>1\). Consider a probe consisting of $N$ subsystems, evolving under a Hamiltonian, given by \(H=H_1+\chi H_2\), where $H_1$ is independent of the parameter and the parameter generator $H_2$ is a sum of genuine $k$-body terms,
\begin{equation}
H_2=\sum_{i_1<\cdots<i_k} h_{i_1\cdots i_k}.
\end{equation}
Each operator $h_{i_1\cdots i_k}$ acts nontrivially only on the specified $k$ subsystems. The total number of such terms is \(\binom{N}{k}\sim \frac{N^k}{k!}\). Its spectral range (seminorm) is bounded,
\begin{equation}
\|g_{i_1\cdots i_k}\|_s \le c,
\end{equation}
where  $c$ is a constant that is independent on $N$.  Let the probe be prepared in a pure initial state $\ket{\psi_0}$ and it evolves for time $t$ under the unitary \(U(\chi,t)=e^{-itH(\chi)}.\) The output state is
\begin{equation}
\ket{\psi(\chi)}=U(\chi,t)\ket{\psi_0}.
\end{equation}
For pure states, the QFI  associated with estimation of the parameter $\chi$ is \(\mathcal F = 4t^2(\Delta G)^2,\) where \(G\) is given as 
\begin{equation}
    G=\sqrt{\langle H(\chi)^2\rangle-\langle H(\chi)\rangle ^2}.
    \label{eq:FQvariance}
\end{equation}
The variance of $G$ is bounded by its seminorm, \((\Delta G)^2 \le \frac{1}{4}\|G\|_s^2,\) where \(\|G\|_s = g_{\max} - g_{\min}\) is the difference between the largest and smallest eigenvalues of $G$. Substituting into Eq.~\eqref{eq:FQvariance}, the general bound is obtained as
\begin{equation}
\mathcal F \le t^2 \|G\|_s^2.
\label{eq:FQseminorm}
\end{equation}
The seminorm of $G$ is bounded above by the sum of the seminorms of the
individual $k$-body terms:
\begin{equation}
\|G\|_s
\le
\sum_{i_1<\cdots<i_k}
\|g_{i_1\cdots i_k}\|_s
\le
c \binom{N}{k}.
\end{equation}
Hence, \(\|G\|_s = O(N^k),\). This scaling is achievable by probe states that place support on the extremal eigenstates of $G$. Therefore, the QFI scales as \(\mathcal F = O(t^2 N^{2k}).\) The quantum Cram\'er-Rao bound implies that for an unbiased estimator,
\begin{equation}
\delta\chi \ge \frac{1}{\sqrt{\mathcal F}} 
\sim \frac{1}{t N^{k}}.
\end{equation}
Thus the ultimate precision limit for estimating $\lambda$ scales as \(\delta\chi_{\min} \sim \frac{1}{N^{k}}.\)

\qed

This scaling cannot be improved by adaptive measurements, collective POVMs, or the inclusion of arbitrary parameter-independent interactions, since none of these operations modifies the spectrum of the generator $G$ nor increases its seminorm. Therefore the bound is fundamental and tight. In the work by Boixo \emph{et. al} provided bounds on the QFI corresponding to a multibody encoding Hamiltonian, where the bound is attainable using a specific genuine multipartite entangled state with the assumption that all eigenvalues of each single-body operator are non-negative~\cite{Boixo07,Boixo08}. Later Bhattacharyya \emph{et al.} considered \(k\)-body Hamiltonian with negative eigenvalues that leads to a distinction in the nature of the optimal probe for odd- and even-body interacting encoding Hamiltonians~\cite{Bhattacharyya25}. In particular, the authors find that for odd values of \(k\), the optimal input probes need to necessarily possess genuine multiparty entanglement (GME) which is not true when \(k\) is even. Also, the usefulness of such nonlinear interaction is observed in several other works~\cite{li2023a,imai2025}. On the other way around the bound on QFI can provide the presence of maximum possible \(k\)-body interaction in the Hamiltonian, which is explored in~\cite{cieifmmonde2024}. Moreover, a one-dimensional interacting Kitaev chain can exhibit super-Heisenberg scaling, which can be re-expressed  as a \(k\)-body interactive model in the spin basis~\cite{Yang22} and the effect of \(k\)-body Lindblad operator has been explored in Ref.~\cite{Beau17}.  

\emph{Environment-affected quantum sensing.} In realistic quantum sensing protocols, the probe system is inevitably coupled to its surrounding environment, and this interaction strongly influences the achievable estimation precision. While ideal noiseless quantum metrology predicts Heisenberg limit for suitably entangled probes of size $N$, decoherence typically degrades this enhancement and restores the standard quantum limit, \(\Delta \chi \geq {1}/{\sqrt{N}}\), or even worse under strong dissipation and dephasing \cite{Escher11,Rafal12}. Since the QFI, \(\mathcal{F}[\rho_\theta]\) depends sensitively on the competition between coherent parameter encoding and dissipative dynamics, environmental effects fundamentally reshape metrological scaling laws.

A major advance in this direction was the realization that noise does not always simply destroy quantum advantage. In particular, Demkowicz-Dobrza\'nski and Maccone have shown that in noisy metrology, entanglement-assisted strategies can outperform sequential unentangled protocols even when both are equivalent in the noiseless case and that passive auxiliarys may become genuinely useful resources~\cite{Rafal14}. This result was later clarified through the extended convexity of QFI, where the total Fisher information was decomposed into quantum and classical contributions arising from noisy mixtures, providing a unified framework for open-system parameter estimation~\cite{Alipur14}. Related studies demonstrated that auxiliary systems can enhance precision under amplitude damping, depolarization, and correlated dephasing channels, establishing the operational usefulness of entanglement-assisted metrology in realistic environments~\cite{Lu15,Tan2015,Rossi2015,Nichols2016} and such entanglement-assisted metrology under dissipation has been verified in a photonic experimental setup~\cite{wang2018}.

Beyond passive protection, active monitoring of the environment can itself restore quantum advantage. Time-continuous measurements of emitted photons or quantum trajectories allow one to condition the probe dynamics on measurement outcomes, thereby recovering lost Fisher information and even restoring Heisenberg scaling in noisy metrology \cite{Albarelli2017,Albarelli2018_b,Cortez2017}. In particular, continuous monitoring of the environment has been shown to recover optimal precision bounds that would otherwise be inaccessible from the reduced density matrix alone, revealing that part of the metrological information is stored in the environment itself \cite{Albarelli2018}. This idea naturally connects to environment-assisted sensing protocols, where surrounding auxiliary spins are used not as a source of decoherence but as signal amplifiers. Such schemes, originally proposed for NV-center magnetometry and trapped-ion clocks, exploit the coupling between the probe and nearby environmental spins to achieve nearly Heisenberg limited sensitivity enhancement \cite{Goldstein2011,Cappellaro2012}. An introduction of general bounds for the parameter estimation error in nonlinear quantum metrology of QMB open systems in the Markovian limit has been proposed in Ref.~\cite{Beau17}. The authors show that given a \(k\)-body Hamiltonian and \(p\)-body Lindblad operators, the estimation error of a Hamiltonian parameter using a GHZ state as a probe scale as \(N^{-[k-p/2]}\), surpassing the shot-noise limit for \(2k>p+1\). Also, there have been numerous works on environmental effects on quantum sensing protocol~\cite{Ozaydin2014,Huang2015,Nichols2016,Li2017,Wang2017,Saleem2023}

Another important direction concerns non-Markovian dynamics and critical open systems. Memory effects and information backflow can temporarily increase distinguishability between nearby quantum states, leading to enhanced QFI and improved estimation precision compared to purely Markovian evolution \cite{Chin12,Macieszczak15}. Similarly, dissipative phase transitions and non-Hermitian criticality can strongly amplify susceptibility because of the closing of the Liouvillian gap, generating enhanced sensing near exceptional points and dissipative steady-state transitions in two-mode system \cite{Bu2023}. These results show that dissipation can itself become a metrological resource rather than merely a limitation.

\emph{Non-Hermitian system as sensors.} Non-Hermitian quantum systems provide a fundamentally new perspective on concepts traditionally formulated within Hermitian quantum mechanics, including quantum phase transitions, topological phases, and the role of symmetry \cite{gong_prx_2018,kawabata_prx_2019,chen_njp_2019,Ueda_review,bergholtz2021,okuma2023}. Over the past two decades, these systems have attracted considerable interest because effective non-Hermitian dynamics naturally arise in several physical contexts. For instance, non-Hermitian Hamiltonians can emerge through an effective description of a subsystem embedded inside a larger, ordinary Hermitian quantum system~\cite{naimark_dilation_2008}, or by introducing a skew-Hermitian component that effectively captures gain and loss of particles or energy in open systems \cite{brody_prl_2012,liu_prap_2020}.  Furthermore, when a quantum system interacts weakly with a noisy environment, its reduced dynamics is well described by the GKSL master equation \cite{open_quan_book,lidar_2020_lecture}. Within this framework, non-Hermiticity naturally appears through an effective non-Hermitian Hamiltonian governing the no-jump evolution \cite{ohlsson_pra_2021,khandelwal_prx_2021,chen_prl_2022,nakanishi_pra_2022}, accompanied by quantum jump operators that account for the continuous monitoring of the system by its surrounding environment \cite{minganti_pra_2020,fleckenstein_prr_2022}. Recent experimental progress across a variety of physical platforms, including photonic systems \cite{ozdemir_np_2019}, electrical circuits~\cite{zhang2023_a} and superconducting circuits \cite{chen_prl_2021}, has enabled the controlled implementation of non-Hermitian dynamics in laboratory settings. 

Non-Hermitian systems have also been developed as a promising platform for quantum sensing, largely due to the presence of exceptional points (EPs), which represent spectral degeneracies unique to non-Hermitian Hamiltonians. At an EP, both eigenvalues and eigenvectors simultaneously coalesce~\cite{bender_ropp_2007}, resulting in unusual spectral and dynamical properties. In quantum metrology, EPs have attracted significant attention because the QFI  may display enhanced scaling in the vicinity of the EP \cite{liu_prl_2016,Chen17,Lau2018,McDonald2020Oct,Chu2020}. The first sensing protocol based on EP physics was proposed in Ref.~\cite{jan2014} and later experimentally realized using optical microresonator platforms~\cite{Chen17}. These pioneering studies triggered extensive theoretical and experimental investigations demonstrating improved sensing performance in non-Hermitian systems compared to their Hermitian counterparts \cite{ep_sensing_1,chen_njp_2019,Lau2018,Hodaei17,ep_sensing_5,naghiloo_nature_ep_sensing_2019}.

Despite these advances, the validity of QFI enhancement near EPs within the framework of critical quantum metrology remains an active subject of debate~\cite{ding_prl_2023,Wiersig26}. In particular, it has been shown that enhanced sensing performance does not necessarily require the presence of EPs and can also occur in non-Hermitian systems away from spectral degeneracies~\cite{Budich20,bao_prap_2022,Sarkar2024Jul,Agarwal2025,agarwal2026}. An especially striking phenomenon appears in non-Hermitian many-body systems, where the QFI can scale exponentially with system size \cite{McDonald2020Oct,koch2022}, indicating a dramatically amplified response to parameter perturbations compared to conventional Hermitian many-body systems. Such enhanced sensitivities have already been observed experimentally across a range of platforms, including superconducting qubit systems~\cite{Chen17,Guo21,10.1063/5.0168372,Xiao24,Yu24,Parto2025Jan,wu2025,Han25,claverorubio2025vibrationalparametricarraystrapped}, among others. 

Non-Hermitian quantum systems have been emerged as one of the most promising platforms for quantum-enhanced sensing, owing to their intrinsically amplified response to weak perturbations. The earliest developments focused on sensing near exceptional points (EPs), non-Hermitian degeneracies where both eigenvalues and eigenvectors coalesce. Near an EP of order $n$, the spectral response to a perturbation $\epsilon$ follows the characteristic nonanalytic scaling
\begin{equation}
\Delta E \sim \epsilon^{1/n},
\end{equation}
which motivated the first generation of EP-enhanced sensing proposals in $\mathcal{PT}$-symmetric cavities and optical microresonators \cite{liu2016,hodaei2017,chen2017}. These ideas were later experimentally validated in optical resonators and dissipative qubits, firmly establishing EPs as a practical sensing resource \cite{chen2017,naghiloo2019}. However, subsequent studies clarified that the apparent divergence of spectral susceptibility near EPs does not automatically translate into unlimited quantum advantage, since quantum and classical noise impose stringent bounds on the achievable precision \cite{Lau2018,wiersig2020}. A major conceptual shift occurred with the realization that non-Hermitian quantum sensing does not fundamentally rely on EPs. In particular, universal sensing enhancement can arise directly from the intrinsic nonorthogonality of non-Hermitian eigenstates even away from spectral singularities \cite{Xiao24,yang2026}. This broadened the scope of non-Hermitian metrology from EP-centric protocols to generic open-system sensing architectures, including dissipative photonic interferometers and time-modulated non-Hermitian probes. A second transformative direction concerns non-Hermitian topology and the non-Hermitian skin effect (NHSE). In topological non-Hermitian sensors, the boundary sensitivity of edge or skin modes leads to exponentially amplified response,
\begin{equation}
\mathcal{F} \sim e^{\alpha N},
\end{equation}
where $N$ is the system size and $\alpha$ is determined by the non-Bloch localization length \cite{Budich20,koch2022}. This mechanism has been extended to fully quantum-noise-limited topological sensors (QUANTOS), where the Fisher information inherits the exponential scaling of the boundary mode sensitivity in the topologically nontrivial phase. The experimental realization of NHSE-induced topological switching in electrical circuits further demonstrated that nonreciprocal boundary accumulation can be harnessed as a robust sensing resource \cite{zhang2023_a,McDonald2020Oct,koch2022}. Likewise, ultracold-atom experiments observing non-Hermitian topology and skin effects provide a highly controllable route toward quantum simulation of NH sensors \cite{guo2021}. Such exponentially enhanced sensing has now been explored in superconducting circuits, dissipative spin systems, and NV-center platforms, significantly expanding the experimental relevance of non-Hermitian metrology \cite{Wang2026}.


This progression highlights how the field has evolved from spectral singularities to a much wider framework based on nonorthogonality, boundary localization, topology, and many-body amplification, making non-Hermitian systems a uniquely versatile platform for precision quantum technologies.

\begin{figure}
    \centering
    \includegraphics[width=\linewidth]{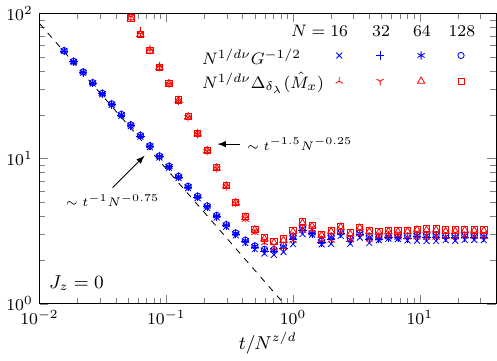}
    \caption{Time dependence of the QFI at the critical point of the XXZ model for a small perturbation in the external magnetic field~\cite{rams2018}. The evolution time is rescaled with the system size, while the QFI (blue symbols) is normalized according to the adiabatic-limit scaling. In the short-time regime, the QFI follows the scaling \( G_{1/2}\sim t\,N^{1-h/(2d)},\) reaching Heisenberg scaling, as indicated by the dashed line. The corresponding estimation precision obtained from measurements of the transverse magnetization operator \(\hat{M}_x\) is shown by the red symbols. For short evolution times, the magnetization-based estimation becomes suboptimal, exhibiting the scaling \( \Delta_{\delta\lambda}(\hat{M}_x;\lambda_c,t)\sim t^{-3/2}N^{-1/4}, \) which nevertheless remains below the shot-noise limit with respect to the system size \(N\). In the long-time regime, \(t\gg N^{z/d}\), both the QFI and the magnetization-based precision approach the expected adiabatic limit.}
    \label{fig:time_scaling}
\end{figure}


\subsubsection{Time as resource in parameter estimation} 

In many-body sensors, time is not merely an interrogation duration butcan itself serve as an active metrological resource through nonequilibrium evolution, critical dynamics, periodic driving, and long-lived temporal order. In the following, we review various aspects of the dynamical sensing protocols.

\emph{Dynamical critical sensing.} It is important to distinguish between the formal scaling of the QFI and the resource accounting of a complete sensing protocol. Near a critical point, the QFI may show super-Heisenberg scaling with system size, namely $\mathcal{F}$ scales as $N^{\beta}$ with $\beta > 2$~\cite{Boixo07,Boixo08,Mirkhalaf20,Yang22,he2023,montenegro2024review,Sahoo25,Mondal25b,cheng2025,Sarkar25,Debnath25b,debnath2026,Agarwal25b,Yousefjani25,Sahoo26}. 
Such scaling is not, by itself, inconsistent with fundamental bounds, because the relevant constraint is set by the norm of the generator and the interrogation time, rather than by N squared alone~\cite{Giovannetti06,Boixo07,Zwierz10,Zwierz12,Pang14,Pang17,rams2018}. In QMB systems, especially those with extensive local energy scales or spatially structured generators, the generator norm itself can acquire a nontrivial system-size dependence. As a result, an apparent super-Heisenberg scaling of the QFI can still fully respect the bound set by the generator norm.

Thus the operationally relevant question is whether this scaling survives once all relevant resources are included, particularly the time required to prepare the many-body state. In adiabatic critical protocols, the closing of the many-body gap near criticality leads to critical slowing down, which can reduce or even eliminate the advantage suggested by the static QFI alone. The practical relevance of apparent super-Heisenberg quantum sensing based on the QFI and the quantum Cramér--Rao bound has been widely investigated, with several works concluding that strategies yielding sub-Heisenberg scaling are not operationally ineffective~\cite{giovannetti2012}. However, in certain systems, including localization-based settings such as Stark-localized models, super-Heisenberg scaling can persist even after accounting for state-preparation time, while still respecting the generator-norm bound~\cite{yang2023,he2023,Yousefjani25,Debnath25b,Sahoo26}. In this context, e.g., Ref~\cite{Gietka2021} argues that although adiabatic critical quantum metrology can offer enhanced scaling through criticality, but once the time required for adiabatic preparation is included, it cannot generally reach the Heisenberg limit, even with shortcuts to adiabaticity. Therefore, super-Heisenberg scaling should not be regarded as intrinsically unphysical. Rather, the relevant generalized Heisenberg limit should be understood in terms of the available generator norm, interrogation time, and state-preparation cost. Considering the state preparation time, one can obtain the following result~\cite{rams2018}: {\it For adiabatic critical quantum metrology, even if the ground-state quantum Fisher information displays super-Heisenberg scaling, the inclusion of the time required to prepare the critical probe state restores the operational precision bound to the Heisenberg limit.} The outline of the proof as follows:

Near criticality, the many-body excitation gap closes as \( \Delta \sim L^{-z}\sim N^{-z/d}\), where \(N=L^d\) with \(d\) being the dimension of the system. Adiabatic preparation of the critical ground state therefore requires a minimum evolution time scaling as
\[
t_{\mathrm prep}\sim \Delta^{-1}\sim N^{z/d}.
\]
The QFI for the parameter \(\chi\) to be estimated generated by the evolution of the local Hamiltonian  satisfies
\[
\mathcal F \le 4t^2 \|\partial_\chi H\|^2.
\]
Since \(\|\partial_\chi H\|\sim N\), it immediately implies \(\mathcal F \lesssim t^2N^2\). Consequently, the quantum Cram\'er-Rao bound yields
\[
(\Delta\chi)^2 \ge \frac1{t^2N^2},
\]
which is precisely the Heisenberg limit. Thus, although criticality enhances the ground-state fidelity susceptibility, the divergent preparation time compensates this enhancement and prevents any super-Heisenberg scaling~\cite{Gietka2021}.

\qed

The above discussion is examplified by considering the one-dimensional spin-$1/2$ XXZ chain in a transverse magnetic field,
\[
H(\lambda) = -\sum_{i=1}^{N} \left(\sigma_i^x\sigma_{i+1}^x + \sigma_i^y\sigma_{i+1}^y + \Delta \sigma_i^z\sigma_{i+1}^z\right) + \lambda \sum_{i=1}^{N} \sigma_i^x ,
\] where $\Delta$ denotes the anisotropy parameter and $\lambda$ is the external field to be estimated. Close to the quantum critical point the fidelity susceptibility at criticality scales as $\chi_F \sim N^{2/(d\nu)}$,
where $d$ is the spatial dimension and $\nu$ is the correlation-length exponent defined through $\xi \sim |\lambda-\lambda_c|^{-\nu}$. Consequently, the estimation error inferred from the ground-state overlap scales as $\delta\lambda \sim N^{-1/(d\nu)}$, which may suggest an apparent super-Heisenberg scaling when $1/(d\nu)>1$. However, preparing the ground state near the critical point requires adiabatic evolution with a timescale set by the inverse energy gap, $\Delta E \sim L^{-z}$, where $z$ is the dynamical critical exponent and $L\sim N^{1/d}$ is the linear system size. The required adiabatic time therefore scales as $t \sim L^{z} \sim N^{z/d}$. Once this preparation time is included as a physical resource, the achievable precision obeys the bound $\delta\lambda \gtrsim 1/N$, restoring the standard Heisenberg scaling (see Fig.~\ref{fig:time_scaling}). Thus, although criticality enhances the sensitivity of the ground state through diverging correlations, it does not ultimately allow one to surpass the fundamental metrological limit when all relevant resources are properly accounted for the system under consideration. 

A recent development in quantum metrology is the recognition that quantum criticality can enhance parameter-estimation precision through nonequilibrium dynamics, without relying on ground-state preparation. A dynamic framework for criticality-enhanced quantum sensing was recently developed by Chu \textit{et al.}~\cite{chu2021}, where the parameter sensitivity is amplified through nonequilibrium evolution near a quantum critical point. Considering a parameter-dependent Hamiltonian $\hat H_\chi=\hat H_0+\chi \hat H_1$, the authors exploit the closed operator algebra generated by $\{\hat H_1,\hat C,\hat D\}$ with $\hat C=-i[\hat H_0,\hat H_1]$ and $\hat D=-\tfrac12[\hat H_\chi,[\hat H_0,\hat H_1]]$, which allows an analytic evaluation of the parameter generator. In this framework the excitation gap scales as $\epsilon\sim\sqrt{\Delta}$, where $\Delta(\chi)$ vanishes at the critical point $\chi_c$, signaling a quantum phase transition. The transformed local generator governing the parameter encoding takes the form
\begin{equation}
h_\chi = \hat H_1 t
+\frac{\cos(\sqrt{\Delta}\,t)-1}{\Delta}\hat C
-\frac{\sin(\sqrt{\Delta}\,t)-\sqrt{\Delta}\,t}{\Delta\sqrt{\Delta}}\hat D ,
\end{equation}
which diverges as $\Delta\rightarrow0$ for $\sqrt{\Delta}\,t\sim\mathcal{O}(1)$, indicating the emergence of critical quantum dynamics. Consequently, the QFI  exhibits a strong enhancement,
\begin{equation}
\mathcal{F}_\chi(t)\simeq
4\,\frac{\left[\sin(\sqrt{\Delta}\,t)-\sqrt{\Delta}\,t\right]^2}{\Delta^3}
\,\mathrm{Var}_{|\Psi\rangle}(\hat D),
\end{equation}
implying a dramatic increase in sensitivity near the critical point. As a concrete example, the authors analyze the one-axis twisting Hamiltonian $\hat H_\chi=\lambda \hat S_z^2+\chi \hat S_x$ describing collective spins called as LMG model, where the closing of the excitation gap at the critical coupling leads to a rapid growth of the QFI during the nonequilibrium evolution. This result demonstrates that critical quantum dynamics can substantially enhance metrological sensitivity without requiring the preparation of the ground state at the critical point, thereby providing a practical route toward criticality-enhanced quantum sensing. Further development followed  in this direction, e.g., in system with atom-photon and optomechanical systems interaction~\cite{Tang2023,zhu2024,hotter2024,Gietka2022,yuan2025}. Several protocols have been proposed to achieve time-scaling advantage in quantum systems~\cite{garbe2022}.

\emph{Floquet driving.} From the perspective of utilization of time as a resource in quantum sensing, 
periodic or Floquet driving schemes are recently being proposed for designing a new class of efficient quantum sensing devices. 

Several works use such driving tools to accelerate the process and to restore the quantum scaling that 
would otherwise be lost due to environmental effects or due to partial accessibility of the systems. This motivation paves a new way to utilize the Floquet engineering, where time-periodic driving is exploited as an active metrological resource. One of the pivotal works 
in this direction demonstrates that the well-known no-go theorems in noisy quantum metrology can be circumvented through periodic driving~\cite{Bai23}. Specifically, 
applying a suitable Floquet drive to atoms in a Ramsey spectroscopy setup enables the recovery of the ideal \( t^2 \) scaling of the QFI. This enhancement occurs when a Floquet bound state is formed between each driven atom and its local environment, effectively mitigating the decoherence-induced degradation of sensitivity that typically leads to an unfavorable \( t^{-1} \) scaling. Furthermore, by incorporating optimal control techniques, this approach can also restore the Heisenberg limit with the number of atoms \( N \), thereby offering a powerful strategy for quantum-enhanced sensing. To investigate the role of Floquet driving in quantum many-body sensing, 
a partially accessible integrable \( XY \) spin-chain system (where only local reduced density matrices are allowed for measurements) is considered and it is shown that the local steady states of the driven system recover the lost metrological advantage \cite{Mishra21}. Specifically, 
by applying a suitably engineered periodic transverse field and exploiting the system's local steady states,  it is shown that whenever the Floquet quasienergy gap closes, the reduced states exhibit not only restored Heisenberg limited sensitivity but, in certain regimes, even super-Heisenberg scaling. Importantly, this enhancement persists across broad regions of the phase diagram rather than being restricted solely to criticality, thereby offering a practical alternative to ground-state critical sensing without requiring full state tomography.


On the other hand, GHZ state preparation can be made faster by Floquet engineering. In Ref.~\cite{Ma25}, innovative strategies are proposed to enhance quantum metrology by generating non-Gaussian states through Floquet-engineered dynamics. In particular,  by applying periodic driving fields, specifically two-axis twisting (TAT) and turn dynamics, to an ensemble of \(N\) particles, it is possible to produce non-Gaussian states that exhibit QFI scaling as \(N^2\), approaching the Heisenberg limit. This scaling is achieved within an optimal time that decreases logarithmically with \(N\), specifically \(t_{\mathrm{opt}} \propto \ln{N}/N\). These results also highlight that  Floquet engineering can substantially reduce the dynamical cost of preparing highly sensitive quantum probes. 


\emph{Time crystal.} Periodic driving naturally connects to the rapidly developing ideas of Floquet time crystal that spontaneously breaks the discrete time-translation symmetry of the Hamiltonian~\cite{Wilczek12,Li12,Else16,Yao17,Zhang17,Choi17}.
Time crystal phases stabilize robust subharmonic temporal response, and the translational-symmetry-breaking turns out to be a potential resource for many-body quantum sensing. Recent works have proposed to exploit the time-crystal phase for engineering robust dynamical sensing devices that can overcome the shot-noise limit~\cite{Yousefjani2025_a,sahoo2025_qb, iemini2024}.

Several theoretical works have shown that time-crystalline order can enhance the QFI due to long-range spatiotemporal correlations and extended interrogation times, enabling sensitivities beyond the standard quantum limit \cite{Montenegro2023,Cabot2023,Yousefjani25}. More precisely, in Ref.~\cite{Yousefjani25},  a clean, disorder-free periodically driven quantum system can host a robust discrete time-crystal (DTC) phase, which maintains long-lived subharmonic oscillations under periodic driving. Crucially, this phase enables highly sensitive detection of the system's spin exchange coupling, with sensitivity sharply dropping when the system transitions out of the DTC phase. Through finite-size scaling analysis, the transition is shown to be of second order. Interestingly, the sensing protocol remains effective across a variety of initial conditions and can even benefit from imperfections in the driving sequence. Further, boundary time crystals-dissipative many-body phases exhibiting persistent oscillations have been shown to possess 
favorable QFI scaling with both the number of spins and measurement time~\cite{Montenegro2023,Cabot2023}. Further studies suggest that such systems can act as sensors for weak AC fields and environmental perturbations, where the metrological performance originates from multipartite correlations and long-lived oscillatory dynamics \cite{dominic2025}. In addition, recent proposals demonstrate that dissipative time crystals in periodically driven Rydberg systems can provide multiple time-crystal phase boundary~\cite{Jiao2025}.

Let us illustrate a recent work~\cite{Ullah25e} in which a periodically driven three-qubit quantum system is considered, and its dynamics are governed by a Floquet protocol. Each driving period \(\mathcal{T}\) consists of two sequential operations: (i) a global transverse magnetic field pulse of strength \(h_x\) applied along the \(x\)-axis for a duration \(\mathcal{T}_1\), followed by (ii) an Ising interaction of strength \(J\) applied for the remaining duration \(\mathcal{T}_2 = \mathcal{T} - \mathcal{T}_1\). The corresponding Floquet operator describing one complete driving cycle is given by
\begin{equation}
    \hat{U}_F = e^{-i \hat{H}_z \mathcal{T}_2}\, e^{-i \hat{H}_x \mathcal{T}_1}.
\end{equation}
Here, the transverse-field Hamiltonian and the Ising interaction under periodic boundary conditions are given by
\begin{align*}
    \hat{H}_x &= h_x \left( \hat{\sigma}^1_x + \hat{\sigma}^2_x + \hat{\sigma}^3_x \right),\\
    \hat{H}_z &= J_1 \hat{\sigma}^1_z \hat{\sigma}^2_z + J_2 \hat{\sigma}^2_z \hat{\sigma}^3_z + J_3 \hat{\sigma}^1_z \hat{\sigma}^3_z,
\end{align*}
respectively, with \(J_\alpha\) (\(\alpha=1,2,3\)) denoting the coupling strengths between the qubits.
In the periodic driving (PD) regime (a phase that exhibits period-doubling oscillations), the system exhibits enhanced sensitivity to the Ising coupling strengths while simultaneously suppressing sensitivity to the transverse magnetic field. Conversely, in non-PD regimes, the dynamics are more favorable for precise estimation of the transverse-field strength.

Apart from DTC, the existence of boundary time-crystal (BTC) is very surprising where without Floquet driving, along with the environment, causes in stabilizing new dynamical phase~\cite{iemini2018}. A BTC is a special type of time-crystal that forms not in the bulk of a system, but at its boundary or edge. It exhibits periodic time-translation symmetry breaking only at the boundary, while the bulk of the system remains in a symmetric or trivial phase. BTCs often arise in open systems coupled to external environments or reservoirs, where dissipation plays a constructive role, stabilizing the oscillations at the boundary.
In that respect, it is found that a sensing protocol can be based on BTC for detecting weak, oscillatory AC fields~\cite{dominic2025}. This setup consists of a one-dimensional spin chain with the bulk coupled strongly to a dissipative environment, while the boundary spin is weakly coupled and coherently driven which is given as
\begin{equation}
\frac{d\rho}{dt} \;=\; -i\omega_0\left[  S^x,\rho\right] \;+ \frac{\kappa}{N/2} \left( 2 S^- \rho S^+ -\{S^+ S^-,\rho\} \right),
\label{eq:btc_lindblad}
\end{equation}
where the $N$ spins are driven coherently with strength $\omega_0$ and decay collectively with a rate $\kappa$ rescaled by $N/2$ and  \( \hat{S}^\alpha = \sum_{i} \frac{\hat{\sigma}^\alpha_i}{2},\) collective spin operators with \( \hat{S}^\pm = \hat{S}^x \pm i \hat{S}^y \). Analyzing the behavior of QFI, the study suggests that such a BTC can function as a sensitive probe for weak oscillatory AC fields where the maximum QFI (\(F^*\)) scales quadratically achieving HL (see Fig.~\ref{fig:btc_sensor}), highlighting the potential of BTCs as metrological resources in open-system quantum sensing. Apart from this, a recent work~\cite{hotter2024} has shown that driven-dissipative critical quantum sensors can combine the advantages of conventional dynamical metrological strategies with critical enhancement~\cite{Ding2022,ilias2022,Ilias2024,cabot2024,yang2023,yang2025}. In particular,  the interplay between eigenstate changes near criticality and dynamical phase accumulation generates
an additional contribution to the Fisher information, leading to the enhancement of parameter sensitivity.
Overall, these studies present time-crystalline phases not just as a theoretical
curiosity, but as a viable resource for robust, quantum-enhanced metrology.

\begin{figure}
    \centering
    \includegraphics[width=\linewidth]{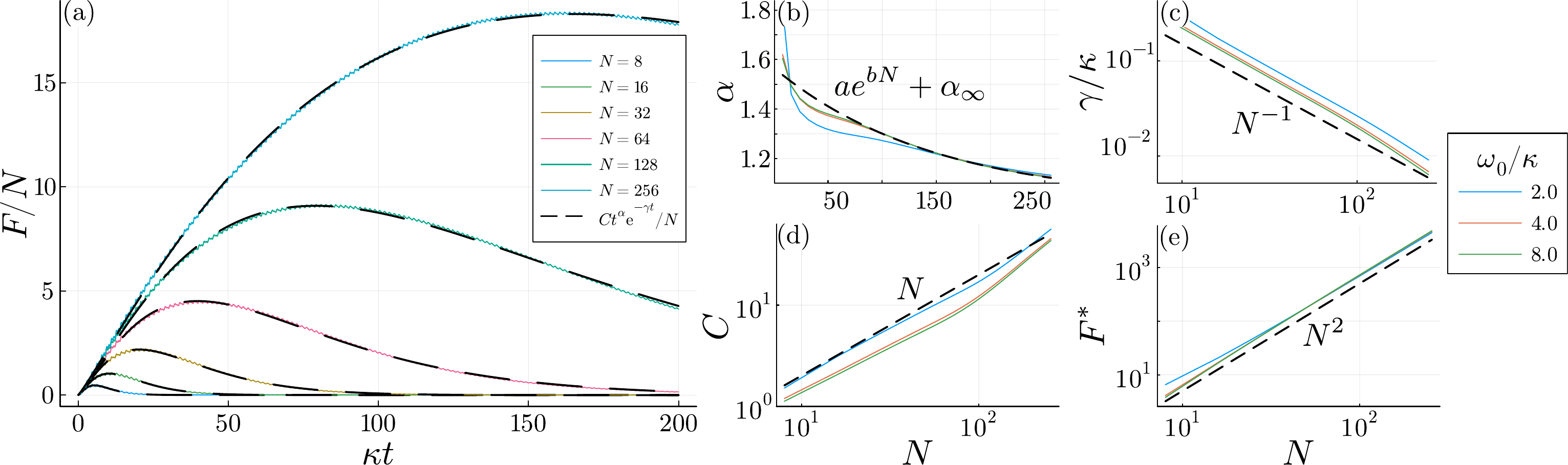}
    \caption{\textcolor{black}{(a) QFI as a function of time \(t\) for different system sizes. The QFI is well described by the fitting form \(F\sim C\, t^{\alpha}e^{-\gamma t}\), where \(C\), \(\alpha\), and \(\gamma\) are fitting parameters. The scaling of the fitting parameters with system size is plotted in (b), (c) and (d). The corresponding maximum QFI, denoted by \(F^{*}\) and shown in panel (e), exhibits a quadratic scaling advantage, indicating quantum-enhanced sensing performance. The figure is taken from Ref.~\cite{dominic2025}.}}
    \label{fig:btc_sensor}
\end{figure}

\emph{Counter-diabatic (CD) sensing.} Following the advancement of adiabatic quantum sensing protocols, an important development has been made through counter-diabatic or transitionless quantum driving, which enables a quantum system to follow the instantaneous eigenstates of a time-dependent Hamiltonian within a finite evolution time. In conventional adiabatic evolution, the Hamiltonian \(H_0(t)\) must vary sufficiently slow in order to suppress diabatic transitions between instantaneous eigenstates. CD driving circumvents this limitation by introducing an auxiliary control Hamiltonian \(H_{\mathrm{CD}}(t)\), such that the total Hamiltonian becomes \(H(t)=H_0(t)+H_{\mathrm{CD}}(t)\). For an instantaneous eigenstate \(|n(t)\rangle\) of \(H_0(t)\), the counter-diabatic term is given by
\[
H_{\text{CD}}(t) = i\hbar \sum_n \left( |\partial_t n(t)\rangle \langle n(t)| - \langle n(t) | \partial_t n(t) \rangle |n(t)\rangle \langle n(t)| \right).
\]
By suppressing nonadiabatic excitations, CD driving provides an efficient route for rapid quantum state preparation and therefore appears naturally suited for quantum sensing protocols, which  require highly sensitive probe states. However, it has been shown recently that shortcut-to-adiabaticity protocols based on CD driving cannot overcome the fundamental metrological scaling bounds imposed by quantum critical dynamics~\cite{Gietka2021}. By investigating models such as the Landau--Zener and quantum Rabi models,  it is demonstrated that although CD driving accelerates the preparation of near-critical probe states, the associated time cost still limits the achievable QFI. Consequently, the Heisenberg limit cannot be surpassed through CD-assisted critical metrology, and under realistic constraints such as finite evolution time and decoherence, noncritical sensing strategies may even outperform critical protocols.  

Finally, it is worth emphasizing that in the presence of \(k\)-body interaction without decoherence, nonlinearity can enhance the precision scaling of the QFI  to $t^2N^{2k}$~\cite{Boixo07,Luis07,Boixo08,Choi08,Napolitano16,Gross10,Hall12,Joo12,Sewell14}.  Moreover, when $k$-body Hamiltonian dynamics is affected by $l$-body noise, the condition of linear dependence limits the QFI scaling to at most $tN^{2k -l}$, demonstrating how multipartite decoherence can suppress the metrological advantages offered by nonlinear quantumm evolution~\cite{Beau17,rafal17}.

Furthermore, several additional theoretical and experimental developments in the direction of quantum many body sensing have been reported in Refs.~\cite{Lenef97,Wildermuth05,Canuel06,Wildermuth06,Aigner08,Esteve2008,Debs11,Zhou12,Steinke13,Eto13,Ng13,Muntinga13,Aguilera14,Hudelist2014,Carraz14,Bina16,Muessel14,Liu16,Jasperse17,Cheiney18,Apellaniz18,Barrett19,Condon19,Trimeche19,Yang2020,Gersemann20,Szigeti2020,carcy2021,Ufrecht21,Alvarez22,Mishra22,Chen22,iemini2024,Castanet24,Saleem24,Ostermann24,Bate25,Mihailescu25,LeDesma25,Stolzenberg25,Cassens25,Zhang2026,Wang2026_a,Stas2026}. 
 
\subsection {Quantum imaging and illumination}
\label{sec:qimageqillum}

Quantum imaging and illumination represent closely related sensing applications in which quantum resources are used to extract spatial, spectral, or object-level information from weak signals. In quantum imaging, nonclassical light improves contrast, resolution, or noise performance in forming images beyond classical limits. Quantum illumination, in turn, exploits quantum correlations to improve target detection in noisy backgrounds, even when the signal returning from the object is very weak. Together, these protocols extend quantum metrology from parameter estimation in idealized interferometers to realistic sensing tasks involving images, spectra, and low-reflectivity objects.

\subsubsection{Quantum imaging} 
\label{subsec:imaging}

In classical imaging, resolution is limited by diffraction. For a circular aperture, the Rayleigh criterion gives the minimum resolvable angular separation as
\begin{equation}
\delta\theta_R \simeq 1.22\,\frac{\lambda}{D},
\end{equation}
where $\lambda$ denotes the wavelength of the incident light and $D$ is the aperture diameter. The image of a point source is described by the point-spread function (PSF), e.g., for a Gaussian approximation,
\begin{equation}
\mathrm{PSF}(x)=\frac{1}{\sqrt{2\pi}\sigma}\exp\!\left(-\frac{x^2}{2\sigma^2}\right),
\end{equation}
where $\sigma\sim \lambda/\mathrm{NA}$ sets the diffraction-limited width. On the other hand, quantum imaging~\cite{Lugiato02,Shih07,Genovese16} refers to a broad class of imaging techniques that exploit genuinely quantum properties of light, such as entanglement, squeezing, or higher-order correlations, to enhance imaging performance beyond what is achievable with classical imaging and detection schemes. 

Imaging can be formulated as estimating parameters (e.g., source separation $d$) from measurement outcomes with probabilities $p(i|d)$ (see Fig.~\ref{fig:quantum_imaging} for the corresponding schematic illustration). In that case, FI is calibrated, and the Cramér-Rao bound is given as \(\Delta d \ge {1}/{\sqrt{N F(d)}}\) with \(F\) being the FI associated with a specific measurement strategy and $N$ repetitions, but the ultimate bound is set by the QFI, given as
\begin{equation}
\Delta d \ge \frac{1}{\sqrt{N\,\mathcal{F}(d)}},
\end{equation}
where \(\mathcal{F}(d)\) is defined as the QFI. Broadly speaking, quantum imaging is expected to outperform classical imaging in two distinct and conceptually independent ways. First, it can enable the resolution of spatial features below the classical diffraction limit, commonly referred to as sub-Rayleigh or super-resolution imaging. Second, it can improve the scaling of estimation precision with photon number by employing nonclassical states of light.  From a wave-optical perspective, this enhancement can be interpreted as the collective quantum state exhibiting an effective wavelength reduced by a factor of $N$, thereby enabling finer spatial sensitivity.
For direct imaging of two incoherent point sources separated by $d$, the FI behaves as \(F_{\mathrm direct}(d)\propto d^2 \quad \text{as } d\to 0\), leading to a diverging uncertainty $\Delta d\sim 1/d$ which is called as the ``Rayleigh's curse''. In contrast, the QFI remains finite, \(\mathcal{F}(d\to 0)=\mathrm{const},\) indicating that the loss of resolution is measurement-induced rather than fundamental.

\textit{Sub-Rayleigh quantum imaging.}
A central concept in classical optical imaging is Rayleigh’s criterion~\cite{Rayleigh1879,Born99}, which provides a heuristic resolution limit based on diffraction. According to this criterion, two incoherent point sources are considered resolvable only if their separation exceeds the characteristic width of the point-spread function (PSF) of the imaging system. When the separation falls below this diffraction-limited scale, the intensity patterns of the two sources strongly overlap, rendering them difficult to distinguish using conventional image-plane intensity measurements.

A major conceptual advance, in this direction, has been made, in which the resolution problem from the perspective of quantum parameter estimation is examined~\cite{Tsang16}. By modeling the optical field at the image plane as a quantum state whose density operator depends on the source separation,  the QFI associated with estimating the separation is shown to remain finite even in the limit of vanishing distance between two equally bright incoherent sources. In stark contrast, the FI obtained from direct intensity measurements goes to zero in this limit, giving rise to the so-called Rayleigh’s curse. This result shows that the apparent loss of resolvability at small separations is not due to a fundamental absence of information in the optical field, but rather to the inadequacy of conventional measurement strategies that probe only spatial intensity. To provide a concrete and experimentally feasible approach that attains the quantum limit, spatial mode demultiplexing (SPADE), a measurement scheme is proposed~\cite{Tsang16} that resolves the optical field into an appropriate spatial-mode basis before photon detection. SPADE can, in principle, achieve the optimal QFI for estimating the separation, thereby overcoming Rayleigh’s curse. Physically, direct imaging fails at small separations because it is insensitive to the subtle changes in the mode structure of the field that encode separation information, whereas SPADE accesses this information through linear optical preprocessing. This insight has stimulated extensive theoretical~\cite{Kolobov2000,Tsang11,Gottesman12,Lupo16,Rehacek17,Ang17,Tsang15,Tsang17,Larson18,Napoli19,Bisketzi19,Tsang19,Zhou19b,Khabiboulline19,Datta20,Lupo20,Sorelli21,Kurdzialek21,Almeida21,Liang21,Wang21,Bojer22,Karuseichyk22} and experimental~\cite{Paur16,Backlund18,Boucher20,Pushkina21,Sorelli21b,Mazelanik22} works, significantly generalizing the original framework.

\textit{Ghost imaging.} 
Among the earliest examples of quantum imaging is ghost imaging~\cite{Klyshko88a,Klyshko88b,Klyshko88c,Pittman95} in which an image of an object is reconstructed using intensity or coincidence correlations, even though the detector that directly interacts with the object lacks spatial resolution. Typically, a source produces two correlated beams: a probe beam that illuminates the object and is measured by a bucket detector without spatial resolution, while a reference beam that never interacts with the object but is detected by a spatially resolving detector. The image is then reconstructed by correlating the signals from the two detectors. Although ghost imaging,  which was originally proposed using entangled photon pairs, is regarded as a paradigmatic quantum imaging protocol, it was later shown that similar imaging capabilities can be achieved with classically correlated light. Thus, ghost imaging underscores the central role of correlations in image formation while remaining an important bridge between classical and quantum imaging paradigms.
\begin{figure}
    \centering
   \includegraphics[width=1.0\linewidth]{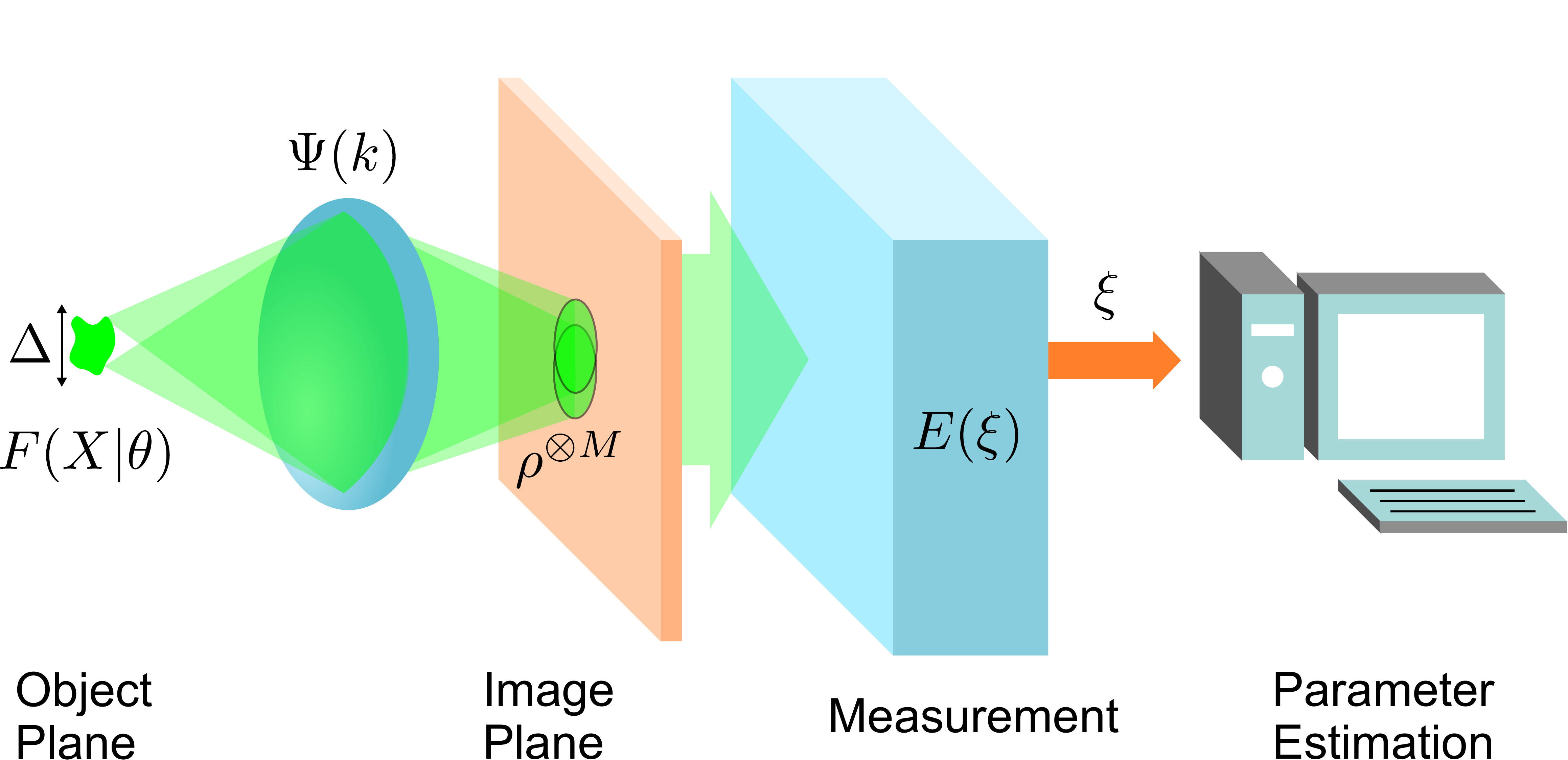}
 \caption{Quantum imaging from a quantum metrology perspective~\cite{Tsang19}.
Schematic of a far-field incoherent optical imaging system. The object is described by an intensity distribution
\(F(X|\theta)\), characterized by parameters \(\boldsymbol{\theta}\) (e.g., a characteristic width or source separation).
The imaging system is defined by its optical transfer function \(\Psi(k)\).
The optical field on the image plane is described by the quantum state \(\rho^{\otimes M}\), corresponding to \(M\) independent temporal modes. A generalized measurement, represented by a POVM \(E(\xi)\), is performed on the field, yielding measurement outcomes \(\xi\), from which the object parameters are estimated.}
\label{fig:quantum_imaging}
\end{figure}

\textit{Sub-shot noise imaging.}
Another important direction in quantum imaging is the suppression of noise below the shot-noise limit. In conventional imaging, photon-number fluctuations impose a fundamental constraint on the detectability of weakly absorbing or weakly scattering objects, especially under low-illumination conditions where increasing the photon flux may be undesirable or impossible. By employing nonclassical states of light, such as squeezed states, it is possible to reduce the relevant noise below the shot-noise level, thereby enhancing image contrast and sensitivity.  For example,  intensity squeezing has been investigated to improve the signal-to-noise ratio in imaging by lowering the noise in photon detection below the shot-noise limit~\cite{Heidmann87,Jedrkiewicz04,Treps02,Treps03,Blanchet08,Boyer08,Brida09,Taylor13,Taylor14,Dowran18}.
Sub-shot-noise imaging thus addresses a complementary aspect of quantum-enhanced imaging performance, focusing not on spatial resolution but on improving signal-to-noise ratios in photon-limited regimes.

\subsubsection {Quantum illumination} 
\label{subsec:illumination}

Quantum illumination (QI) ~\cite{Lloyd08,Shapiro20,Karsa24} is a technique for detecting the presence or absence of a weakly reflecting object (the target) embedded in a noisy thermal background using quantum resources.  It can be regarded as a subset of the broader concept of quantum radar. In the standard setting, a nonclassical bipartite quantum state is prepared, consisting of two modes: the signal and the idler. The idler is kept intact and sent directly to the detector, while the signal is transmitted toward the region of interest. If a target is present (Hypothesis 1 (\(H_1\))), the signal reflects weakly from the target and is further degraded by a thermal background (noise) before reaching the detector; if no target is present (Hypothesis 0 \(H_0\)), the signal passes only through the thermal background. At the detector, an optimal joint measurement is performed on the signal and idler, and from the measurement outcomes, the presence or absence of the target is inferred from the minimum possible error probability. 

\begin{figure}
    \includegraphics[scale=0.3]{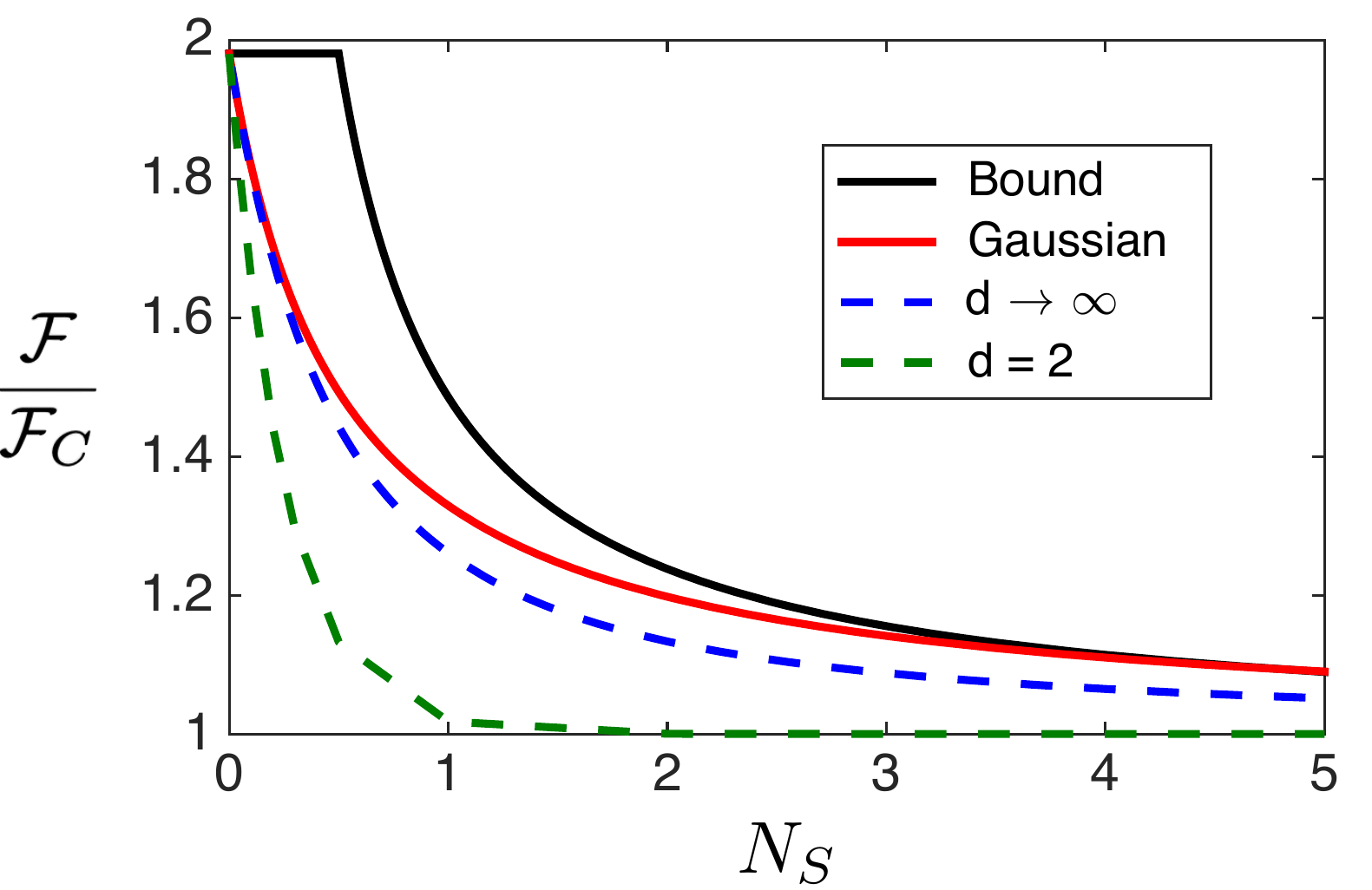}
   \caption{Quantum illumination: QFI-based enhancement of detection performance~\cite{Sanz17}. The horizontal axis represents the signal photon number $N_S$, while the vertical axis shows the ratio $\frac{\mathcal{F}}{\mathcal{F}_C}$, where $\mathcal{F}$ denotes the QFI for a given input state and $\mathcal{F}_C$ is the optimal QFI achievable using over all classical states. Since the upper bound on the optimal error probability in quantum illumination is determined by the QFI, this ratio quantifies the entanglement-induced advantage of the protocol. The QFI gains obtained using Gaussian states and Schrödinger cat states are shown by the red solid and dashed lines, respectively, whereas the optimal gain is indicated by the black solid line. Both classes of states exhibit their maximum advantage in the low-photon regime.}
\label{fig:QI}
\end{figure}

To formalize the QI problem, we assume that the probe light is initially prepared in an arbitrary two-mode entangled pure state,  $\rho_{SI}\!=\!\vert\psi\rangle_{SI}\langle\psi\vert$,
where the subscripts $S$ and $I$ denote the signal and idler modes,
respectively. The advantage of QI lies in the photon correlations between these modes, hence $\vert\psi\rangle_{SI}$ must be nonseparable. In contrast, classical illumination (CI) is realized by discarding the idler mode, so that the probe is described by a single signal-mode state, e.g., $\rho_{S}\!=\!\vert\psi\rangle_{S}\langle\psi\vert$. The ambient environment is modeled as a thermal state,
\begin{equation}
\rho_{B}=\sum_{n=0}^{\infty}\varrho_{n}\vert n\rangle_{B}\langle n\vert,\label{eq:thermalstate}
\end{equation}
with $\varrho_{n}\!=\!N_{B}^{n}/\!\left(N_{B}+1\right)^{n+1}$ representing
the thermal distribution with an average photon number $N_{B}$. Thus, the overall initial state is given by $\rho_{SI}\!\otimes\!\rho_{B}$.
The low-reflectivity target is modeled as a beam splitter represented by the unitary operator
\begin{eqnarray}
U\!\left(\eta\right) & = & \exp\!\left[\eta\!\left(a_{S}^{\dagger}b-a_{S}b^{\dagger}\right)\right],\label{eq:beam-splitter}
\end{eqnarray}
where $a_{S}$ and $b$ are the annihilation operators for the signal
and environment modes, respectively. When the reflectivity ($\eta\!\ll\!1$) is low, the reflectivity of the target  reads as $R\!=\!\sin^{2}\eta\!\sim\!\eta^{2}$. When
the initial state $\rho_{SI}\!\otimes\!\rho_{B}$ evolves under $U\!\left(\eta\right)$
and the signal mode is traced out, the resulting state is
\begin{equation}
\begin{aligned}
\rho_{RI}^{(1)}
&\equiv \rho_{RI}(\eta) \\
&= \operatorname{Tr}_{S}\!\left[
U(\eta)\,\rho_{SI}\!\otimes\!\rho_{B}\,U^{\dagger}(\eta)
\right].
\end{aligned}
\label{eq:eta-density-matrix}
\end{equation}
This state, defined under $H_{1}$ (object present), is the reduced density matrix for the reflected and idler modes. In the absence of the object $\left(\eta=0\right)$, the state becomes
\begin{equation}
\rho_{RI}^{(0)}
\equiv
\left.\rho_{RI}(\eta)\right|_{\eta=0}
=
\rho_{I}\otimes\rho_{B},
\end{equation}
with $\rho_{I}={ \Tr}_{S}\!\left(\rho_{SI}\right)$.
The new notations
$\rho_{0}\!\equiv\!\rho_{RI}^{\left(0\right)}$ and $\rho_{1}\!\equiv\!\rho_{RI}^{\left(1\right)}$ are introduced for convenience. An important observation from this formulation is that determining whether the object is present reduces to a state discrimination problem between $\rho_0$ and $\rho_1$.

The performance of any quantum illumination protocol is typically characterized using the \emph{quantum Chernoff bound (QCB)}~\cite{Audenaert07}. Helstrom~\cite{Helstrom69} originally derived the optimal measurement strategy and the corresponding minimum error probability for distinguishing between two quantum states, $\rho_0$ and $\rho_1$, prepared with prior probabilities $p_0$ and $p_1$ (with $p_0 + p_1 = 1$), when only a single copy of the state is available for measurement. The quantum Chernoff bound generalizes this result to the asymptotic setting, where many copies of the states are provided and joint measurements are performed on the tensor product of $M$ copies. In this regime, the minimum error probability decays exponentially with the number of copies as 
\[
P_{\text{err,min}} \;\sim\; e^{-M \zeta_{\text{QCB}}},
\]  
where 
\[
\zeta_{\text{QCB}} \coloneqq - \min_{0 \leq s \leq 1} \, \log \, \Tr\!\left(\rho_0^s \rho_1^{\,1-s}\right)
\]  
is known as the \emph{quantum Chernoff exponent} or \emph{collective decay constant}.

\textit{Connection with QFI:} In the QI scheme, an upper bound on the optimal error probability, $P^{\text{I,II}}_{\text{err}} (\mathcal{F})$, can be obtained as a function of the QFI corresponding to the reflectivity parameter, \(\eta\) for any input state~\cite{Sanz17} which is given by 
\begin{equation}
\label{QFI-bound-illumination}
    P^{\text{I,II}}_{\text{err}} (\mathcal{F}) \sim \exp\Big[-\frac{\eta^2 \mathcal{F} M}{8}\Big].
\end{equation}
 Moreover, using the QFI associated with the reflectivity parameter and the above formula, it has been shown that an entangled transmitter for a fixed number of photons can reduce the error probability by up to \(3\) dB compared to the optimal classical strategy in the low-reflectivity limit~\cite{Sanz17} (see Fig.~\ref{fig:QI}).
The quantum advantage is maximal in the low-photon regime and decreases at least inversely with the number of signal photons. This QFI-based bound is obtained by estimating the reflectivity amplitude of the target within the quantum metrology framework.

It is worth noting that the Chernoff bound is a tighter bound compared to the QFI-based bound. However, note that the Chernoff bound is obtained by optimizing over global measurements, whereas the bound in the QI protocol based on QFI (in Eq.~\eqref{QFI-bound-illumination}) is obtained from local measurements. Given that the implementation of a global measurement is a nontrivial task in experiments, both the bounds in QI are useful.

There exists another bound on the minimum error probability in the QI protocol, which is also obtained from local measurements. It is based on signal-to-noise ratio and is given by 
\begin{equation}
    P_{\text{err}} (R) = \frac{1}{4} \exp(-MR^2/2).
\end{equation}
$R \coloneqq \frac{\mu_1 - \mu_0}{\sigma_1 +\sigma_0}$  is the known SNR 
with $\mu_x \coloneqq \langle \hat{O} \rangle_{\rho_x}$ and $\sigma_x \coloneqq \langle \hat{O}^2 \rangle_{\rho_x} - \langle \hat{O} \rangle_{\rho_x}^2$. Here $\rho_{0} (\rho_1)$ corresponds to the state when the target is absent (present). $\hat{O}$ denotes the operator corresponding to the collective measurement acting on the joint signal-idler state reflected from the target to discriminate between these two states $\rho_0$ and $\rho_1$.  This SNR-based bound indicates that a higher SNR corresponds to a lower error probability.
This bound is obtained by using the state discrimination theory of discriminating two states, $\rho_0$ and $\rho_1$.  Thus, optimizing the strategy for distinguishing between $\rho_0$ and $\rho_1$ reduces to finding the optimal measurement $\hat{o}$ that maximizes the SNR.

The connection between the state discrimination theory and parameter estimation theory has also been established by unveiling the relationship between SNR and QFI under the assumption of low object reflectivity when measurements are local~\cite{Zhong25}. 
This connection provides a relation between $P_{\text{err}} (R)$ and $P_{\text{err}} (\mathcal{F})$ which is given by 
\begin{equation}
    P_{\text{err}} (\mathcal{F}) \leq P_{\text{err}} (R)
\end{equation}
and the equality is achieved only when 
\begin{equation*}
    \frac{R}{\eta} = \frac{\sqrt{\mathcal{F}}}{2}.
\end{equation*}
Note that the equality condition provides a clear operational criterion for identifying the optimal measurement that gives the optimum value of $P_{\text{err}} (R)$ when optimized over all possible measurements. Moreover, there exist hierarchies,
\begin{equation}
    P_{\text{err}} (\mathcal{F}) \leq P_{\text{err}} ({\mathscr{F}}) \leq P_{\text{err}} (R), 
\end{equation}
where $\mathscr{F}$ denotes the CFI. This hierarchy encapsulates the successive optimizations needed to saturate the ultimate error probability bound $P_{\text{err}} (\mathcal{F})$. The
first equality is achieved when $\mathscr{F} = \mathcal{F}$, which requires optimization over all projective measurements, and the
second equality holds when $ \frac{R}{\eta} = \frac{\sqrt{\mathcal{F}}}{2}$, necessitating an
optimal estimator, such as the maximum likelihood estimator. Thus, this connection not only enriches the theoretical link between state
discrimination and parameter estimation but also provides a clear operational criterion for identifying optimal measurements in target detection tasks. Subsequent studies have further extended the QI protocols, including adaptive schemes, non-Gaussian Probes, and multimode entangled transmitters.~\cite{Casariego22,Gupta2024,Wei24,Zhao25}.

\subsection {Further connections with quantum information protocols}
\label{subsec:future-connections}

The applications of quantum metrology stretch wider to further areas of quantum information. In this subsection, we discuss the connection between quantum metrology and other well-known areas of quantum information, such as quantum teleportation and quantum walks.

\subsubsection{Quantum teleportation of unknown parameter}

Quantum teleportation~\cite{Bennett93, Pirandola15} is a quantum information protocol that enables the transfer (and not transport) of an unknown quantum state from one location (attended by ``Alice'') to a distant one (operated by ``Bob''), by using shared entanglement and classical communication, with its performance being typically quantified by the fidelity between the input and the teleported state. However, there exist physical scenarios where the goal is not to transmit the entire state but rather specific information about parameters encoded in the input state. In such cases, the reliability of the protocol can be quantified using the QFI corresponding to the parameter(s) of the teleported state. Under the influence of amplitude damping noise acting on Bob's part of the shared entangled state~\cite{Xiao16},  the combination of a prior partial measurement by Alice and the partial measurement reversal by Bob, performed before the final unitary operations, is shown to significantly enhance the teleportation QFI, effectively mitigating, almost completely, the detrimental effects of amplitude damping noise.
Furthermore, several works~\cite{Metwally17,El19,Guo19,Jafarzadeh20,Seida20a,Seide20b,Li23b,Huang24} have explored quantum teleportation of transmitting the information about specific parameter(s) encoded in the input state, with the performance quantified by the QFI associated with the parameter(s) of the teleported state.

\subsubsection{Quantum walks as metrological probes} 

Quantum walks~\cite{Aharonov93,Kempe03,Venegas-Andraca12,Portugal13,Xia20b}, 
the quantum counterparts of classical random walks, exist in two main forms: (i) the discrete-time quantum walks (DTQWs)~\cite{Aharonov93} and (ii) the continuous-time quantum walks (CTQWs)~\cite{Farhi98,Mulken11}. 

In DTQW, the dynamics are governed by two subsystems: a walker and a quantum coin. At each time step, a ``coin toss" (mathematically, a unitary coin operation) is applied to the coin. Depending on the resulting coin state, the walker’s position is updated. Both time and  position are discrete. A key feature of DTQWs is that the standard deviation of the walker’s position grows ballistically with the number of steps, in contrast to the diffusive spread of a classical random walk. Variants such as the split-step quantum walks (SSQWs) have also been widely studied, with applications in quantum algorithms, quantum computation, and the simulation of diverse quantum phenomena. Within the context of quantum metrology, most DTQW-based studies have focused on estimating the coin parameter(s) using single- and multiparameter estimation frameworks~\cite{Rajauria20,Annabestani22,Cavazzoni24,Moradi26}.
For instance, Ref.~\cite{Cavazzoni24} has shown that the dimensionality of the coin space serves as a resource for enhancing the precision of coin-parameter estimation in DTQWs. The study considers various choices for the quantum coin, including general unitary operators and the generalized Grover coin. More recently, DTQW-based metrology has been extended to quantum magnetometry, topology-optimized probing, and split-step-walk-enhanced estimation protocols~\cite{Shukla24,Cavazzoni24b,Moradi26}.

By contrast, CTQWs
feature only one degree of freedom and describe the continuous-time evolution of a quantum particle across a discrete set of positions governed by a Hamiltonian. Similar to DTQWs, CTQWs display ballistic spreading with time.
In the quantum metrology setting, research on CTQWs has primarily concentrated on estimating Hamiltonian parameters based on the probability of finding the particle at various sites over time for a given initial state~\cite{Seveso19,Gianani23}. 
For example, CTQWs on several classes of graphs, including complete, cycle, complete bipartite, hypercube, and star graphs are analyzed, aiming to estimate the tunneling amplitude between nodes present in Hamiltonian using quantum parameter estimation techniques based on the probabilities of finding the particle at various nodes over time~\cite{Seveso19}. Precisely, it studies the scaling of the optimal QFI with the number of nodes and  the corresponding optimal measurements and initial states.
Complete position measurements are often optimal, except in the case of star graphs, where no advantage over local measurements is found. Even incomplete or nearly local position measurements are able to extract nonzero information, with their efficiency (the ratio of FI to QFI) closely tied to the fraction of graph nodes under experimental control. In this case as well, the star graph again stood out as an exception, since monitoring only the central node yields the same information as monitoring all nodes individually.
In addition,  the estimation problem of multiple Hamiltonian parameters in CTQWs has been studied through a deep neural network trained on the probabilities of finding the particle at various sites and times~\cite{Gianani23}.  Apart from that, Ref.~\cite{Zatelli20} recast scattering in a continuous-time quantum walk as a quantum metrology task for estimating the height of a lattice impurity.

\subsection{Quantum sensing in dark matter search}
\label{subsec:particle-physics}

An important frontier of quantum sensing lies in gravitation, high-energy physics, and beyond-standard-model searches, where quantum metrological tools are employed to detect extremely weak signals such as gravitational waves and ultralight dark-matter interactions.  While detection of gravitational wave are covered in Sec.~\ref{subsec:inferometric-metrology} and~\ref{subsec:photonic-sensors}, below we provide a brief review on dark matter sensing. Quantum sensing has also been investigated as a promising route toward dark-matter detection. Ultralight dark-matter candidates such as axions, axion-like particles, and dark photons can generate extremely weak forces, phase shifts, or excess noise signals that are detectable using quantum-enhanced metrological techniques. Mechanical quantum sensors based on optomechanical resonators, levitated particles, torsion balances, and interferometric accelerometers can operate close to the standard quantum limit, allowing coherent accumulation of weak dark-matter-induced perturbations over long timescales~\cite{Carney2021}. Quantum resources such as squeezing, entanglement, and backaction-evading measurements can substantially enhance the QFI and thereby improve sensitivity to weak dark-matter couplings~\cite{zheng2016,malnou2019,Dixit2021,Backes2021,brady2022}. A major recent development has been the formulation of dark-matter detection as a quantum channel estimation and quantum noise sensing problem~\cite{Shi2023}. In microwave axion haloscopes, the weak axion-photon interaction appears as an excess bosonic noise signal, and entanglement-assisted sensing protocols based on two-mode squeezed vacuum states were shown to outperform both classical detection schemes and single-mode squeezing strategies~\cite{Shi2023}. More recently, distributed quantum sensor networks have been proposed for directional dark-matter searches~\cite{Fukuda2025}. In such protocols, coherent phase differences between spatially separated sensors encode information about the velocity and direction of the dark-matter ``wind,'' enabling simultaneous estimation of both coupling strength and directional properties without sacrificing intrinsic sensitivity. Together, these developments establish quantum sensing as a promising bridge between quantum metrology, gravitation, and particle physics.

\section{\textbf{Quantum sensing in physical platforms}}
\label{sec:experiment}

Having explored the foundational aspects of quantum metrology together with a wide range of sensing protocols,  we now turn to the physical platforms that implement these schemes in laboratories. Throughout the previous sections, 
we have discussed a range of sensing protocols for the precise estimation of several physical parameters, including AC and DC fields, temperature, frequency, coupling strengths, chemical potential, and other quantities of interest. These examples illustrate how quantum resources can be exploited not only to improve measurement precision but also to design parameter-specific sensing strategies adapted to different physical settings. This naturally calls for the integration of theoretical principles with experimentally viable platforms. The ultimate goal is therefore the  realization of quantum sensors across diverse physical architectures, including photonic systems, ultracold atoms and molecules, trapped ions, solid-state spin defects, and superconducting circuits. Each platform, however, comes with its own unique advantages and constraints: photonic systems are naturally suited for interferometry, imaging, communication-compatible sensing, and continuous-variable protocols, but often face challenges related to loss and efficient detection; ultracold atoms and trapped ions offer long coherence times and exquisite controllability, though they typically require complex experimental infrastructure; solid-state spin defects such as nitrogen-vacancy centres provide nanoscale spatial resolution and operation under relatively practical conditions, but may be limited by material disorder, decoherence, and readout efficiency; superconducting circuits allow strong coupling, fast control, and quantum-limited microwave measurements, while requiring cryogenic environments and careful noise mitigation schemes. Therefore, the unification of quantum-metrological theory with platform-specific strengths and limitations is essential for identifying where genuine quantum advantage can be robustly realized in practical sensing applications, such as precision measurement, imaging, spectroscopy, atomic clocks, magnetometry, thermometry and field sensing.

\begin{figure*}
\includegraphics[width=\linewidth]{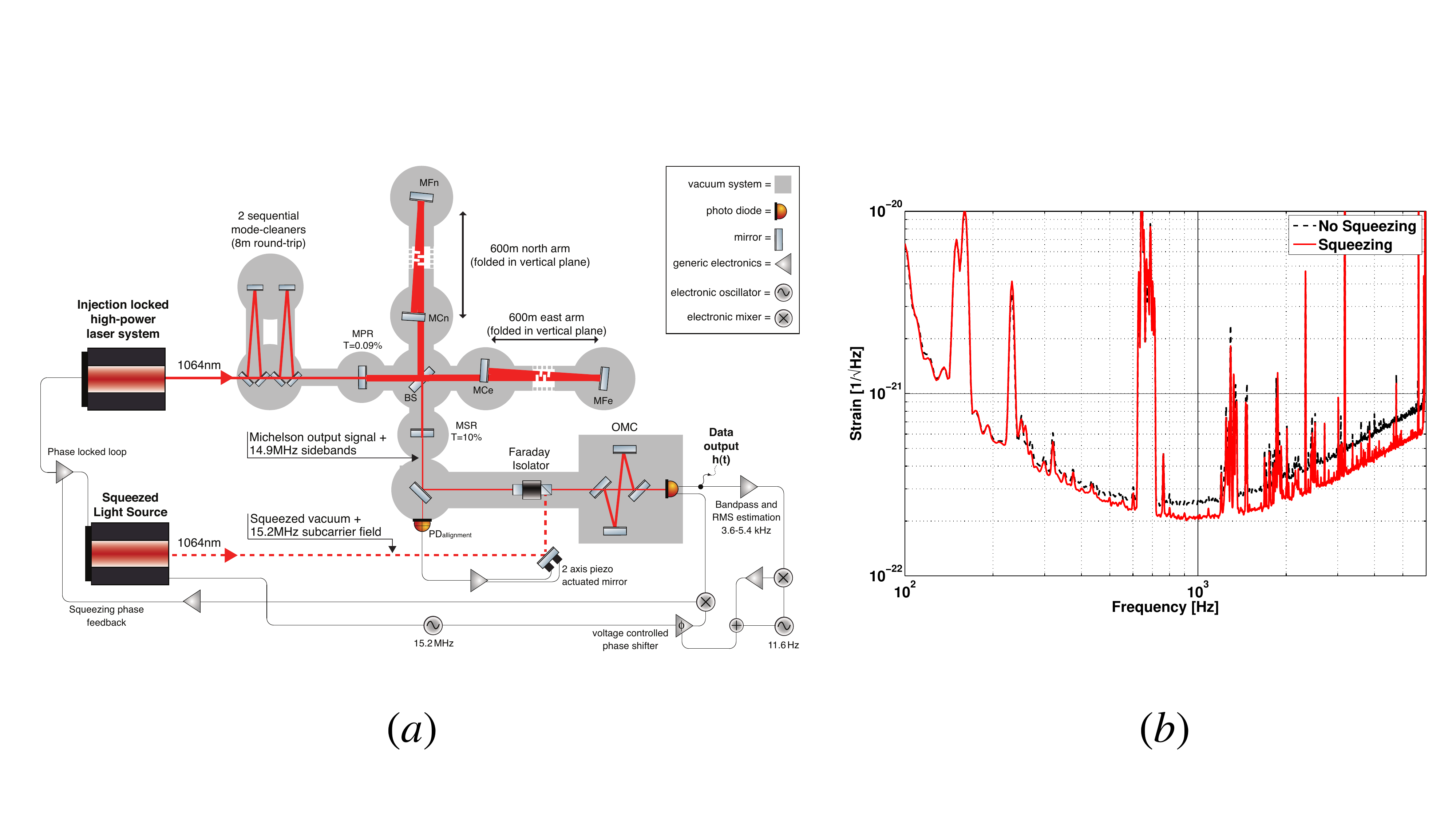}
   \caption{(a) A simplified optical layout of the squeezed-light enhanced gravitational wave detector GEO 600, which consists of the conventional GEO 600 observatory and the additional squeezed-light source. The observatory has two singly folded arms with a total optical length of 2400 m. A gravitational wave passing from most directions will shorten one arm, while the length of the perpendicularly orientated arm is increased, and vice versa in the next half-cycle of a passing wave, producing a periodic power change of the output light that is detected by a photodiode. (b) Strain amplitude spectral densities of GEO 600 with and without squeezing being applied is plotted~\cite{Grote2013}.}
\label{fig:LIGO1}
\end{figure*}

\subsection{Photonic sensors} 
\label{subsec:photonic-sensors}
Photonic platforms~\cite{Knill01,Brien09,Lemzini18,Slussarenko19,Wang20,Polino20,Ramakrishnan23,Riedel26} 
use single photons or continuous optical modes as carriers of quantum information, which can be manipulated using optical elements such as beam splitters, phase shifters, nonlinear crystals, waveguides, interferometers, single-photon detectors, and integrated photonic circuits. These platforms offer several advantages for quantum tasks. First, photons exhibit extremely low decoherence, making them ideal for quantum communication and long-distance entanglement distribution. Second, photonic systems can operate at room temperature, which simplifies experimental implementation. Third, their compatibility with optical fiber networks and integrated-chip architectures makes them attractive for scalable quantum networks and compact sensing devices. Due to these advantages of optical platforms, optical set-ups are well suited for phase estimation~\cite{Caves1981,noauthor11,Polino20},
imaging~\cite{Genovese16,Moreau19}, 
spectroscopy~\cite{Schlawin16,Dorfman16}, and interferometry~\cite{Caves1981,Abadie11,Aasi13,Schnabel17},
where nonclassical states of light, such as single photons, entangled photons, and squeezed states, can be used to reduce quantum noise below classical limits.

In optical interferometry, the injection of squeezed vacuum states into large-scale interferometers has enabled noise reduction below the shot-noise limit, establishing a practical route toward quantum-enhanced sensing in gravitational-wave detection~\cite{Caves1981}. 
This idea has become one of the most successful examples of quantum-enhanced sensing in a macroscopic measurement device. In the context of gravitational-wave detection, squeezed-light injection is first implemented at the observatory scale in GEO600, demonstrating operation beyond the quantum shot-noise limit~\cite{Abadie11, Grote2013} (see Fig. \ref{fig:LIGO1}). 
Subsequently, Aasi \textit{et al.}~\cite{Aasi13} has reported the large-scale implementation of quantum squeezing to enhance the sensitivity of a gravitational-wave detector. In this work, a squeezed vacuum state of light is injected into the dark port of the LIGO interferometer to reduce quantum shot noise, which limits high-frequency sensitivity. By lowering the quantum fluctuations in the measured quadrature of the optical field, the experiment achieves a broadband noise reduction of about $2-3$ dB in the shot-noise-dominated frequency range, corresponding to a measurable improvement in strain sensitivity, and thus demonstrates a fully realized experimental realization on a kilometer-scale interferometer.  It represents an important milestone in quantum metrology, demonstrating enhanced gravitational-wave sensitivity using a squeezed vacuum optical system integrated into a large-scale interferometer, and paving the way for the routine use of squeezing in advanced gravitational-wave observatories.
Furthermore, tabletop CV optical experiments using homodyne detection and adaptive feedback have demonstrated optimal phase-estimation strategies, highlighting the role of Gaussian-state engineering and measurement optimization in realistic noisy settings~\cite{Yonezawa2012,Andersen2016CVReview}.

Additionally, optical platforms have provided direct experimental demonstrations of quantum-enhanced sensing across several representative tasks. In phase estimation, entangled multiphoton states and optimized photonic probes have been used to beat the SQL, with demonstrations ranging from four-photon interferometry to \(\mathrm{N00N}\)-state and adaptive phase-estimation protocols~\cite{Nagata07,Afek10,Daryanoosh18}.
In quantum imaging, spatial correlations of photon pairs and twin beams have enabled sub-shot-noise imaging and enhanced imaging of weakly absorbing or low-light samples, showing that quantum correlations can improve the signal-to-noise ratio beyond classical illumination strategies~\cite{Brida10,Moreau19,Zhang24c}. 
In quantum spectroscopy, experiments with entangled photon pairs and quantum-light sources have begun to probe molecular and material responses through entangled two-photon absorption, coincidence-based spectroscopy, and ultrafast Raman schemes, illustrating how quantum correlations can provide new time-frequency and signal-scaling advantages in spectroscopic measurements~\cite{Eshun22,Moretti23,Fan24}.
These experimental developments establish photonic systems as a practical testbed for translating quantum-metrological resources into phase, image, and spectroscopic measurements. Together, these experimental developments on optical setups establish a unified framework in which squeezing, entanglement, and optimized quadrature measurements form the backbone of modern quantum-enhanced metrology. 

However, photonic platforms have certain crucial limitations. In particular, photon-photon interactions are intrinsically weak, which makes the realization of deterministic two-qubit gates challenging. As a result, many entangling operations in linear optical systems are typically probabilistic and often require auxiliary photons, adaptive measurements, and feed-forward control. Losses, imperfect sources, finite detector efficiencies further constrain scalability. Although integrated photonic circuits have significantly improved stability, miniaturization, and circuit complexity, scaling photonic architectures toward scalable quantum sensors
remains experimentally demanding.


\subsection{Ultracold atoms, molecules and ion-based sensors} 

Complementary to photonic sensors, where quantum enhancement is primarily achieved through nonclassical states of light and optimized optical measurements, ultracold atoms, molecules, and trapped ions provide a broad class of highly controllable massive-particle platforms that can be exploited for precision measurements ~\cite{Leibfried03,Bloch08,Lewenstein07,Carr09,Cronin09,Lewenstein12,Bohn17,Pezze18,Safronova18,Bruzewicz19}. Although they differ in their microscopic structures and in the details of experimental realizations, they share several common features that make them particularly attractive for quantum metrology: long coherence times, well-resolved internal transitions, high degrees of isolation from environmental noise, and precise control using laser, microwave, and electromagnetic trapping fields. Neutral atoms are especially well suited for clocks, interferometers, optical lattices, and many-body sensing; trapped ions provide exceptional state preparation and readout fidelity together with strong confinement and controllable interactions; and ultracold molecules offer rich internal structure and large electric dipole moments.

Ultracold atomic systems are particularly versatile because low-temperature trapping techniques allow dilute gases of bosons and fermions to be prepared in highly controllable quantum states in optical dipole traps, magnetic traps, optical lattices~\cite{Lewenstein07} and leads to the quantum simulations of paradigmatic many-body Hamiltonians, such as Bose Hubbard~\cite{Jaksch98,Greiner02}, Fermi Hubbard H~\cite{Hofstetter02,Jordens08,Schneider08} systems and spin-chains~\cite{Duan03,Trotzky08,Simon11}. 
The flexibility of this platform has enabled the exploration of a broad range of equilibrium and nonequilibrium phenomena, including superfluid-Mott-insulator transitions~\cite{Jaksch98,Greiner02},
BEC-BCS crossover physics~\cite{Nozieres85,Regal04,Zwierlein05}, localization~\cite{Anderson58,Billy08,Roati08}, 
quantum scar~\cite{Bernien17,Turner18a, Turner18b,Bluvstein21}, many-body dynamics, and thermalization~\cite{Trotzky12}, Kibble-Zurek mechanism~\cite{Pyka2013}. More recently, synthetic gauge fields~\cite{Lin09,Aidelsburger13,Miyake13}, 
spin-orbit coupling~\cite{Lin11},
Floquet engineering~\cite{Struck11,Goldman14,Jotzu14,Meinert16,Eckardt17}, tunable long-rangle interaction that can be used for preparing metrologically useful entangled states \cite{Flores25,kim2009,Islam2013,Richerme2014,Jurcevic2014,Piltz2016,monroe2021} and synthetic dimensions~\cite{Celi14,Arguello-Luengo24} have extended the scope of ultracold platforms toward  geometrically engineered matter, providing new avenues for realizing topological and other exotic quantum phases~\cite{Mancini15,Stuhl15,Salamon20a,Salamon20b,Salamon22}.

Cold and ultracold molecules complement atomic gases by offering additional rotational, vibrational, hyperfine, and electronic degrees of freedom, together with large electric dipole moments. These features make molecular platforms especially attractive for precision spectroscopy~\cite{Carr09,Safronova18},
controlled quantum chemistry~\cite{Ospelkaus10,deMiranda11},
dipolar many-body physics. Trapped ions provide a complementary architecture with long-lived internal states, strong Coulomb-mediated interactions, and high-fidelity state preparation and readout~\cite{Leibfried03,Haffner08,monroe2021}. 
Moreover, ultracold atoms and trapped ions also provide highly controllable architectures for quantum information processing and high fidelity quantum gates~\cite{Cirac95,Monroe95,Jaksch2000,Saffman10,Bruzewicz19}. Hence, cold atoms, molecules and ions, all together, provide powerful settings for quantum sensing, since atom interferometry, phase transitions, quantum correlations, topology, localization, quantum scar, etc., can be exploited for designing various kinds of quantum sensing devices~\cite{Agarwal25b}.

A broad range of quantum-sensing tasks, such as estimation of timekeeping, weak forces, magnetic and electric fields, acceleration, gravity, tests of relativity, searches for variation of fundamental constants, dark-matter searches and rotation measurements, are amenable to the atomic, molecular, and ionic platforms. Neutral atom lattice clocks and trapped-ion clocks are the most precise timekeeping devices~\cite{Chou10,Bloom14,Ludlow15}.  Light-pulse atom interferometers~\cite{Kasevich1991,Cronin09,Bongs2019,Salvi23,Brown25} have been exploited for precision measurements of acceleration, gravity, rotations. Cold atomic ensembles, spinor BECs, vapor-cell magnetometers, lattice gases and ion lattices are used for magnetic-field sensing and magnetic microscopy~\cite{Vengalattore2007,Budker2007,Kitching2011}. Rydberg atoms, polar molecules and trapped ions are extremely sensitive to electric fields~\cite{Ni2010,Sedlacek2012,Holloway2014,Gilmore2021}. BECs, trapped atoms and ions are excellent force and displacement sensors~\cite{Biercuk2010,Blms2018,Gilmore2021}. Atomic, molecular and ionic spectroscopy provide especially powerful and promising settings for investigating parity violation, variation of constants, and dark-matter searches~\cite{Safronova2018,Baron2014,acms2018}. They directly enable quantum-state-controlled chemistry and dipolar collision studies~\cite{Ospelkaus2010,deMiranda2011}. Apart from that, as mentioned before, cold gases in optical lattices and trapped-ion array experiments have already established the necessary ingredients, such as criticality, correlations, entanglement, and many-body spin dynamics, for many-body-enhanced sensing. The explicit demonstrations of all these exotic quantum phenomena promise the emergence of new kinds of robust QMB-based quantum sensing devices.

However, these platforms also have certain constraints. In particular, gate operations are relatively slow, which limits the speed of digital quantum computing. Furthermore, the experimental infrastructure is large and complex, typically requiring ultra-high vacuum systems and sophisticated laser cooling setups. 

\begin{figure}
    \centering
    \includegraphics[width=\linewidth]{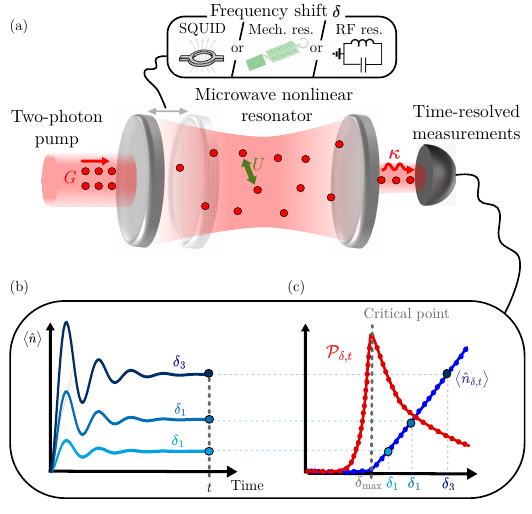}
    \caption{A schematic illustration of a frequency-estimation protocol based on a nonlinear microwave resonator with Kerr nonlinearity~\cite{Beaulieu25}. The cavity frequency, and hence the cavity--pump detuning \(\delta\), is modified through changes in the effective cavity length, which can be experimentally controlled using SQUID-based architectures. The system dynamics are governed by the Hamiltonian and master equation \( H = \delta \hat a^\dagger \hat a + \frac{U}{2}\hat a^\dagger \hat a^\dagger \hat a \hat a + \frac{G}{2}(\hat a^\dagger \hat a^\dagger + \hat a \hat a), \) and \( \frac{\partial \rho}{\partial t} = -i[H,\rho] + \frac{\kappa}{2} \left( 2\hat a \rho \hat a^\dagger - \{\hat a^\dagger \hat a,\rho\} \right), \)
respectively. Panel (b) shows the reconstruction of the intracavity photon number through measurements of photon loss from the cavity field. Panel (c) illustrates the corresponding photon-number fluctuations (red dots), which become maximal near the critical point associated with the second-order dissipative finite-component phase transition of the device.}
\label{fig:superconduting_sensing}
\end{figure}

\subsection{Superconducting qubit-based sensors}

Superconducting qubits~\cite{makhlin2001,You2005,devoret2004,Clarke2008,Devoret2013,Wendin2017,Krantz2019,Kjaergaard2020} provide another complementary solid-state platform for quantum sensing and metrology ~\cite{Clarke2008,Devoret2013,Danilin2024,Danilin2018,Wang2019,CastellanosBeltran2008,Bergeal2010,Abdo2011,Macklin2015}.
These systems are realized in superconducting electrical circuits operating at millikelvin temperatures that incorporate nonlinear elements such as Josephson junctions. Various circuit design goals can be achieved by combining inductors, capacitors, and
Josephson junctions. Superconducting qubits can be fabricated using standard microfabrication techniques, enabling on-chip integration and the development of scalable quantum processors. Further, this architecture allows strong and controllable interactions between qubits and microwave photons.  In this context, the microwave photons are not external optical modes, but on-chip electromagnetic modes of superconducting resonators or transmission lines. Circuit quantum electrodynamics (QED) is therefore a native superconducting-circuit architecture: resonators can be used for dispersive readout, microwave control, and bus-mediated coupling among many transmons, while in resonator arrays, the same modes may become the simulated photonic or polaritonic degrees of freedom~\cite{Blais2021}. Combined with circuit QED, fast microwave control, tunable couplers, parametric driving, and engineered dissipation, superconducting circuits offer a highly programmable setting for implementing effective Hamiltonians. Moreover, fast gate operations and high-fidelity control through microwave pulses make them suitable for implementing quantum gates and complex quantum circuits, and hence advocate them as an attractive  platform for quantum computing, quantum simulation and quantum metrology tasks.

This controllability makes the superconducting systems suitable for realizing a wide range of effective QMB models, along with various equilibrium and nonequilibrium quantum phenomena. They include  Jaynes-Cummings and quantum-Rabi physics~\cite{Wallraff2004,Niemczyk2010,Braumuller2017},
spin chains~\cite{Heras2014,Salathe2015} and Hubbard models~\cite{Leib2010,Angelakis2007,Mirhosseini2018,Greentree2006,Ma2019,Karamlou2024}. Using superconducting qubits, Roushan \emph{et al.} have simultaneously demonstrated  synthetic magnetic fields and strong particle interactions~\cite{Roushan2016}. Further works for generating synthetic magnetic fields, topological phases and for engineering flat bland and other exotic band structures have been carried out~\cite{Carusotto2020,tan2018,Wang2016}. Apart from that, various quantum phenomena, such as the Kibble-Zurek mechanism~\cite{Gong2016}, dynamical quantum phase transition~\cite{Xu2020}, driven-dissipation phase transition~\cite{Carmichael2015,Fink2017}, quantum walks~\cite{Yan2019,Gong2021} can be probed in superconducting platforms. Moreover, superconducting processors have been used to emulate many-body localization and Stark localization, for example, the works carried out by Guo \emph{et al.}, Xu \emph{et al.} 
and others~\cite{Xu2018,Guo2021_a,Yao2024}. In principle, superconducting systems offer an advanced physical system for mimicking a large pool of effective quantum many-body models and for simulating a broad class of quantum phenomena~\cite{Blais2021,Gu2017,Carusotto2020,LeHur2016,Yao2024,Lamata2018}.

Historically, superconducting circuits were first developed and widely used as precision measurement devices rather than as quantum-computing platforms. In particular, Josephson junctions based superconducting quantum interference devices (SQUIDs) became among the most sensitive detectors of magnetic flux and magnetic field, with applications in magnetometry, biomagnetism, materials characterization, and low-noise readout~\cite{squid2004,Kleiner2004,Fagaly2006,makhlin2001,Ilichev2007,Bal2012,Danilin2024}.  The same Josephson nonlinearity is later repurposed to engineer leading solid-state architecture for quantum information processing and computing~\cite{makhlin2001,You2005,Clarke2008,Devoret2013,Blais2021}. Josephson photon counters, artificial $\Lambda$-systems, and transmon-based sensors have been developed for detecting individual photons with very low energy~\cite{Chen2011,Inomata2016,Opremcak2018,Wang2021}. Inomata \emph{et al.} has demonstrated the detection of a propagating single microwave photon with an artificial $\Lambda$-system~\cite{Inomata2016}, while
quantum microwave radiometry using photon-induced dephasing of a superconducting qubit is developed~\cite{Wang2021}. The Axion and hidden-photon dark-matter searches look for extremely weak microwave signals, and correspondingly, superconducting sensing has become a strong motivation for the dark matter searches~\cite{Dixit2021,Backes2021,Jawell2023,Braggio2025}. Further, it is demonstrated that quantum-limited microwave amplification is essential for superconducting sensing because weak microwave fields must be measured before thermal and amplifier noise dominate~\cite{Yurke1989,CastellanosBeltran2007,Bergeal2010,Macklin2015,Roy2016}. Superconducting microwave cavities and SQUID-based resonators can be coupled to mechanical degrees of freedom, via which force sensing can be performed~\cite{Teufel2011}. A criticality enhanced, SQUID based leaky-cavity resonator has been developed for precise measurment of photon number with underlying Kerr medium~\cite{Beaulieu25} (see Fig.~\ref{fig:superconduting_sensing}). Superconducting microwave optomechanics is used to detect and squeeze mechanical motion~\cite{Bothner2020} while a flux-tunable SQUID system is employed for remote sensing of a levitated superconductor~\cite{Schmidt2024}. Moreover, these qubits can also be used as spectroscopic probes because their phase, transition frequency, and relaxation or dephasing rates respond sensitively to weak fields, noise spectra, or coupled excitations~\cite{Sung2019,Wolski2020,Wang2022}.Furthermore, superconducting-circuit platforms have already demonstrated the essential ingredients for QMB-enhanced sensing, including strong light-matter coupling, engineered spin and Hubbard dynamics, synthetic gauge fields, driven-dissipative criticality, quantum walks, and localization-like nonequilibrium dynamics; these capabilities make them a promising solid-state route toward the scalable, chip-integrated newly emerging many-body quantum sensors.

At the same time, superconducting architectures present their own experimental and technological challenges. In particular, they typically suffer from shorter coherence times compared to atomic platforms due to interactions with the solid-state environment. In addition, maintaining the required cryogenic environment demands sophisticated dilution refrigeration systems. Also,
improving coherence, reducing noise, and achieving large-scale quantum devices remain important technological challenges for superconducting platforms.

\subsection{Solid-state quantum sensors: defects, spins, nuclei, and nanostructures}

Solid-state quantum sensors, in which quantum degrees of freedom are embedded in the host material, constitute a broad class of sensing platforms  and  are used to probe external perturbations. These include electronic spins associated with point defects and color centers, nuclear spins addressed through magnetic-resonance techniques, and confined electronic states in semiconductor nanostructures such as quantum dots, which together provide complementary routes toward local, scalable, and material-integrated quantum sensing of several quantities of interest. Solid-state quantum sensors offer a complementary set of advantages: unlike many cold-atom, trapped-ion, or superconducting platforms, these solid-state sensors can operate under ambient or near-ambient conditions, without requiring ultrahigh vacuum, laser cooling, or dilution refrigeration. Their active quantum degrees of freedom, such as defect spins, nuclear spins, or mesoscopic electronic states, can often be embedded directly in or placed very close to the target material, which is often crucial for local and nanoscale sensing of magnetic fields, electric fields, temperature, strain, pressure, and noise environments. In addition, these solid-state platforms are naturally compatible with microfabrication, chip integration, scanning-probe geometries, and operation in realistic condensed-matter, chemical, and biological environments. These features make them especially attractive as deployable sensing devices, although their performance is often limited by material disorder, surface noise, inhomogeneous broadening, and shorter coherence times compared with photonic and near-isolated systems~\cite{Chipaux15,Degen17}.

\subsubsection{Nitrogen-vacancy center (NV-center) sensors}

 \begin{figure}
\includegraphics[scale=1.3]{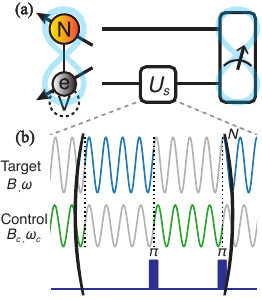}
\caption{Multiparameter sensing using a single NV centre~\cite{Isogawa26}. (a) Schematic diagram of the quantum-control sensing protocol which can estimate of both amplitude and frequency of linearly polarized AC magnetic fields. The electronic  and nitrogen nuclear spins are sensor and  auxiliary qubits respectively. In the scheme, Bell state is initially prepared, the unitary acts on the electron spin and finally Bell measurement is performed. 
(b) The sequence of interactions between the target and control fields with the sensor qubit is illustrated schematically. These interactions are applied in an alternating manner, while dynamical decoupling is realized through periodic \(\pi\)-pulse sequences, thereby defeating the 
unwanted Ising-type inetraction between the sensor and the auxiliary qubits. 
}
    \label{fig:2602.17648}
\end{figure}

Among solid-state sensors, NV-centers in diamond are especially prominent because of high controllability, optical readout, room-temperature operation, and nanoscale spatial resolution.
The negatively charged NV center hosts an optically addressable spin-triplet ground state that can be initialized and read out through spin-dependent fluorescence and coherently manipulated by microwave fields~\cite{Doherty13}.
NV-centers are particularly sensitive to magnetic fields, electric fields, temperature, strain, pressure, and nearby nuclear or electronic spins, and thereby are suitable for quantum sensing applications~\cite{Lesage12,Ledbetter12,Steinert13,Doherty13,Schirhagl14,Rondin14,Chipaux15,Degen17,Galazzo23}.

Single NV centers and NV ensembles have been used for nanoscale magnetometry and magnetic imaging \cite{Taylor08,Maze08,Balasubramanian08,Grinolds13,Jensen14,Degen17},
electric-field sensing~\cite{Dolde11}, 
and thermometry~\cite{Clevenson15}. In this context, Isogawa \textit{et al.}~\cite{Isogawa26} proposes and experimentally demonstrates a quantum control scheme for multiparameters that fully resolves the singularity in the QFIM, enabling simultaneous estimation of the amplitude and frequency of linearly polarized AC fields, shown in Fig.~\ref{fig:2602.17648}.  More specifically, their approach removes the singularity in the QFIM by modifying the time evolution directions of the corresponding generators, while preserving the optimal scaling behavior for each parameter. Consequently, the protocol achieves sensitivities close to the fundamental quantum limits in the long-time (high-frequency) regime. Further, near saturation of the quantum Cram{\'e}r–Rao bound in the phase estimation has been experimentally demonstrated in a  solid-state spin system  provided by a NV-center in diamond~\cite{Yu22}. Further works on NV-center-based quantum sensing include nanoscale thermometry, including in living cells~\cite{Kucsko13}, 
nanoscale nuclear magnetic resonance (NMR) and detection of external nuclear spins~\cite{Mamin13}, imaging of current flow and magnetic textures in materials~\cite{Tetienne15,Staudacher15,Chang17,Dovzhenko18}, 
and also biological and chemical sensing with nanodiamonds~\cite{McGuinness11,Fujiwara20}.
All these results establish  NV-centers as a particularly versatile platform for local quantum metrology, bridging atomic-scale spin control with practical nanoscale sensing.

\subsubsection{Magnetic-resonance sensing with nuclear and electronic spins}

As discussed in the preceding subsection, NV centers in diamond can be a versatile defect-spin platform for local quantum sensing. Magnetic resonance sensing is a closely related field that makes it possible to resolve the spatial texture of the nuclear or electronic spins, along with detecting a field amplitude. Nuclear-spin and magnetic-resonance approaches exploit the high spectral resolution and long coherence times of spin degrees of freedom for probing molecular structure and local spin dynamics.  In conventional nuclear magnetic  resonance (NMR) and electron spin resonance (ESR), the signals are usually detected from macroscopic spin ensembles and is constrained with severe sensitivity limitations for nanoscale samples. It is possible to overcome this limitation by using NV centers as atomically localized magnetic-resonance detectors: the NV electronic spin can be optically initialized and read out,
controlled by microwave pulses, and used to detect the fluctuating magnetic fields generated by nearby nuclear or electronic spins, as established in high-sensitivity early nanoscale NV-NMR experiments~\cite{Taylor08,Zhao12,Mamin13,Staudacher13}. These experiments demonstrate high-sensitivity diamond magnetometry, multipulse NV sensing, and remote nuclear-spin detection.

Subsequent experiments extended NV-detected magnetic resonance to single-spin sensitivity, enabling three-dimensional imaging of individual dark spin, nanoscale magnetic-resonance imaging with chemical contrast, and the detection of molecular dynamics near diamond surfaces~\cite{Grinolds14,Muller14,Haberle15,Staudacher15}. More advanced implementations have further combined quantum-logic protocols and high-field NV sensing to perform NMR detection and spectroscopy of individual proteins and nanoscale chemical species, while ensemble-NV sensors have facilitated high-resolution magnetic-resonance spectroscopy of statistically polarized nuclear-spin ensembles~\cite{Lovchinsky16,Aslam17,Glenn18}. Therefore, , NV-detected magnetic resonance  forms a bridge between conventional NMR/ESR spectroscopy and local solid-state quantum sensing, preserving chemical and spectral details while enabling nanoscale spatial resolution, and high to small-volume sensitivity under ambient or near-ambient conditions.

\subsubsection{Semiconductor quantum dots and other solid-state sensors}

Semiconductor quantum dots and mesoscopic solid-state devices~\cite{Kouwenhoven97,Kouwenhoven01,Hanson07,Burkard23}
are another important route to implement local quantum sensing, based on the confinement and control of individual charges and spins in nanoscale electronic structures. They are generally endowed with strong light–matter interaction enabling efficient single-photon generation, potential for spin-photon interfaces in hybrid quantum networks, compatibility with semiconductor fabrication for scalable integration, fast optical control of qubits, while their performances are influenced by charge noise and solid-state decoherence, typically requirement of cryogenic operation, shorter life time and material imperfections prohibiting uniformity and reproducibility.

The key gradients behind their sensing performance are Coulomb blockade, discrete energy levels, gate-controlled  tunnel coupling, and spin-to-charge conversion~\cite{Hanson07,Burkard23,Kouwenhoven01,korotkov1999,Petta05}. Experiments establish nearby quantum point contacts and sensor dots as ultrasensitive electrometers for detecting Coulomb-blockade charging and real-time single-electron tunneling~\cite{Field93,Schoelkopf98,Schleser04,Gustavsson06,Ihn09,Gustavsson09}.
These charge-sensing ideas are subsequently extended to spin-sensitive detection through spin-to-charge conversion, including single-shot electron-spin readout~\cite{Elzerman04},  
spin-dependent tunnel-rate readout~\cite{Hanson05}, 
coherent control of coupled electron spins~\cite{Petta05},  and fast RF charge and spin sensing~\cite{Barthel10}. In principle, readout of single spins provides a major progress towards solid state-based quantum technologies. Single-shot, time-resolved readout of an electron spin in silicon~\cite{Morello10} is reported with readout fidelity better than \(90\% \) and  a spin lifetime of $\sim 6$ seconds at a magnetic field of $1.5$ Tesla. Further, notable works include recent gate-based and scalable semiconductor sensing schemes~\cite{West2019,Nakajima2021,GonzalezZalba2015}. Overall, these advances show the potential of semiconductor quantum dots and mesoscopic devices as local electrometers, spin sensors, charge-noise probes, microwave detectors, and components of scalable quantum-classical measurement interfaces, , thereby supplying a crucial building block for solid-state quantum sensing and quantum information technologies~\cite{Reilly2015}.

\section{\textbf{Outlook}}
\label{sec:summary}

While the current spurt of activity in quantum metrology and sensing is rather recent, the underlying physical and mathematical foundations have been established  for quite a long while. 
However, recent advances in quantum technologies on experimental and theoretical fronts have led to unprecedented precisions in estimation of parameters in physical systems in diverse arenas. 

Although a lot of new techniques have been discovered in recent years, a lot remains to be uncovered. In particular, we have limited understanding as yet of the resources necessary for attaining the best precision allowed by quantum mechanics. This is especially blurred in the case of encoding parameters of non-unitary dynamics and open quantum systems. Noisy environments also pose a significant challenges, both with respect to modeling the relevant environment for a given physical platform and for finding the optimal sensing strategy under realistic noisy condition. 

Single-parameter sensing devices has achieved remarkable success but  have limited fundamental and commercial utility. Multiparameter quantum sensing, however, possesses fundamental as well as experimental difficulties, and  it remains  unclear what resources are necessary for the best precision or what probes allow nontrivial quantum-enhanced precision for a given set of parameters. Even for single-parameter sensing, the roles of quantum coherence, entanglement and other nonclassical resources, especially in high-dimensional  the probe states are less understood. Specifically,  an interesting direction in the case of qudits is to explore the role of the relative phase of the optimal probe state on the precision.

Another useful line of study is to find out if costly resources like entanglement in the probe states can be replaced by more economical ones like asymmetry to reach the competitive precision levels. The key point here is that while the best precision level is always great to achieve, near-best ones may also be enough for most jobs and may require an inexpensive and less demanding resources. 
The role of interactions among the probe parts of an estimation process is another relatively less-studied area, while being important, since in many physical substrates, the parts of the probe interact with each other. 
Closely related is the need to understand which many-body physical characteristics  are essential to achieve enhanced precision in estimating  physical quantities connected to many-body cooperative phenomena.

Last but not least, choosing the correct physical platform for estimating a particular set of physical parameters while keeping an eye on the resources necessary - in the probes, the encoder, and the measurement, and on the quality and quantity of noise acting on the entire process, is of crucial importance. It is important, e.g., to know if hybrid quantum systems can be advantageous in reaching high precision in a relatively noise-insulated way by combining the best of two worlds with which the hybrid system is created. 

\acknowledgements
We acknowledge the facility at Harish-Chandra Research Institute and all the collaborators who have worked on this topic.  ASD acknowledges support from the project entitled ``Technology Vertical - Quantum Communication'' under the National Quantum Mission of the Department of Science and Technology (DST)  (Sanction Order No. DST/QTC/NQM/QComm/$2024/2$ (G)). TKK and PG acknowledge ``INFOSYS scholarship for senior students''. The research of TKK was carried out and financed within the framework of the second Swiss Contribution MAPS (Grant No. 230870).

\bibliography{reference}
\end{document}